\documentclass[twocolumn,pra, aps,superscriptaddress,floatfix]{revtex4}

\usepackage{lineno}
\usepackage{mathptmx}
\usepackage{subfigure}
\usepackage{dcolumn}
\usepackage{amsmath,amssymb}
\usepackage{bm}
\usepackage{color}
\usepackage{overpic}
\usepackage{latexsym}
\usepackage{epstopdf}
\usepackage{color}
\usepackage[english]{babel}
\usepackage{amsmath}
\usepackage{latexsym}
\usepackage{stmaryrd}
\usepackage{bigints}
\usepackage{psfrag,graphicx} 
\usepackage{epsf} 
\usepackage{subfigure} 
\usepackage{amsmath} 
\usepackage{amssymb} 
\usepackage{amsfonts}
\usepackage{bm}
\usepackage{natbib}
\usepackage{epstopdf}
\usepackage{appendix}
\usepackage{relsize} 
\usepackage{amsmath}

\definecolor{mygrey}{gray}{0.35}
\definecolor{myblue}{rgb}{0.2,0.2,0.8}
\definecolor{myzard}{cmyk}{0,0,0.05,0}
\definecolor{mywhite}{rgb}{1,1,1}
\definecolor{myred}{rgb}{1,0.,0.3}
\definecolor{sigreen}{rgb}{0.4,0.8,0.4}

\usepackage[colorlinks=true,citecolor=myblue,linkcolor=myred]{hyperref}
\def\be{\begin{equation}}
\def\ee{\end{equation}}
\def\ba{\begin{align}}
\def\enda{\end{align}}
\def\bi{\begin{itemize}}
\def\ei{\end{itemize}}

 \def\ee{\mathord{\rm e}}
 
 \def\ii{\mathord{\rm i}}

\def\half{\textstyle\frac{1}{2}}

\def\fourth{\textstyle\frac{1}{4}}

 \def\ee{\mathord{\rm e}}
 
 \def\ii{\mathord{\rm i}}

\def\half{\textstyle\frac{1}{2}}

\def\fourth{\textstyle\frac{1}{4}}

\renewcommand{\ii}{{\rm i}}
\renewcommand{\ee}{{\rm e}}

\def\beq{\begin{equation}}
\def\beq{\begin{equation}}
\def\eeq{\end{equation}}

 \newcommand{\ket}[1]{|#1\rangle}
 \newcommand{\bra}[1]{\langle #1|}

\begin{document}


\title[Short Title]{Gross-Neveu-Wilson model and correlated symmetry-protected topological phases }

\author{A. Bermudez}
\affiliation{Departamento de F\'isica Te\'orica, Universidad Complutense, 28040 Madrid, Spain}
\author{E. Tirrito}
\affiliation{ICFO-Institut de Ciencies Fotoniques, The Barcelona Institute of Science and Technology, 08860 Castelldefels (Barcelona), Spain}
\author{M. Rizzi}
\affiliation{Johannes Gutenberg-Universit\"at, Institut f\"ur Physik, Staudingerweg 7, 55099 Mainz, Germany}
\author{M. Lewenstein}
\affiliation{ICFO-Institut de Ciencies Fotoniques, The Barcelona Institute of Science and Technology, 08860 Castelldefels (Barcelona), Spain}
\affiliation{ICREA, Llu\'is Companys 23, 08010 Barcelona, Spain}
\author{S. Hands}
\affiliation{Department of Physics, College of Science, Swansea University, Singleton Park, Swansea SA2 8PP, United Kingdom}

\begin{abstract}
We show that a Wilson-type discretization of the Gross-Neveu model, a fermionic $N$-flavor quantum field theory  displaying asymptotic freedom and chiral symmetry breaking, can serve as a playground to explore correlated symmetry-protected phases of matter using techniques borrowed from high-energy physics. A large-$N$ study, both in the Hamiltonian and Euclidean formalisms, yields a phase diagram with trivial, topological, and symmetry-broken phases separated by critical lines that meet at a tri-critical point. We benchmark these predictions using tools from condensed matter and quantum information science, which show that the large-$N$ method captures the essence of the phase diagram even at $N=1$. Moreover, we describe a cold-atom scheme for the quantum simulation of this lattice model, which would allow to explore the single-flavor phase diagram.

\end{abstract}

\maketitle
\setcounter{tocdepth}{2}
\begingroup
\hypersetup{linkcolor=black}
\tableofcontents
\endgroup

\section{Introduction}
\label{sec:introduction}

The understanding and classification of all possible phases of matter is one of the most important challenges of contemporary condensed-matter physics~\cite{sachdev}, and high-energy physics~\cite{hands_phases_qcd},  finding also important implications in quantum information science~\cite{yoshida}. Such a complex  quest can benefit enormously from the complementary perspectives and tools developed by these different communities, calling for a cross-disciplinary dialogue that can lead to a very interesting collaborative  approach.
The theory of spontaneous symmetry breaking~\cite{landau_sb} and critical phenomena~\cite{wilson_kogut} are representative examples, where such an open dialogue has 
provided  fundamental insight to unveil generic and universal properties in the classification of  various phases of matter, and the transitions between them. However, these examples 
do not exhaust all possible phenomena~\cite{wen_book},  encouraging further efforts to provide  a  general classification  encompassing other  exotic orders.  

Some of these studies were initially stimulated by the community working on quantum chaos~\cite{chaos_book}, which looked for a complete classification of various random matrix ensembles depending on the symmetries, leading to the so-called ten-fold way~\cite{az_ten_fold}.  The ten-fold way turned out to be a fundamental tool for the classification of  non-interacting phases of matter~\cite{table_top_insulators}  which, in contrast to symmetry-broken phases, can exist within the same symmetry class~\cite{chern_tknn,haldane,kane_mele}. In this case, transitions between different phases of matter can only occur via  gap-closing continuous phase transitions, but there is neither  symmetry-breaking, nor any underlying local order parameter. In contrast, these new phases  are characterized by a topological invariant, the value of which  changes abruptly across the symmetry-preserving critical point. This leads to the notion of symmetry-protected topological (SPT) phases~\cite{spt_review}, which includes the fermionic topological insulators and superconductors, but also other SPT phases of bosons and spins. 

From a quantum-information perspective, the recent progress in the so-called tensor networks~\cite{mps} has triggered the interest within this community in the general question of classifying topological phases of matter for generic interacting systems~\cite{classification_ti}, including static and dynamical situations~\cite{zhang}. Note  that, despite the considerable progress, a complete classification  has so far  been accomplished  only  for (1+1)-dimensional systems~\cite{classification_1d}. At such reduced dimensionalities,  there is essentially a single  gapped phase, which is trivial (i.e. it can be transformed  into an uncorrelated product state by  local unitaries) unless additional discrete symmetries are taken into account. Such symmetries may protect the phases, such that the states belonging to different symmetry sectors cannot be transformed into one  another using local symmetry-preserving operations.   
A detailed understanding and characterization of the properties of these SPT phases, in the presence of interactions and strong correlations, is an open  question of current interest. As  argued in this work, these phases are 
not only relevant in condensed-matter systems, but also arise in the context of high-energy physics for certain lattice formulations of quantum field theories (QFTs).

In this paper, we focus on strongly-correlated SPT phases of a paradigmatic model of  high energy physics: the Gross-Neveu model~\cite{GN_model}. This QFT describes  Dirac fermions with $N$ flavors interacting via quartic interactions in 1 spatial and 1 time dimension,  and was originally introduced as a toy model that shares several fundamental features with quantum chromodynamics. We consider a Wilson-type discretization of the QFT~\cite{wilson_fermions}, and term  the lattice version as the Gross-Neveu-Wilson model.   Despite extensive studies of the GN model,  a detailed characterization of strongly-correlated SPT phases has not been discussed in detail, to the best of our knowledge, neither in the large, nor in the finite $N$ limit. The present work has the ambition of filling this gap using  methods of contemporary theoretical physics and numerical simulations. Moreover, we present a scheme for the experimental realization of  this discretized QFT using cold-atom quantum simulators. In this way, we hope that the  Gross-Neveu model will get upgraded from a toy model used to understand some essential features of more realistic high-energy QFTs, into a cornerstone in the classification of correlated topological phases of interest in condensed matter and quantum information, which can also be explored in a realistic experiment of atomic, molecular and optical physics.

We now summarize our main results, and how they are organized in this paper: In Sec.~\ref{sec:GNWmodel}, we discuss generalities of the Gross-Neveu-Wilson model viewed from the complementary perspectives of high-energy, condensed-matter, and cold-atom physics. This section is intended to bridge the specific knowledge gaps between these different communities, in our effort to provide a self-contained cross-disciplinary study.    In Sec.~\ref{sec:SPT_GNW}, we study  the occurrence of correlated SPT phases in the model using tools common to high-energy physics. We discuss the phase diagram from the large-$N$ expansion, including both a continuous time approach (i.e. Hamiltonian field theory on the lattice), and a discretized time approach (leading to Euclidean field theory on the
lattice). This detailed study has allowed us to identify important details of the Euclidean approach, which must be carefully considered in order to understand the phase diagram of the model. In particular, we provide a  neat picture  where trivial gapped phases and correlated SPT phases, well-known in the condensed-matter and quantum information communities,  and parity-breaking Aoki phases, well-known to the lattice field theory community, coexist in a rich phase diagram.  Moreover, the large-$N$ approach is exploited to indicate the existence of tri-critical points where these three different phases are joined. We benchmark the large-$N$ predictions using  tools common to the condensed-matter and quantum-information communities, i.e. tensor-network techniques based on  matrix-product-state variational ansatzs. As discussed in the text,  these quasi-exact numerics for a $N=1$ realization of the Gross-Neveu-Wilson model confirm the large-$N$ prediction of the phase diagram, and provide additional information that complements the large-$N$ high-energy-inspired understanding of the model. Finally, we present a proposal for a potential experimental realization of the Gross-Neveu-Wilson model  with ultra-cold atoms. In this way, relativistic models of high-energy physics could be explored with table-top non-relativistic experiments at ultra-low temperatures by focusing on  low-energies and long wavelengths.

\section{The Gross-Neveu model: high-energy physics, condensed matter, and cold atoms}
\label{sec:GNWmodel}

\subsection{The Gross-Neveu quantum field theory: continuum version and Wilson lattice approach}
\label{sec:GN_qft}

The Gross-Neveu model is a relativistic QFT describing $N$  species (flavors) of a massless Dirac field, which live    in a (1+1)-dimensional spacetime and interact via  four-fermion terms~\cite{GN_model}. This model originates from its higher-dimensional counterparts, the so-called Nambu-Jona-Lasinio models~\cite{NJL_model,NJL_review}, which were introduced as  alternatives to   non-Abelian gauge  theories~\cite{yang_mills}. Pre-dating quantum chromodynamics (QCD)~\cite{qcd}, these models  offer a simplified framework  to study   essential features of the strong interaction, such as dynamical  mass generation  by chiral symmetry breaking. In addition to these features, the lower-dimensional  Gross-Neveu model was introduced post-QCD as a tractable  QFT displaying asymptotic freedom in a renormalizable framework. In contrast to some of its  higher-dimensional cousins, this feature permits to derive rigorous results concerning the renormalization group and  the convergence of perturbation theory~\cite{gawedzki}.

 In the continuum, this model is described by the following normal-ordered Hamiltonian $H=\int{\rm d}x:\!\mathcal{H}\!\!:$ with 
\beq
\label{eq:GN_continuum}
 \mathcal{H}=-\sum_{n=1}^N\!\overline{\psi}_n(x)\ii\gamma^1\partial_x  \psi_n(x)-\frac{g^2}{2N}\!\left(\sum_{n=1}^N\overline{\psi}_n(x) \psi_n(x)\!\right)^2.
\eeq
Here, $\psi_n(x),\overline{\psi}_n(x)=\psi^\dagger_n(x)\gamma^0$ are two-component spinor field operators for the $n$-th fermionic species, and $\gamma^0=\sigma^z$, $\gamma^1=\ii\sigma^y$ are the  gamma matrices, which can be   expressed in terms of Pauli matrices for  a (1+1)-dimensional Minkowski spacetime, leading to the chiral  matrix $\gamma^5=\gamma^0\gamma^1=\sigma^x$~\cite{note_gamma}. Therefore, the Gross-Neveu model describes a collection of $N$ copies of  massless Dirac fields  coupled via the quartic interactions.

The first term in Eq.~\eqref{eq:GN_continuum} corresponds to the kinetic energy of the  massless Dirac fermions, where we use natural units $\hbar=c=1$, whereas the second term describes  two-body interactions between pairs of fermions that scatter off each other with a strength $g^2/N$. This model has a   global, discrete chiral symmetry $\psi_n(x)\to\gamma^5\psi_n(x)$, $\forall x$, as follows  directly from the anti-commutation  relations of the Dirac matrices. Additionally,  a global  $U(N)$  internal symmetry  becomes apparent by introducing  $\Psi(x)=(\psi_1(x),\cdots,\psi_N(x))^{\rm t}$, after rewriting the Gross-Neveu Hamiltonian density as 
\beq
\label{eq:GN_cont_capital_psi}
 \mathcal{H}=-\overline{\Psi}(x)\ii\gamma^1\partial_x  \Psi(x)-\frac{g^2}{2N}\left(\overline{\Psi}(x) \Psi(x)\right)^2,
 \eeq
which is invariant under the transformation   $\Psi(x)\to u\otimes\mathbb{I}_2\Psi(x)$, $\forall x$, with   the unitary matrix $u\in U(N)$.
We note that the fields have classical mass dimension $d_{\psi}=1/2$, while the interaction couplings are dimensionless   $d_{g}=0$.

In the limit where the number of flavors  $N$ is very large, D. J. Gross and A. Neveu showed that this model yields a renormalizable QFT displaying asymptotic freedom, i.e. the interaction strength $g^2$ is a relevant perturbation in the infra-red (IR), but becomes  weaker at high energies in the ultra-violet (UV) limit~\cite{GN_model}. Moreover,   even if the discrete chiral symmetry prevents the fermions from acquiring a mass to all orders in  perturbation theory, they showed that a mass can be  dynamically generated through the spontaneous breaking of this chiral symmetry, which can be captured    by  large-$N$ methods. 
In contrast to the Higgs mechanism, where masses can be generated by introducing additional scalar fields that undergo spontaneous symmetry breaking themselves, here a physical  mass (i.e. gap) is generated dynamically as a non-perturbative consequence of  the four-fermion interactions. These results are exact in the $N\to\infty$ limit, and it is possible to calculate the leading corrections for a finite, but still large, $N$. 

A different strategy  to explore such non-perturbative effects is the so-called lattice field theory (LFT),  which discretizes the fermion fields on a  uniform lattice 
$ \Lambda_{\rm s}=a\mathbb{Z}_{N_{\rm s}}=\{{x}:x/a=n\in\mathbb{Z}_{N_{\rm s}}\}$, where $N_{\rm s}$ is the number of lattice sites, and $a$ is the lattice spacing~\cite{lattice_book}.  A naive discretization of the derivative of the Dirac operator  yields the  Hamiltonian $H_{\rm N}=a\sum_{{ x}\in\Lambda_{\rm s}}\!\! :\!\mathcal{H}_{\rm N}\!\!:$, which describes a system of interacting  fermions hopping between neighboring sites of a one-dimensional lattice 
\beq
\label{eq:naive_GN_lattice}
\mathcal{H}_{\rm N}=\!\left(\!\overline{\Psi}( x)\frac{-\ii \gamma^1}{2a} \Psi\!\left(x+a\right)+{\rm H.c.}\!\!\right)-\frac{g^2}{2N}\!\left(\overline{\Psi}(x) \Psi(x)\right)^2\!.
\eeq
Here, the   lattice  fields   fulfill  the desired  anti-commutation algebra in the continuum limit   $\{\Psi_\mu^{\phantom{\dagger}}({x}), \Psi_{\nu}^{{\dagger}}({x}')\}=\frac{1}{a}\delta_{\mu,\nu}\delta_{{x},{x}'}\to\delta_{\mu,\nu}\delta({x}-{x}')$ as $a\to 0$, where $\mu,\nu\in\{1,\cdots,2N\}$. 

Unfortunately, this naive discretization also leads to spurious  fermion doublers which, for $g^2=0$,   correspond to  massless Dirac fields  appearing as long-wavelength excitations around the corners of the Brillouin zone~\cite{nn_theorem}. In the present case, the Brillouin zone is  ${\rm BZ}_{\rm s}=\{k=2\pi n/N_{\rm s}\}=(-\pi/a,\pi/a]$ such that,  in addition to the target massless Dirac field around $k=0$, a single  doubler arises around the corner $k=\pi/a$~\cite{susskind_1d}. Note that, as soon as the interactions   are switched on $g^2>0$, there will be scattering processes where the doubler affects the properties of the massless Dirac field, such that the continuum limit may differ from the desired QFT~\eqref{eq:GN_continuum}. Among several possible
 strategies to cope with the presence of such  fermion doublers~\cite{lattice_book}, K. Wilson considered introducing a momentum-dependent mass term, the so-called Wilson mass, that sends all the doublers to the cutoff of the lattice field theory~\cite{wilson_fermions}. In this way, one expects that these heavy fermions will not influence the universal long-wavelength properties of the continuum limit.
 
 For the  Hamiltonian QFT  of interest~\eqref{eq:GN_continuum}, this can be accomplished by introducing an additional Wilson term in the naive discretization~\eqref{eq:naive_GN_lattice} leading to $H_{\rm W}=a\!\sum_{{ x}\in\Lambda_{\rm s}}\! :\!\mathcal{H}_{\rm W}\!\!:$,   where 
\beq
\label{eq:Wilson_lattice}
\mathcal{H}_{\rm W}=\mathcal{H}_{\rm N}+\left(\overline{\Psi}({ x})\frac{r }{2a} \big( \Psi({ x})-\Psi\!\left({ x}+a\right)\big)+{\rm H.c.} \!\right)\!,
\eeq
which will be referred to as the Gross-Neveu-Wilson (GNW) model in this work. Here, $r\in[0,1]$ is the so-called Wilson parameter. In the continuum limit, and for $g^2=0$, the mass of the  doubler around $k=\pi/a$ becomes $m_\pi=2r/a$, while the Dirac fermion around $k=0$ remains massless $m_0=0$. We will set $r=1$ henceforth, such that the doubler mass coincides  with the  UV energy cutoff of the QFT. On the other hand, the Dirac field around $k=0$ remains massless, and one expects that the IR limit will be governed by the desired chiral-invariant QFT.

 This situation gets more involved when the interactions are switched on $g^2>0$, as the additional Wilson terms~\eqref{eq:Wilson_lattice} break explicitly the discrete chiral symmetry (i.e. $r\overline{\Psi}({ x}){\Psi}({ x})\to-r\overline{\Psi}({ x}){\Psi}({ x})$ under the discrete chiral transformation since $\gamma^5\gamma^0\gamma^5=-\gamma^0$). Accordingly,   the vanishing mass $m_0=0$ of the Dirac fermion around $k=0$ is no longer protected by the discrete chiral symmetry, and it can become finite even for  perturbative interactions in contrast to the continuum model. Since one is interested in recovering the  QFT~\eqref{eq:GN_continuum} for massless Dirac fermions, it is thus necessary to approach the continuum limit using a different strategy. The idea is to introduce an additional   mass term in the lattice Hamiltonian~\eqref{eq:Wilson_lattice} leading to $\tilde{H}_{\rm W}=a\!\sum_{{ x}\in\Lambda_{\rm s}}\! :\!\tilde{\mathcal{H}}_{\rm W}\!:$, where we have introduced
\beq
\label{eq:Wilson_GN_lattice}
\tilde{\mathcal{H}}_{\rm W}=\mathcal{H}_{\rm W}+m\overline{\Psi}({ x}) \Psi({ x}),
\eeq
and $m$ is a bare mass parameter. By tuning this mass as a function of the interaction strength $m(g^2)$, one must search for a critical line $m=m_{\rm c}(g^2)$ where the renormalized mass of the Dirac fermion around $k=0$ vanishes $\tilde{m}_0=0$, such that the correlation length fulfills $\xi\gg a$ (i.e. a second-order quantum phase transition). In this case,   the physical quantities of interest become independent of the underlying lattice, and one expects to recover the desired continuum QFT. The key question is to analyze if such continuum scale-invariant limit corresponds to the  chiral-invariant  Gross-Neveu model~\eqref{eq:GN_continuum}, or if a     QFT of a different nature emerges in the IR limit. The answer to this question may depend on the possible phases of the lattice field theory~\eqref{eq:Wilson_lattice} and the different critical lines in between them. Therefore, addressing this question requires a detailed non-perturbative approach using for instance large-$N$ methods on the lattice, or Monte-Carlo methods from lattice field theory. In this work, we will present a detailed large-$N$ analysis of the lattice GNW model, applying it to the prediction of its phase diagram, and benchmarking it with  numerical simulations for the $N=1$ single-flavour case.

\subsection{Symmetry-protected topological phases for interacting fermions}
\label{sec:ti}

A wide variety of phases transitions can be understood according to Landau's theory of spontaneous symmetry breaking~\cite{landau_sb}, which exploits the notion of symmetry and local order parameters to classify various phases of matter. Nowadays, we understand that Landau's theory does not exhaust all possibilities, as one can find  different phases of matter within the same symmetry class that can only be connected via phase transitions where the symmetry is not broken. These so-called symmetry-protected topological (SPT) phases  cannot be described by local order parameters,  but require instead the use of certain topological invariants to characterize their groundstate (e.g. topological insulators and superconductors~\cite{review_top_insulators,Bernevig}). These topological invariants  are in turn related to observables  displaying  quantized values  that are robust with respect to perturbations that respect these symmetries (i.e. the topological numbers can only change via a gap-closing  phase transition). Accordingly, these new phases of matter can be organized within different symmetry classes, as occurs for the fermionic topological insulators~\cite{table_top_insulators,10_fold_ryu}.  Despite having a gapped bulk, these insulators  display a quantized conductance (e.g. integer quantum Hall effect~\cite{iqhe_exp})  related to a topological invariant (e.g. first Chern number~\cite{chern_tknn}). A bulk-edge correspondence allows to understand this topological robustness by the appearance of current-carrying edge excitations through a band-inversion process,  corresponding to mid-gap states that are exponentially localized within the boundaries of the material (e.g.   one-dimensional edge modes where fermions  cannot back-scatter due to disorder~\cite{edge_iqhe}).

The connection between SPT phases and LFTs is very natural for  three-dimensional time-reversal-invariant topological insulators~\cite{review_top_insulators}. Here, the band-inversion process yielding the topological phase leads to an odd number of massless Dirac fermions localized within the boundaries of the material. As emphasized in~\cite{wilson_atoms,strong_coupling_wilson_3d_U(1)}, this band inversion can be understood in terms of lower-dimensional versions of  domain-wall fermions~\cite{domain_wall}, whereby an odd number of Wilson doubler masses change their sign, contributing each with a two-dimensional massless Dirac fermion localized at the boundary. In fact, we note that the Wilson-like terms in Eq.~\eqref{eq:Wilson_lattice} arise very naturally in the low-energy
description of topological insulating materials in various dimensionalities~\cite{Bernevig}.

Let us now discuss how these topological effects also appear in the  non-interacting limits of the GNW  model~\eqref{eq:naive_GN_lattice}-\eqref{eq:Wilson_GN_lattice}. Here, the band inversion would occur when tuning the bare mass to lie within $m\in(m_{\pi},m_{0})$, where we recall that $m_0=0$ and  $m_{\pi}=2/a$ correspond to the masses of the Wilson fermions.  To understand the SPT phase in this LFT, we consider periodic boundary conditions,  such that the Hamiltonian in momentum space is
$
\tilde{H}_{\rm W}=\sum_{n=1}^{N}\sum_{k\in{\rm BZ}_{\rm s}}\psi^{\dagger}_n(k)\mathsf{h}_{k}(m)\psi^{\phantom{\dagger}}_n(k),
$
where we have introduced the flavor-independent  single-particle Hamiltonian
\beq
\label{eq:single-particle_h}
\mathsf{h}_{k}(m)=\left(m+\frac{1-\cos ka}{a}\right)\gamma^0-\frac{\sin ka}{a}\gamma^5.
\eeq
By a straightforward diagonalization, one finds 
\beq
\tilde{H}_{\rm W}=\sum_{n=1}^{N}\sum_{k\in{\rm BZ}_{\rm s}}\sum_{\eta=\pm}\epsilon_\eta(k)\psi_{n,\eta}^\dagger(k)\psi_{n, \eta}^{\phantom{\dagger}}(k),
\eeq
 where  $\psi_{n,\eta}^{{\dagger}}(k), \psi_{n,\eta}^{\phantom{\dagger}}(k)$ are the creation-annihilation operators of a fermionic excitation with flavor $n$ in the energy band
 \beq
\epsilon_\pm(k)=\pm\frac{1}{a}\sqrt{\big(ma+1-\cos ka\big)^2+\sin^2ka}.
\eeq

This band structure has a non-zero gap for $m>0$, yielding an insulating phase. In order to show that this insulator is topological, and an instance of a SPT phase, we note that this band structure has an associated topological invariant that can be defined through the Berry connection  $\mathcal{A}_{n}(k)=\ii \bra{\epsilon_{n,-}(k)}\partial_k\ket{\epsilon_{n,-}(k)}$, where we have introduced the single-particle negative-energy states   $\ket{\epsilon_{n,-}(k)}=\psi_{n,-}^\dagger(k)\ket{0}$. In our case~\eqref{eq:single-particle_h}, the  Berry connection can be expressed as
\beq
\label{eq:connection}
\mathcal{A}_n(k)=\frac{1}{2}\frac{(1+ma)\cos ka-1}{1+(1+ma)^2-2(1+ma)\cos ka},
\eeq
which allows to construct a topological invariant,  the so-called Zak's phase~\cite{zak_phase}, as the    integral of the Berry connection over the  Brillouin zone. From Eq.~\eqref{eq:connection},  the total Zak's phase $\varphi_{\textrm{Zak}}=\sum_{n}\int_{\textrm{BZ}_{\rm s}}\!\!\textrm{ d}k\mathcal{A}_n(k)$ can be expressed as
\beq
\label{eq:zak}
\varphi_{\textrm{Zak}}=N\pi\big(\theta(2+ma)-\theta(ma)\big),
\eeq
where $\theta(x)$ is  Heaviside's step function. We note that, as occurs with the Chern number and the transverse conductivity in the quantum Hall effect~\cite{chern_tknn}, the topological Zak's phase can be related to an observable: the electric polarization~\cite{polarization,polarization_zak}. 

Since  the groundstate is constructed by filling all negative-energy states $\ket{\textrm{gs}}=\prod_{k\in{\textrm{ BZ}_{\rm s}}}\ket{\epsilon_-(k)}$, the above integral over the whole Brillouin zone~\eqref{eq:zak}  characterizes the  topological features of the  LFT groundstate. Accordingly, this LFT hosts  a SPT phase in the parameter regime   $m_{\pi}<m<m_{0}$ for $N$ odd, which coincides with the band-inversion regime introduced above. This regime can be interpreted as the result of a mass-inversion process, whereby the mass of some of the Wilson fermions  gets inverted. This becomes apparent after rewriting the Zak's phase in terms of  the $N$ Wilson  masses 
\beq
\label{eq:wilson_masses}
\tilde{m}_0=m, \hspace{2ex}\tilde{m}_\pi=m+2/a.
\eeq 
Indeed, a non-trivial topological invariant  (i.e. $\varphi_{\textrm{Zak}}/2\pi\not\in\mathbb{Z}$) can only be achieved  when an odd number of fermion doubler pairs display a different mass sign  
\beq
\label{eq:zak_masses}
\varphi_{\textrm{Zak}}=\half N\pi\big({\rm sgn}(\tilde{m}_\pi)-{\rm sgn}(\tilde{m}_0)\big).
\eeq

We note that this SPT phase can be identified with a  one-dimensional topological insulator in the so-called chiral-orthogonal  $\mathsf{BDI}$ class~\cite{table_top_insulators,10_fold_ryu,Bernevig}, which would display zero-energy modes localized at the edges of the chain for open boundary conditions.
 Note that this chiral symmetry class is not related to the standard notion of chirality in QCD, which is indeed broken by the GNW model. Instead, it is related to the ten-fold Cartan's classification of symmetric spaces, and its connection to single-particle Hamiltonian via the time-evolution operator~\cite{table_top_insulators}. For the non-interacting GNW single-particle Hamiltonian~\eqref{eq:single-particle_h},  we find that time-reversal $\mathsf{T}$ yields $T^\dagger h_{-k}(m)^*T=h_{k}(m)$ where  $T=-\ii\sigma^x\sigma^y=\gamma^0$, and  charge-conjugation $\mathsf{C}$ leads to $C^\dagger h_{-k}(m)^*C=-h_{k}(m)$ where $C=\ii\sigma^z\sigma^y=\gamma^5$~\cite{note_gamma}. The combination of these two anti-unitary symmetries is called chiral, or sub-lattice, symmetry $\mathsf{S}=\mathsf{T}\mathsf{C}$, and yields $S^\dagger h_{k}(m)S=-h_{k}(m)$ with $S=\gamma^0\gamma^5=\gamma^1$. To avoid confusion with the chiral symmetry of high-energy physics, which is a fundamental ingredient  in  low-energy effective descriptions of  QCD, and pivotal in our previous discussion of the GNW model, we will refer to the $S$ as the sublattice symmetry. Since $T^2=C^2=S^2=+1$, the corresponding GNW  topological insulator with an odd number $N$ of fermion flavors~\eqref{eq:zak_masses} is in the $\mathsf{BDI}$ class.  
 
 We note that the ten symmetric spaces that classify the topological insulators/superconductors also correspond to the target spaces of  effective non-linear-sigma model  describing the long-wavelength properties of the edge/boundary. When  such a non-linear-sigma model includes a  topological term, the edge modes are robust and evade Anderson localization in the presence of symmetry-preserving disorder~\cite{table_top_insulators}. This perspective allows us to understand the difference of $N$ even/odd in the GNW model. For $N$ even, there can be a symmetry-preserving disorder that couples the different flavors of the  edge states, leading to scattering/localization and destroying the $\mathsf{BDI}$ topological protection. On the other hand, for $N$ odd, at least one of the edge modes will remain robust against inter-flavor scattering, and thus evade Anderson localization. We note that similar parity effects can occur also in models with more than one fermion doubler in the regime where an even number of Wilson masses gets inverted, as occurs for higher-dimensional time-reversal topological insulators~\cite{wilson_atoms}. 

 In contrast to the LFT perspective described in Sec.~\ref{sec:GN_qft}, where one is mainly interested in searching for the second-order quantum phase transitions to recover a continuum limit described by the QFT of interest~\eqref{eq:GN_continuum}; the study of symmetry-protected topological phases focuses on the topological gapped phases away from criticality. Interestingly,  even in the non-interacting regime, the emerging QFTs governing  their response to external fields turns out to be very different from the original discretized QFT,   and can be described in terms of  topological quantum field theories (e.g. Chern-Simons or axion QFTs)~\cite{TI_field_theories}. A generic question of current interest in the study of SPT phases  is to explore the interplay of topological features and strong-correlation effects as  interactions between the fermions are switched on~\cite{ti_interactions_reviews}.

For the GNW lattice model~\eqref{eq:naive_GN_lattice}-\eqref{eq:Wilson_GN_lattice}, the interactions do not modify the symmetry class as $\overline{\Psi}\Psi\to \overline{\Psi}\Psi$, and $\overline{\Psi}\gamma^5\Psi\to \mp \overline{\Psi}\gamma^5\Psi$ under  time-reversal and charge-conjugation transformations, respectively~\cite{note_gamma}. Accordingly,  the quartic terms in Eq.~\eqref{eq:GN_cont_capital_psi}, or its chiral extension introduced in Eq.~\eqref{eq:GN_cont_chiral__capital_psi} below, do not modify the aforementioned $\mathsf{BDI}$ symmetry class. A question of  potential interest for both the SPT and LFT communities is the precise determination of the  critical lines of  the lattice model $m_{\rm c}(g^2)$  for non-perturbative interactions. From this knowledge, the LFT community can  explore the nature of the continuum QFT in the vicinity of the critical line, while the SPT community may study how the topological phase is modified in presence of interactions. As argued above, a possible tool to study non-perturbative effects could be   large-$N$ methods, or Monte-Carlo methods in Euclidean lattice field theory. We remark, however, that the standard Euclidean approach where time is also discretized~\cite{lattice_book} can lead to  qualitative differences  of the phase diagram in the $(m,g^2)$ plane. Discretizing time introduces additional fermion doublers, which may lead to additional critical lines that are not present in the Hamiltonian  approach~\eqref{eq:naive_GN_lattice}-\eqref{eq:Wilson_GN_lattice} with continuous time~\cite{hamiltonian_lgt}. Although this is not relevant when one is only interested in  the nature of the continuum QFT, it will be of relevance for topological insulators where one is interested in the finite region of phase space with the topological gapped phase. In this work, we will show that special care in the Euclidean lattice formulation is required in order to recover the relevant phase diagram. 

For one-dimensional models, the  study of lattice field theories in the   Hamiltonian  approach can be efficiently accomplished using variational methods based on Matrix Product States (MPS)~\cite{mps_lft}.  In this work, we shall confront predictions of the large-$N$ approximation with  results from MPS numerical methods for the study of topological insulating phases in the GNW  model with Wilson fermions.

\subsection{Cold-atom quantum simulators of high-energy physics}
\label{sec:cold_atom}

As an alternative to  Monte-Carlo numerical methods in lattice field theory, one may follow R. P. Feynman's insight~\cite{qs_feynman}, and develop schemes to control a quantum-mechanical device such that its dynamics reproduces faithfully  that of the model of interest (i.e. quantum simulation). From this perspective, a very appealing application of the future fault-tolerant quantum computers will be their ability to function as universal quantum simulators~\cite{qs_lloyd} that can address  complicated quantum many-body problems  relevant for different disciplines of physics and chemistry. Prior to the development of quantum error correction and large-scale fault-tolerant quantum computers, one may consider building special-purpose quantum simulators that are  designed to tackle a 
particular family of models. This is the case of cold-atom quantum simulators of  lattice models~\cite{qs_cold_atoms_1,qs_cold_atoms_2}, where  neutral atoms are laser-cooled to very low temperatures  in deep optical lattices~\cite{cold_atoms_review}.

In the continuum, neutral-atom systems are typically described by a Hamiltonian QFT, albeit a non-relativistic one~\cite{cold_atoms_review}, with $H=\int d^3x:(\mathcal{H}_0+\mathcal{V}_{\rm I}):$ containing
\beq
\label{eq:cold_atoms_ho}
\mathcal{H}_0=\sum_{n,\sigma}\Psi^\dagger_{n,\sigma}(\boldsymbol{x})\!\left(\!\!\left(\frac{-\nabla^2}{2m_n}+\epsilon^{n}_{\sigma}\!\!\right)\!\delta_{\sigma,\sigma'}+V^{n}_{\sigma,\sigma'}\!(\boldsymbol{x})\!\right)\Psi^{\phantom{\dagger}}_{n,\sigma'}(\boldsymbol{x}),
\eeq
where  $\Psi_{m,\sigma}^\dagger(\boldsymbol{x}),\Psi_{m,\sigma}^{\phantom{\dagger}}(\boldsymbol{x})$ are   field operators that create-annihilate an atom of the $n$-th species in the internal state $\sigma$. This Hamiltonian contains
 {\it (i)} the kinetic energy for a multi-species gas of Alkali atoms of mass $m_n$, where $n\in\{1,\cdots N_{\rm sp}\}$ labels the atomic species/isotope; {\it (ii)} the internal energy $\epsilon^{n}_{\sigma}$ of the atomic groundstate manifold, which typically consists of various hyperfine levels characterized by the quantum numbers associated to the total angular momentum  $\sigma\in\{F,M_{F}\}$; and {\it (iii)} the single-particle terms $V_n^{\sigma,\sigma'}\!\!(\boldsymbol{x})$, which contain the trapping potential that confines the $n$-th atomic species and, possibly, additional radiation-induced terms that drive transitions between the different atomic levels $\sigma\to\sigma'$. In particular, we shall be interested in periodic trapping potentials  due to the ac-Stark-shift of pairs of retro-reflected laser beams, which will depend on the atomic species, but not on the particular hyperfine level (i.e. state-independent optical lattices). We also consider laser-induced Raman transitions via highly off-resonant excited states. Altogether, this leads to 
\beq
\label{eq:on-site_potentials}
\begin{split}
V^n_{\sigma,\sigma'}\!(\boldsymbol{x})=&\sum_{\nu=x,y,z}\!\!\left(V^n_{0,\nu}\sin^2(k_{{\rm L},\nu}x_\nu)+\half m_n\omega^2_{n,\nu}x_{\nu}^2\right)\delta_{\sigma,\sigma'}\\
&\!\!\!+\sum_l\Omega^{n,l}_{\sigma,\sigma'}\cos(\boldsymbol{\Delta k}_{l}\cdot \boldsymbol{x}-\Delta\omega_{l}t+\varphi_{l}),
\end{split}
\eeq
where $V^n_{0,\nu}$ is the ac-Stark shift for the $n$-th atomic species stemming from  the retro-reflected beams with wave-vector  $k_{{\rm L},\nu}$ along the $\nu$-axis.  Additionally, $\omega_{n,\nu}$ is the frequency of a residual harmonic trapping due to the intensity profile of the lasers. Finally, $\Omega^{n,l}_{\sigma,\sigma'}$ is the two-photon Rabi frequency for the Raman transition induced by the $l$-th pair of laser beams with wave-vector (frequency) difference $\boldsymbol{\Delta k}_{l}$ ($\Delta\omega_{l}$), and phase $\varphi_{l}$.

 In addition, at sufficiently low temperatures, the  neutral atoms also interact by contact scattering processes leading to
\beq
\label{eq:cold_atoms_v}
\mathcal{V}_{\rm I}=\frac{1}{2}\sum_{\sigma,\sigma'}\sum_{n,n'} U^{n,n'}_{\sigma,\sigma'}\Psi^\dagger_{n,\sigma}(\boldsymbol{x})\Psi^\dagger_{n,'\sigma'}(\boldsymbol{x})\Psi^{\phantom{\dagger}}_{n',\sigma'}(\boldsymbol{x})\Psi^{\phantom{\dagger}}_{n,\sigma}(\boldsymbol{x}),
\eeq
where the interaction strengths $U^{n,n'}_{\sigma,\sigma'}$ depends on the $s$-wave scattering lengths $a_{\sigma\sigma'}$ of the corresponding channels, some of which can be controlled by Feshbach resonances~\cite{cold_atoms_review}. We also note that fully-symmetric interactions between all species can be  achieved by  using alkali-earth atoms~\cite{su_N_review}, which could be an interesting property for the experimental realization of higher number of flavors $N$ in the Gross-Neveu-Wilson model.

As announced above, in the regime of deep optical lattices $V^n_{0,\nu}\gg E^n_{\rm R}=\boldsymbol{k}^2_{\rm L}/2m_n$, one can introduce the basis of so-called Wannier functions, which are localized to the minima of the potential, and show that this non-relativistic QFT yields a family of Hubbard-type models with tunable parameters~\cite{hubbard_atoms,hubbard_exp}. Therefore, by doing controlled table-top experiments with   cold atoms, it becomes possible to explore the physics of strongly-correlated electrons in solids, which has opened a  fruitful avenue of research in quantum simulations of condensed-matter models~\cite{qs_cold_atoms_1,qs_cold_atoms_2}. More recently, several works have explored the possibility of extending this cold-atom Hubbard toolbox~\cite{hubbard_toolbox} to the  quantum simulation of high-energy physics, including relativistic QFTs~\cite{aqs_qft_fermions_2d,wilson_atoms,aqs_qft_fermions_interactions,dqs_qft_phi4,O_N_models,aqs_phi_4},   gauge field theories~\cite{dqs_lattice_gauge_theories,aqs_qft_gauge_abelian_qlinks, aqs_qft_gauge_U(1)_qlinks,dqs_qft_SU(2)_qlinks,compresed_qs_gauge}, theories for coupled Higgs and gauge fields~\cite{gauge_higgs_atoms,gauge_higgs_meurice}, and also theories of relativistic fermions interacting with Abelian/non-Abelian gauge fields~\cite{aqs_QED,aqs_qft_fermions_U(1)_qlinks,aqs_qft_fermions_SU(2)_qlinks}.

In this work, we shall be concerned with a cold-atom realization of the Gross-Neveu  model using a Wilson-fermion discretization~\eqref{eq:naive_GN_lattice}-\eqref{eq:Wilson_GN_lattice}. We note that there are cold-atom proposals to implement this QFT~\eqref{eq:GN_continuum} with optical superlattices lattices by  a different discretization~\cite{aqs_qft_fermions_interactions}, via the so-called staggered fermions~\cite{susskind_1d,hamiltonian_lgt}. Since we are interested in the connection of this model with correlated SPT phases, we will instead focus on the Wilson-fermion approach of Eqs.~\eqref{eq:naive_GN_lattice}-\eqref{eq:Wilson_GN_lattice}. Building on previous proposals for the quantum simulation of Wilson fermions~\cite{wilson_atoms,creutz_hubbard,wilson_kuno,wilson_hauke}, we present in this work a simplified scheme to realize the  GNW model using a two-component single-species Fermi gas confined in a one-dimensional optical lattice with laser-assisted tunneling.

\section{ Correlated symmetry-protected topological phases in the  Gross-Neveu-Wilson model}
\label{sec:SPT_GNW}
\subsection{Phase diagram from the large-$N$ expansion}

As advanced in the previous sections, our goal is to  determine the  critical lines of  the GNW  model~\eqref{eq:naive_GN_lattice}-\eqref{eq:Wilson_GN_lattice} as a function of the coupling strength  $m(g^2)$ for non-perturbative interactions.
We start by developing a large-$N$ expansion for the partition function $Z=\textrm{Tr}\left\{\textrm{exp}\left(-\beta \tilde{H}_{\rm W}\right)\right\}$, where $\beta=1/T$ is the inverse temperature for  $k_{\rm B}=1$. In the continuum, large-$N$ methods were first employed by Gross and  Neveu to prove that the groundstate of their eponymous model~\eqref{eq:GN_continuum} displays a  non-zero vacuum expectation value $\sigma_0=\langle \overline{\Psi}(x)\Psi(x)\rangle\neq 0$ $\forall x$,  as soon as a non-vanishing interaction $g^2>0$ is switched on~\cite{GN_model}. In this way,   the discrete chiral symmetry $\Psi(x)\to \mathbb{I}_N\otimes\gamma^5\Psi(x)$ gets spontaneously broken, since $\sigma_0=\langle \Psi^\dagger(x)\gamma^0\Psi(x)\rangle\to-\langle \Psi^\dagger(x)\gamma^0\Psi(x)\rangle=-\sigma_0$  is no longer fulfilled when the vacuum expectation value is developed. 

This non-perturbative result can be obtained using functional techniques to calculate an effective action for an auxiliary bosonic $\sigma(x)$ field, which condenses due  to the formation of particle--anti-particle pairs, and   acquires a non-zero expectation value  $\langle \sigma(x)\rangle=\sigma_0\neq 0$ in  the chirally-broken phase. On the lattice~\eqref{eq:naive_GN_lattice}-\eqref{eq:Wilson_lattice},  similar results are recovered in the continuum limit~\cite{Eguchi:1983gq,large_N_wilson_GN}, provided that the additional bare mass~\eqref{eq:Wilson_GN_lattice} is adjusted to recover the discrete chiral symmetry.

Let us now comment on a generalization of the GNW model, where the above discrete chiral symmetry  is upgraded to  a continuous  one $\Psi(x)\to\mathbb{I}_N\otimes\ee^{\ii \theta\gamma^5}\Psi(x)$,  $\forall \theta\in[0,2\pi)$~\cite{GN_model}. This requires a modified four-fermion term
\beq
\label{eq:GN_cont_chiral__capital_psi}
 \mathcal{H}=-\overline{\Psi}(x)\ii\gamma^1\partial_x  \Psi(x)-\frac{g^2}{2N}\!\left(\!\big(\overline{\Psi} \Psi\big)\!^2-\big(\overline{\Psi}\gamma^5 \Psi\big)\!^2\!\right)\!\!.
 \eeq
  In this case, in addition to the $\sigma(x)$ field,  it is natural to introduce an additional  bosonic field $\Pi(x)$, obtaining an effective action for both  fields  in the large-$N$ limit. In Ref.~\cite{aoki_phases_finite_T}, S. Aoki showed that the large-$N$  results with lattice Wilson fermions lead to a richer phase diagram displaying new regions where a discrete parity  symmetry $\Psi(x)\to\eta\mathbb{I}_N\otimes\gamma^0\Psi(-x)$, where $|\eta|^2=1$, can also be spontaneously broken $\Pi_0=\langle \overline{\Psi}(x)\ii\gamma^5\Psi(x)\rangle\neq 0$ $\forall x$. In this case, the particle--anti-particle pairs lead to the  so-called pseudoscalar condensate $\langle\Pi(x)\rangle=\Pi_0\neq 0$, which necessarily breaks the parity transformation of the corresponding fermion bilinear due to the vacuum expectation value $\Pi_0=\langle \overline{\Psi}(x)\ii\gamma^5\Psi(x)\rangle\to-\langle \overline{\Psi}(-x)\ii\gamma^5\Psi(-x)\rangle=-\Pi_0$ .
 Interestingly, these results on the chiral GNW model were used to conjecture that these so-called Aoki phases would also appear in the phase diagram of lattice quantum chromodynamics~\cite{aoki_phases_finite_T}. However, in this context, these Aoki phases are considered as unphysical lattice artifacts  not present in the continuum QFT.

In this section, we discuss the role of such Aoki phases in the   GN model~\eqref{eq:GN_cont_chiral__capital_psi} with a Wilson-type discretization~\eqref{eq:naive_GN_lattice}-\eqref{eq:Wilson_GN_lattice}, and their interplay with the topological insulating phases discussed in the previous sections. In the context of symmetry-protected topological phases, such Aoki phases are not artifacts, but become instead physical  phases of matter that shall delimit the region of the phase diagram that hosts a correlated SPT phase. Moreover, from the perspective of a cold-atom implementation, these phases might also be observed in future table-top experiments. We also note that the appearance  of  Aoki phases is not  restricted to the GNW model, but  also occurs in strong-coupling calculations of $U(1)$ Wilson-type lattice gauge theories~\cite{strong_coupling_wilson_2d_U(1), strong_coupling_wilson_3d_U(1)}, which can be used to model the strongly-interacting limit of higher-dimensional topological insulators with long-range Coulomb interactions.

We remark that, in the limit of a single fermion flavor $N=1$, which is the relevant case for the cold-atom implementation,   the four-fermion interactions  of Eq.~\eqref{eq:naive_GN_lattice} can be rewritten as 
\beq
\label{eq:int}
V_{g}=-\!\sum_{x\in\Lambda_{\rm s}}\!\!a\frac{g^2}{4N}\!:\!\!\left(\!\!\bigg(\overline{\Psi}(x)\Psi(x)\bigg)^{\!\!\!2}-\bigg(\overline{\Psi}(x)\gamma^5 \Psi(x)\bigg)^{\!\!\!2}\right)\!\!:,
\eeq
which follows from a so-called Fierz identity in the language of relativistic QFTs.
Accordingly, besides a change in the coupling constant $g^2\to g^2/2$, there is no further distinction between the  $N=1$ GNW model with discrete or continuous symmetry, such that the  previous Aoki phases could in principle  also occur in this limiting case. However, since their prediction is based on the $N\to\infty$ results,  we will have to benchmark large-$N$ methods with other non-perturbative approaches valid for $N=1$ (e.g. MPS numerical simulations or a potential cold-atom quantum simulation). Regarding the first approach, and give a detailed comparison of the large-$N$ predictions with the MPS results of the phase diagram.

\subsubsection{Continuous  time: Hamiltonian   field theory on the lattice}
\label{sec:con_time_large_N}

Let us first discuss the large-$N$ phase diagram of the GNW  model using a functional-integral representation of the partition function with a continuum Euclidean (i.e. imaginary) time $\tau$.  Introducing fermionic coherent states by means of  mutually anti-commuting    Grassmann variables $\Psi_k(\tau), {\Psi}^\star_k(\tau)$, which are  defined at each point of the Brilllouin zone $k\in\textrm{BZ}_{\rm s}$ and  for each imaginary time $\tau\in(0,\beta)$~\cite{comment_grassmann}, one can readily express the finite-temperature partition function as $\mathsf{Z}=\int[{\rm d}{\Psi}^\star{\rm d}{\Psi}]\ee^{-S_{\rm W}[{\Psi}^\star,{\Psi}]}$, where the Euclidean action is 
\beq
\label{eq:action_cont_time}
S_{\rm W}=\int_0^\beta\!\!{\rm d}\tau\left(\sum_{k\in{\rm BZ_s}}\!{\Psi}^\star_k(\tau)\left(\partial_\tau+\mathsf{H}_{k}(m)\right){\Psi_k}(\tau)+V_{g}({\Psi}^\star,{\Psi})\right).
\eeq
Here, $\mathsf{H}_{k}(m)=\mathbb{I}_N\otimes\mathsf{h}_{k}(m)$ is defined in terms of the single-particle Hamiltonian in Eq.~\eqref{eq:single-particle_h}.  Moreover,  
 $V_{g}({\Psi}^\star,{\Psi})$ results from substituting the fermion field operators  by the Grassmann variables  in the normal-ordered interaction~\eqref{eq:int}, which leads to quartic interactions
 \beq
 V_{g}=-\!\sum_{x\in\Lambda_{\rm s}}\!\!a\frac{g^2}{4N}\!\!\left(\!\!\bigg({\Psi}^\star_x(\tau)\gamma^0\Psi_x(\tau)\bigg)^{\!\!\!2}-\bigg({\Psi}^\star_x(\tau)\gamma^1 \Psi_x(\tau)\bigg)^{\!\!\!2}\right)\!,
 \eeq
  Let us note that the propagator associated to the free part of the action displays two poles at $k\in\{0,\pi/a\}$ when $-ma\in\{0,2\}$, which correspond to the aforementioned   Dirac fermions around the corners of the Brillouin zone.

The first step in the large-$N$ approximation is to introduce two auxiliary real scalar fields $\sigma(x),\Pi(x)$ with classical mass dimension $d_\sigma=d_\Pi=1$, such that the partition function can be expressed as a new functional integral over both Grassmann and real auxiliary fields $\mathsf{Z}=\int[{\rm d}{\Psi}^\star{\rm d}{\Psi}{\rm d}{\sigma}{\rm d}{\Pi}]\ee^{-\tilde{S}_{\rm W}[{\Psi}^\star,{\Psi},\sigma,\Pi]}$. Therefore, the new Euclidean action must fulfill  $\int[{\rm d}{\sigma}{\rm d}{\Pi}]\ee^{-\tilde{S}_{\rm W}[{\Psi}^\star,{\Psi},\sigma,\Pi]}=\ee^{-S_{\rm W}[{\Psi}^\star,{\Psi}]}$  up to an irrelevant constant, such that the thermodynamic properties of the system are not modified by the introduction of the auxiliary fields. The idea is to chose a particular action where the four-fermion terms can be understood as effective interactions carried by the auxiliary bosonic fields. Moreover, assuming that these fields are homogeneous, the new action becomes
\beq
\label{eq:action_fermion_boson}
\tilde{S}_{\rm W}=\!\!\int_0^\beta\!\!\!\!{\rm d}\tau\!\left(\sum_{k\in{\rm BZ}_{\rm s}}\!{\Psi}^\star_k(\tau)\!\left(\partial_\tau+\tilde{\mathsf{H}}_{k}\right)\!{\Psi_k}(\tau)+N\frac{N_{\rm s}a}{g^2}(\sigma^2+\Pi^2)\!\!\right)\!,
\eeq
where $\tilde{\mathsf{H}}_{k}=\mathbb{I}_N\otimes\tilde{\mathsf{h}}_{k}$, and the fermionic single-particle Hamiltonian now depends on the auxiliary bosonic  fields
\beq
\label{eq:ham_background_boson_fields}
\tilde{\mathsf{h}}_{k}=\left(m+\sigma+\frac{1-\cos ka}{a}\right)\gamma^0-\frac{\sin ka}{a}\gamma^5+\ii\Pi\gamma^1.
\eeq
Essentially, the $\sigma$ field modifies the mass term of the Dirac fermion, and a vacuum expectation value of the former would thus renormalize the fermion mass, resembling the dynamical mass generation of the continuum model.

The second step in the large-$N$ approximation is to integrate over the fermionic Grassmann fields, obtaining an effective action  for the auxiliary bosons $\mathsf{Z}=\int[{\rm d}{\sigma}{\rm d}{\Pi}]\ee^{-NS_{\rm eff}[\sigma,{\Pi}]}$. This step can be readily performed  since the Grassmann integral is Gaussian, which leads to
\beq
\label{eq:eff_action}
S_{\rm eff}=\beta  L\!\left(\frac{1}{g^2}\left(\sigma^2+\Pi^2\right)\!-\!\!\int_{k,\omega}\!\!\!\log\left(\omega^2+\epsilon_k^2(m+\sigma,\Pi)\right)\!\!\right),
\eeq
where $L=N_{\rm s}a$ is the length of the chain, and we have introduced an abbreviation for the integral over momentum and  Matsubara frequencies $\int_{k,\omega}=\int_{\rm BZ_s}\frac{{\rm d}k}{2\pi}\int_{-\infty}^{\infty}\frac{{\rm d}\omega}{2\pi}$,   assuming already the zero-temperature limit which is the regime of interest of this work. Here, the energies of the new  fermionic single-particle Hamiltonian $\epsilon_k(m+\sigma,\Pi)$ have been expressed in terms of the function
\beq
\label{eq:sp_energy}
\epsilon_k(x,y)=\frac{1}{a}\sqrt{\big(xa +1-\cos ka\big)^2+\sin^2ka+(y a)^2}.
\eeq

When the number of fermion flavors is very large $N\to\infty$, the partition function  $\mathsf{Z}=\int[{\rm d}{\sigma}{\rm d}{\Pi}]\ee^{-NS_{\rm eff}[\sigma,{\Pi}]}$  with the effective action~\eqref{eq:eff_action} yields a groundstate  obtained from the saddle point equations $\left.\partial_\sigma S_{\rm eff}\right|_{(\sigma_0,\Pi_0)}=\left.\partial_\Pi S_{\rm eff}\right|_{(\sigma_0,\Pi_0)}=0$. Non-vanishing values of  $\sigma_0,\Pi_0$ are related to the breaking of the discrete chiral or  parity symmetries discussed above. For instance, the boundary  of the aforementioned Aoki phases  can be obtained from the self-consistent solution of these saddle-point equations imposing $\Pi_0=0$. Using contour techniques for the frequency integrals, and substituting $ka\to k$ in the momentum integrals, we can express the pair of saddle-point equations as follows
 \begin{align}
 \label{eq:gap_eqs}
 \frac{1}{g^2}&=\frac{\mathsf{K}(\theta_0)}{\pi(1+\eta_0)},\hspace{3ex}
 ma=\frac{(1-\eta_0)^2}{2\eta_0}-\frac{g^2}{2\pi }\frac{(1+\eta_0)}{\eta_0}\mathsf{E}(\theta_0).
 \end{align}
 Here, we have used the complete elliptic integrals  of the  first and second kind
 \beq
 \label{eq:comp_ell_int}
 \mathsf{K}(x)=\int_0^{\frac{\pi}{2}}\!\!\frac{{\rm d}k}{\sqrt{1-x\sin^2k}}, \hspace{2ex}\mathsf{E}(x)=\int_0^{\frac{\pi}{2}}\!\!{\rm d}k\sqrt{1-x\sin^2k},
 \eeq
  as well as the following parameters
 \beq
 \label{eq:param_gap_eqs}
 \eta_0=1+ma+\sigma_0a,\hspace{2ex}\theta_0=\frac{4\eta_0}{(1+\eta_0)^2}.
 \eeq

\begin{figure}[t]
  \includegraphics[width=1\columnwidth]{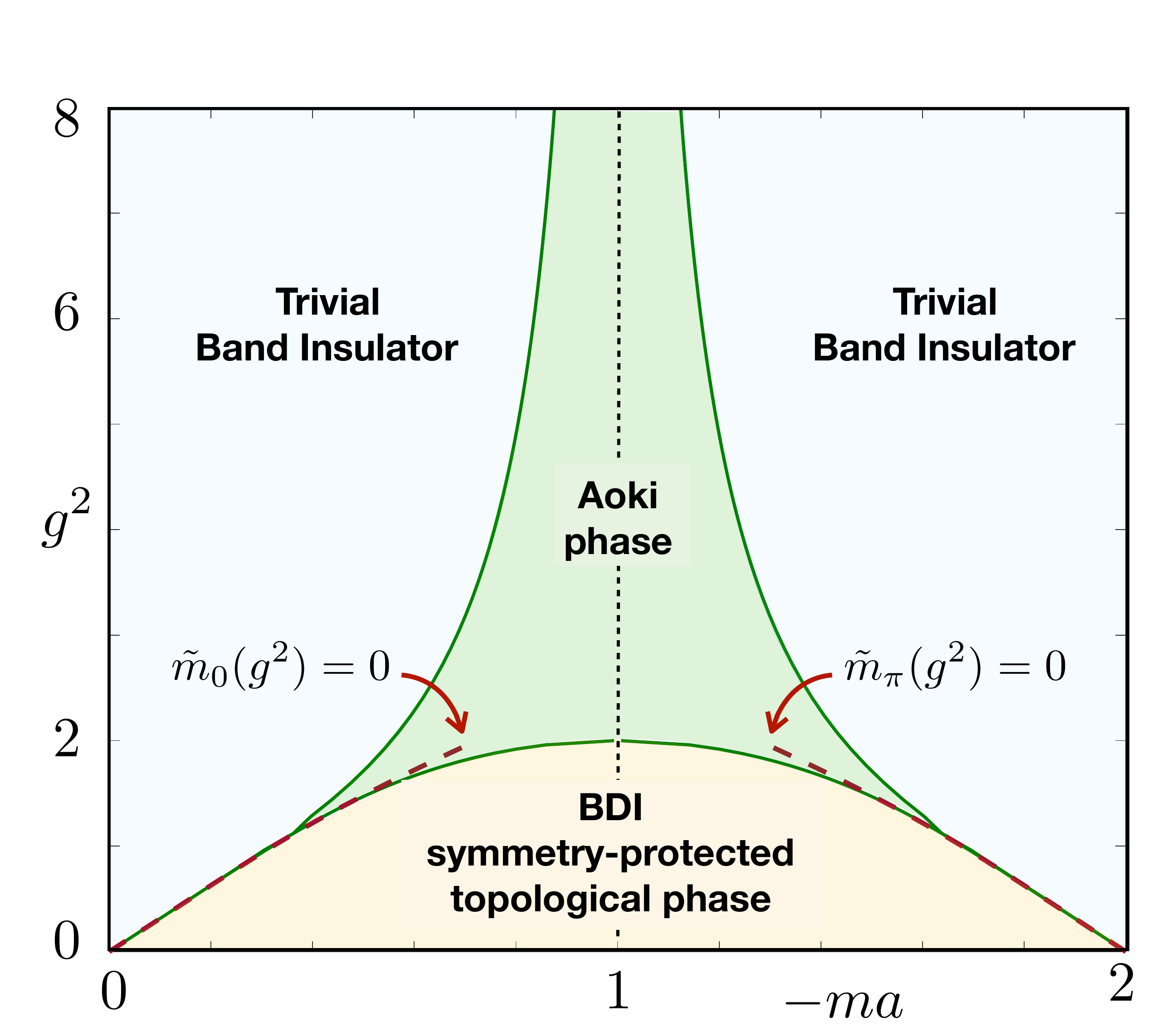}\\
  \caption{ {\bf Symmetry-protected topological phases in the lattice Gross-Neveu model from large-$N$ methods:} The two green solid lines correspond to the critical lines $m_{\rm c}(g^2)$ obtained from the numerical solution of the gap equations~\eqref{eq:gap_eqs}, while the red dashed lines corresponds to the analytical solution obtained from the vanishing of the dynamically-generated~\eqref{eq:saddle_point_solution} Wilson masses~\eqref{eq:ren_w_mass}. The identification of the different regions as a trivial and topological insulators, or as the Aoki phase, follows our discussion below Eq.~\eqref{eq:ren_w_mass}. }
  \label{Fig:large_N_aoki}
\end{figure}

 In general, the solution of the pair of gap equations~\eqref{eq:gap_eqs} must be performed numerically, and leads to the critical lines that delimit the Aoki phase (i.e. solid green lines in Fig.~\ref{Fig:large_N_aoki}).  These lines can be interpreted as different flows of the bare mass $m_{\rm c}(g^2)$ that determine the second-order phase transitions where a scale-invariant QFT should emerge. Note that this figure displays a clear  reflection symmetry with respect to the axis $-ma=1$. In fact, using the expression of the elliptic integrals in terms of hypergeometric functions~\cite{special_functions}, it follows that $\mathsf{K}(x)=(1-x)^{-1/2}\mathsf{K}(x/(x-1))$, and $\mathsf{E}(x)=(1-x)^{1/2}\mathsf{E}(x/(x-1))$, which can be exploited to show that the gap equations~\eqref{eq:gap_eqs} can be rewritten as  
  \begin{align}
 \frac{1}{g^2}&=\frac{\mathsf{K}\left(\frac{\theta_0}{\theta_0-1}\right)}{\pi(1-\eta_0)},\hspace{2ex}
 ma=\frac{(1-\eta_0)^2}{2\eta_0}-\frac{g^2}{2\pi }\frac{(1-\eta_0)}{\eta_0}\mathsf{E}\!\left(\!\frac{\theta_0}{\theta_0-1}\!\right)\!.
 \end{align}
 These gap equations can now be related to  the original ones in Eq.~\eqref{eq:gap_eqs} under the following transformation
 \beq
 \label{eq:symmetry_gap}
 ma\to-2-ma,\hspace{2ex}\sigma_0\to-\sigma_0, 
 \eeq
 which corresponds to the aforementioned reflection symmetry about $-ma=1$, and leads to $\eta_0\to-\eta_0$ and $\theta_0\to\theta_0/(\theta_0-1)$. Accordingly, there should only be three distinct phases in the regime $-ma\in[0,2]$, with the Aoki phase being completely absent for $ma>0$ and $ma<-2$.
 
  To make a connection to the continuum results~\cite{GN_model}, and interpret this phase diagram in light of the symmetry-protected topological phases of Sec.~\ref{sec:ti}, we note that a solution to the gap equations~\eqref{eq:gap_eqs} can be found analytically in the regime of small interactions and masses $g^2,|ma|\ll 1$. In this case, one can assume that  $\eta_0=1+\delta\eta_0$ with $|\delta\eta_0|\ll1$, and perform a Taylor expansion of Eq.~\eqref{eq:gap_eqs} to find that the $\sigma$-field acquires the following non-zero vacuum expectation value
 \beq
 \label{eq:saddle_point_solution}
 \tilde{\sigma}_0=\frac{g^2}{\pi a}+\frac{8}{a}\ee^{-2\pi/g^2}.
 \eeq
  The first contribution stems for the perturbative renormalization of the bare mass $ma\approx-g^2/2\pi$, while the $1/g^2$ behavior of the second contribution  highlights  that the large-$N$ expansion captures non-perturbative effects,  recalling the  chiral symmetry breaking by dynamical mass generation of  the continuum case~\cite{GN_model}. We also note that, as the UV cutoff is removed $a\to 0$, the interaction strength must decrease $g^2\to 0$ to maintain a finite scalar condensate~\eqref{eq:saddle_point_solution}, which shows that the continuum GNW model is an asymptotically free QFT.

As announced above, such a vacuum expectation value~\eqref{eq:saddle_point_solution}  then leads to a  small renormalization of the Wilson masses~\eqref{eq:wilson_masses}, $\tilde{m}_k\to \tilde{m}_k(g^2)$, valid in  the regime $g^2,|m|\ll 1$. We can thus ascertain that the large  mass of the fermion doubler will only be perturbed slightly, remaining thus at the cutoff,  and  maintaining  ${\rm sgn}(\tilde{m}_{\pi}(g^2))=+1$. Conversely,   the sign of the light-fermion mass ${\rm sgn}(\tilde{m}_0(g^2))$ may indeed change as the interactions $g^2$ are increased. According to Eq.~\eqref{eq:zak_masses}, we can write the topological invariant in this regime as
 $
\varphi_{\textrm{Zak}}=\half N\pi\big\{1-{\rm sgn}\big(\tilde{m}_0(g^2)\big)\big\}$, such that the region hosting  a correlated $\mathsf{BDI}$ topological-insulating groundstate  corresponds to  the parameter region with $\tilde{m}_0(g^2)<0$. 

In order to locate this region, we substitute the saddle-point solution ~\eqref{eq:saddle_point_solution} into Eq.~\eqref{eq:action_fermion_boson},  and  perform a long-wavelength approximation $|k|<\Lambda\ll1/a$, yielding the effective free-fermion action
\beq
\label{eq:cont_action}
\tilde{S}_{\rm W}=\int_0^\beta\!\!{\rm d}\tau\sum_{|k|<\Lambda}\!{\Psi}^\star_k(\tau)\left(\partial_\tau+\tilde{\mathsf{H}}_{k}(\tilde{\sigma}_0)\right){\Psi_k}(\tau),
\eeq
 up to an irrelevant constant. Here, we have introduced $\tilde{\mathsf{H}}_{k}(\tilde{\sigma}_0)=\mathbb{I}_N\otimes \tilde{\mathsf{h}}_{k}(\tilde{\sigma}_0)$, where the single-particle Hamiltonian for a massive Dirac fermion is
\beq 
\tilde{\mathsf{h}}_{k}(\tilde{\sigma}_0)\approx \gamma^0\left(\tilde{m}_0+\tilde{\sigma}_0 \right)-\gamma^5k,
\eeq 
which allows us to identify the renormalized  Wilson mass  
\beq
\label{eq:ren_w_mass}
\tilde{m}_0=m\to\tilde{m}_0(g^2)= m+\tilde{\sigma}_0.
\eeq 

The  leftmost red dashed line of Fig.~\ref{Fig:large_N_aoki} corresponds to the points where this renormalized mass vanishes $\tilde{m}_0(g^2)=0$. We note that this analytical solution matches the lower critical line obtained by the numerical solution of the gap equations~\eqref{eq:gap_eqs} remarkably well, even considerably beyond the perturbative regime $g^2,|ma|\ll 1$. Following Eq.~\eqref{eq:ren_w_mass}, the area below this line fulfills  $-m>\tilde{\sigma}_0$, such that the interacting Dirac fermion has a negative renormalized mass  $\tilde{m}_0(g^2)<0$, leading to $\varphi_{\textrm{Zak}}=N\pi$ and to an SPT phase for $N$ odd. 

An analogous behavior can be found in  the regime $g^2,|ma+2|\ll 1$, where the light fermion is  around $k=\pi/a$, while the heavy one corresponds to $k=0$ (i.e. the Wilson fermions interchange their roles). Using the previous symmetry~\eqref{eq:symmetry_gap} to locate the critical line $\tilde{m}_\pi(g^2)=0$ in this parameter regime, we can readily predict the value of this renormalized mass  $\tilde{m}_\pi=m+2/a\to\tilde{m}_\pi(g^2)= m+2/a-\tilde{\sigma}_0$. The vanishing of this mass  leads to the rightmost red dashed line of Fig.~\ref{Fig:large_N_aoki}, which again agrees very well with the numerical solution of the gap equations. Since the heavy fermion around $k=0$ has a large negative mass,  the topological invariant becomes $
\varphi_{\textrm{Zak}}=\half N\pi\big\{{\rm sgn}\big(\tilde{m}_\pi(g^2)\big)+1\big\}$, and one can identify the symmetry-protected phase displaying $\varphi_{\textrm{Zak}}=N\pi$ for $N$ odd, with the parameter region fulfilling $\tilde{m}_\pi(g^2)>0$, and thus $-m<2-\tilde{\sigma}_0$ (i.e. shaded yellow area below the dashed line).

 At larger couplings and intermediate masses, we must resort to the numerical solution of the gap equations, and search for a region of phase diagram that can be adiabatically connected to these two areas hosting a topological phase. This is precisely the shaded yellow lobe of Fig.~\ref{Fig:large_N_aoki}, which is separated from other phases by a gap-closing line. The area above  these lines, given by $\tilde{m}_0(g^2)>0$, and $\tilde{m}_\pi(g^2)<0$,  determines a regime where both renormalized Wilson masses have the same sign, such that  the gapped phase has no topological features, corresponding either to a trivial band insulator (grey area in Fig.~\ref{Fig:large_N_aoki}), or to the aforementioned Aoki phase where the  $\mathbb{Z}_2$ parity symmetry  $\Psi(x)\to\mathbb{I}_N\otimes\gamma^0\Psi(x)$  is spontaneously broken (green area in Fig.~\ref{Fig:large_N_aoki}).

\subsubsection{Discretized    time: Euclidean   field theory on the lattice}

We now move on to the discussion of the large-$N$ phase diagram of the GNW lattice model using a discretized Euclidean time $x_0=\tau$. This is the most common formalism  in lattice field theory computations~\cite{lattice_book}, and can become the starting point to apply other  methods such as  Monte-Carlo numerical techniques. As emphasized below, it will be important to understand the connection between the lattice and Hamiltonian approaches,  requiring a careful treatment of the continuum-time limit to understand lattice artifacts that can change qualitatively the shape of the phase diagram.

In Euclidean LFT, both  space- and time-like coordinates $\{x_\nu\}_{\nu=0,1}$ are discretized 
into an Euclidean lattice $\Lambda_{ E}=\{\boldsymbol{x}:x_0/a_0=n_\tau\in\mathbb{Z}_{N_{\tau}}, \hspace{1ex} x_1/a_1=n_{\rm s}\in\mathbb{Z}_{N_{\rm s}} \}$, where $N_{\tau}(N_{\rm s})$ is the number of lattice sites in the time (space) -like direction, and $a_0$ ($a_1$) is the corresponding lattice spacing. Therefore, a similar discussion to the one around Eqs.~\eqref{eq:naive_GN_lattice}-\eqref{eq:Wilson_GN_lattice} must also be applied to the Euclidean time derivative appearing in the action~\eqref{eq:action_cont_time}, such that  nearest-neighbor hoppings along the time-like direction  also appear. Introducing fermionic coherent states on the Euclidean lattice, and their corresponding Grassmann variables $\Psi_{\boldsymbol{x}},\overline{\Psi}_{\boldsymbol{x}}$,  the finite-temperature partition function can be expressed as $\mathsf{Z}=\int[{\rm d}\overline{\Psi}{\rm d}{\Psi}]\ee^{-S_{\rm W}^E[\overline{\Psi},{\Psi}]}$, where the Euclidean action is 
\beq
\label{eq:euclidean_lattice_action_dimensionfull}
S_{\rm W}^E=a_0a_1\!\!\sum_{\boldsymbol{x}\in\Lambda_E}\!\!\left(\mathcal{S}_{0}^E[\overline{\Psi},{\Psi}]+\mathcal{V}_{g}^E[\overline{\Psi},{\Psi}]\right).
\eeq
 Here, the action is divided into: {\it (i)} the free quadratic term
\beq
\mathcal{S}_{0}^E=\overline{\Psi}_{\boldsymbol{x}}\left(m+\sum_\nu\frac{1}{a_\nu}\right){\Psi}_{\boldsymbol{x}}-\sum_\nu\sum_{s=\pm}\overline{\Psi}_{\boldsymbol{x}}\left(1-\frac{s\hat{\gamma}_\nu}{2a_\nu}\right){\Psi}_{\boldsymbol{x}+sa_\nu{\bf e}_\nu},
\eeq
which is expressed in terms of the Euclidean gamma matrices $\hat{\gamma}_0=\gamma^0$, $\hat{\gamma}_1=\ii\gamma^1$,  and the unit vectors $\{{\bf e}_\nu\}$ of a rectangular lattice;  and {\it (ii)}  the interacting quartic term 
\beq
\label{eq:int_euc}
\mathcal{V}_{g}^E[\overline{\Psi},{\Psi}]=-\frac{g^2}{4N}\!\left(\left(\overline{\Psi}_{\boldsymbol{x}}\Psi_{\boldsymbol{x}}\right)^2-\left(\overline{\Psi}_{\boldsymbol{x}}\hat{\gamma}_5 \Psi_{\boldsymbol{x}}\right)^2\right),
\eeq
which is expressed in terms of the  chiral matrix  $\hat{\gamma}_5=\gamma^5$.

\vspace{1ex}
{\it (i)  Lattice approach with dimensionless fields:} Let us note that, in the lattice Wilson approach~\cite{detar_book}, it is customary to work with dimensionless fields $\boldsymbol{{\psi}}_{\boldsymbol{x}}=\sqrt{a_0+a_1}{\Psi}_{\boldsymbol{x}}$, and rewrite the  action as follows
$
S_{\rm W}^E=S_{\rm W,0}^E+\mathcal{V}_{\tilde{g}}^E[\overline{\boldsymbol{{\psi}}},{\boldsymbol{{\psi}}}]$.  The free part
\beq
\label{eq:dimensionless_action}
S_{\rm W,0}^E=\!\!\sum_{\boldsymbol{x}\in\Lambda_E}\!\!\left({\overline{\boldsymbol{{\psi}}}}_{\boldsymbol{x}}\left(\tilde {m}+1\right)\boldsymbol{{\psi}}_{\boldsymbol{x}}-\sum_{\nu,s}\kappa_\nu\overline{\boldsymbol{{\psi}}}_{\boldsymbol{x}}\left(1-s\hat{\gamma}_\nu\right)\boldsymbol{{\psi}}_{\boldsymbol{x}+s{\bf e}_\nu}\right),
\eeq
is expressed in term of dimensionless tunnelings $\kappa_\nu$, and the dimensionless mass $\tilde{m}$. Similarly, the interacting term  is obtained from Eq.~\eqref{eq:int_euc} by substituting the fields  $\Psi\to\boldsymbol{{\psi}}$ and the coupling constant $g\to\tilde{g}$ by the dimensionless ones.

Since the Grassmann variables  must fulfill periodic (anti-periodic) boundary conditions along the space (time) -like directions, one can move into momentum space $\boldsymbol{{\psi}}_{\boldsymbol{k}},\overline{\boldsymbol{{\psi}}}_{\boldsymbol{k}}$, where the dimensionless quasi-momenta  belong to the Euclidean Brillouin zone ${\rm BZ}_{ E}=\{\boldsymbol{k}: k_0=2\pi(n_\tau+1/2)/N_\tau ,\hspace{1ex} k_1=2\pi n_{\rm s}/N_{\rm s} \}=(0,2\pi]^2$. Then, one can  rewrite the action  as 
\beq
S^{E}_{\rm W}=\sum_{\boldsymbol{k}\in{\rm BZ}_E}\overline{\boldsymbol{{\psi}}}_{\boldsymbol{k}}\mathsf{S}_{\boldsymbol{k}}(\tilde{m}){\boldsymbol{{\psi}}_{\boldsymbol{k}}}+\sum_{\boldsymbol{x}\in\Lambda_E}\!\mathcal{V}^{E}_{\tilde{g}}(\overline{\boldsymbol{{\psi}}},{\boldsymbol{{\psi}}}).
\eeq
where we have introduced $\mathsf{S}_{\boldsymbol{k}}(\tilde{m})=\mathbb{I}_N\otimes \mathsf{s}_{\boldsymbol{k}}(\tilde{m})$, together with the single-flavor action
\beq
\label{eq:sp_action}
\mathsf{s}_{\boldsymbol{k}}(\tilde{m})=\left(\tilde{m}+1-2\sum_{\nu}\kappa_\nu\cos k_\nu \right)\mathbb{I}_2+2\ii\sum_\nu \kappa_\nu\sin k_\nu \hat{\gamma}_\nu.
\eeq
Let us note that, in contrast to the continuum-time free action~\eqref{eq:action_cont_time}, this Euclidean action leads to a propagator with four poles at $\boldsymbol{k}\in\{(0,0), (0,\pi), (\pi,0), (\pi,\pi)\}$ when the bare mass equals $-\tilde{m}\in\{0,4\kappa_0,4\kappa_1,4(\kappa_0+\kappa_1)\}$, each of which corresponds to a long-wavelength Dirac fermion. Accordingly, there is an additional  doubling due to the discretization of the the Euclidean time direction (i.e. the extra fermions with $k_0=\pi$ shall be referred to as time doublers).

  \begin{figure*}[t]
  \includegraphics[width=2.05\columnwidth]{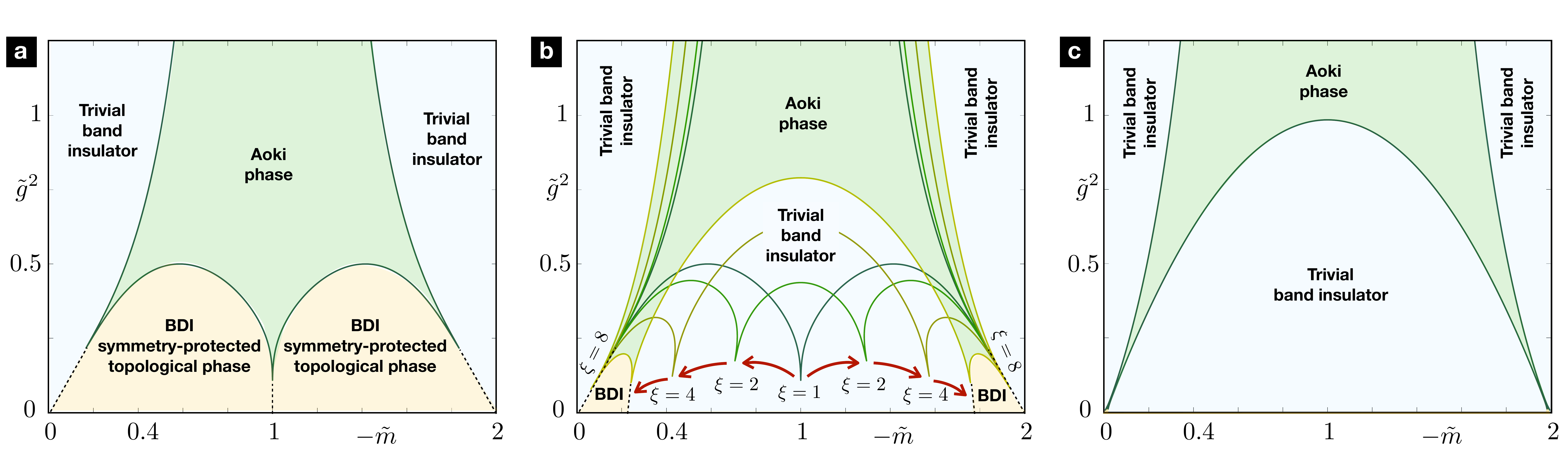}\\
  \caption{ {\bf Symmetry-protected topological phases in the  Gross-Neveu model on anisotropic Euclidean lattices:} Large-$N$ phase diagram obtained from the self-consistent solution of Eqs.~\eqref{eq:euclidean_gap_eqs_adim_1}-\eqref{eq:euclidean_gap_eqs_adim_2} on: {\bf (a)} Isotropic Euclidean lattice $\xi=1$ with $N_{\tau}=N_{\rm s}=512$ sites (i.e. square lattice). In comparison to the continuum-time phase diagram of Fig.~\ref{Fig:large_N_aoki}, one observes that the region of the symmetry-protected topological phase (shaded yellow area) is deformed into a pair of lobes, each of which  contains an $\mathsf{BDI}$  topological insulator   in the space-like dimension. The Aoki phase (green area) adopts a trident-like shape. {\bf (b)} Sequence of anisotropic Euclidean lattices with $N_{\rm s}=N_{\tau}/\xi=512$ sites, and $\xi=1,2,4,8$ (i.e. rectangular lattices). The solid lines represent how the boundary of the Aoki phase gets distorted as the anisotropy increases. The two $\mathsf{BDI}$  topological-insulating lobes shrink and move in opposite directions, leaving an intermediate lobe hosting a trivial band insulator. {\bf (c)} Anisotropic Euclidean lattice with $\xi=128$, showing the fate of the topological phases in the continuum limit.}
  \label{Fig:large_N_aoki_euclidean_dimensionless}
\end{figure*}

At this point, the discussion  parallels that of the  Hamiltonian formalism  of Sec.~\ref{sec:con_time_large_N} via the corresponding steps for the large-$N$ approximation. First, the auxiliary dimensionless lattice fields  $\tilde{\sigma}_{\boldsymbol{x}}, \tilde{\Pi}_{\boldsymbol{x}}$ are introduced, such that the action can be rewritten as
$\mathsf{Z}=\int[{\rm d}\overline{\boldsymbol{{\psi}}}{\rm d}{\boldsymbol{{\psi}}}{\rm d}{\tilde{\sigma}}{\rm d}{\tilde{\Pi}}]\ee^{-\tilde{S}^E_{\rm W}[\overline{\boldsymbol{{\psi}}},{\boldsymbol{{\psi}}},\tilde{\sigma},\tilde{\Pi}]}$, where 
\beq
\label{eq:action_fermion_boson_euclidean}
\tilde{S}^E_{\rm W}=\!\sum_{\boldsymbol{k}\in{\rm BZ}_E}\!\!\overline{\boldsymbol{{\psi}}}_{\boldsymbol{k}}\tilde{\mathsf{S}}_{\boldsymbol{k}}(\tilde{m}+\tilde{\sigma},\Pi){\boldsymbol{{\psi}}_{\boldsymbol{k}}}+N\frac{N_{\rm s}N_{\tau}}{\tilde{g}^2}(\tilde{\sigma}^2+\tilde{\Pi}^2).
\eeq
Here, we have assumed again that the auxiliary fields are homogeneous, introducing $\tilde{\mathsf{S}}_{\boldsymbol{k}}(\tilde{\sigma},\tilde{\Pi})=\mathbb{I}_N\otimes \tilde{\mathsf{s}}_{\boldsymbol{k}}(\tilde{\sigma},\tilde{\Pi})$, such that the new single-flavor action can be obtained from Eq.~\eqref{eq:sp_action} using  $\tilde{\mathsf{s}}_{\boldsymbol{k}}(\tilde{m}+\tilde{\sigma},\Pi)={\mathsf{s}}_{\boldsymbol{k}}(\tilde{m}+\tilde{\sigma})+\ii\hat{\gamma}_5\tilde{\Pi}$. The second and third steps are the same, since the action is quadratic in Grassmann fields, and the saddle-point solutions control the large-$N$ limit. In this case, the gap equations can be expressed as 
\begin{widetext}

\begin{align} 
\label{eq:euclidean_gap_eqs_adim_1}
\frac{N_{\rm s}N_\tau}{\tilde{g}^2}&=\sum_{\boldsymbol{k}\in{\rm BZ}_E}\frac{1}{\left(1-2\sum_\nu\kappa_\nu\cos k_\nu+\tilde{m}+\tilde{\sigma}\right)^2+\sum_\nu4\kappa_\nu^2\sin^2 k_\nu+\tilde{\Pi}^2},  \\ 
\label{eq:euclidean_gap_eqs_adim_2}
-\frac{\tilde{m}N_{\rm s}N_\tau}{\tilde{g}^2}&=\sum_{\boldsymbol{k}\in{\rm BZ}_E}\frac{1-2\sum_\nu\kappa_\nu\cos k_\nu}{\left(1-2\sum_\nu\kappa_\nu\cos k_\nu+\tilde{m}+\tilde{\sigma}\right)^2+\sum_\nu4\kappa_\nu^2\sin^2 k_\nu+\tilde{\Pi}^2},
\end{align} 
\end{widetext}
which are equivalent to those derived in~\cite{Eguchi:1983gq} upon a different definition of the microscopic couplings.

We have solved this system of non-linear equations  for different Euclidean lattices with $N_\tau=\xi N_{\rm s}$, setting $N_{\rm s}=512$ sites in the space-like direction, and using $\xi=\frac{a_1}{a_0}\in\{1,2,4,\cdots, 128\}$ to approach the time-continuum limit $\xi\to\infty$ (see Fig.~\ref{Fig:large_N_aoki_euclidean_dimensionless}). Let us note that the dimensionless tunnelings can be expressed in terms of the anisotropy parameter as  $\kappa_0=\xi/2(1+\xi)$, and $\kappa_1=1/2(1+\xi)$. At this point, it is worth mentioning that the number of lattice sites in the time-like direction $N_\tau$ is also modified in the LFT community to explore non-zero temperatures. In that case, however, the $\kappa_\nu$ parameters remain constant as $N_\tau$ is varied (i.e. the Euclidean lattice is rectangular, but the unit vectors remain the same).

In Fig.~\ref{Fig:large_N_aoki_euclidean_dimensionless}{\bf (a)}, we represent the solution of the gap equations for the isotropic lattice $\xi=1$,  such that $\kappa_0=\kappa_1=\fourth$. We note that the characteristic trident-shaped phase diagram is in qualitative agreement with the results of S. Aoki~\cite{aoki_phases_finite_T}. In order to interpret this phase diagram in terms of the symmetry-protected topological phases, let us recall the distribution of the poles described below Eq.~\eqref{eq:sp_action}. 
At $\tilde{g}^2=0$, we observe that  the critical points separating the different phases correspond to  $-\tilde{m}\in\{0,1,2\}$, which lie exactly at the aforementioned poles signaling the massless Dirac fermions. For $-\tilde{m}\in(0,1)$, the only  Dirac fermion with a negative mass is that around  $\boldsymbol{k}=(0,0)$, while the other 3 doublers have a positive mass. According to the Euclidean generalization of Eq.~\eqref{eq:zak_masses}, namely
\beq
\label{eq:zak_masses_Euclidean}
\begin{split}
\varphi_{\textrm{Zak}}=\half N\pi&\bigg({\rm sgn}\left(\tilde{m}_{(0,\pi)}\right)-{\rm sgn}\left(\tilde{m}_{(0,0)}\right)\\
&+{\rm sgn}\left(\tilde{m}_{(\pi,0)}\right)-{\rm sgn}\left(\tilde{m}_{(\pi,\pi)}\right)\bigg),
\end{split}
\eeq
we see that $\varphi_{\textrm{Zak}}=N\pi$ for $-\tilde{m}\in(0,1)$,  corresponding to the $\mathsf{BDI}$ topological insulator for $N$ odd. For $-\tilde{m}\in(1,2)$, the Wilson fermions around $\boldsymbol{k}=(0,\pi)$ and $\boldsymbol{k}=(\pi,0)$ also invert their masses, leading to  $\varphi_{\textrm{Zak}}=-N\pi$, and yielding again an $\mathsf{BDI}$ topological insulator for $N$ odd. These two areas, extend on to the neighboring lobes of Fig.~\ref{Fig:large_N_aoki_euclidean_dimensionless}{\bf (a)} using a similar reasoning as the one presented around Eq.~\eqref{eq:cont_action}. Therefore, the whole region below the trident that delimits the parity-broken Aoki phase corresponds to the $\mathsf{BDI}$ topological insulator. We note, however, that the black dashed lines in this figure, and subsequent ones, do not follow from the solution of the large-$N$ gap equations, but  are included as a  useful guide to the eye  to delimit the SPT phases. In Sec.~\ref{sec:mps_benchmark} below, we will show that they indeed correspond to a critical line delimiting the SPT phase of a carefully-defined  time-continuum limit.

Let us start exploring how this phase diagram changes as the time-continuum limit is approached, and compare the results to those of  Fig.~\ref{Fig:large_N_aoki}. In Fig.~\ref{Fig:large_N_aoki_euclidean_dimensionless}{\bf (b)}, we represent the phase boundaries for an increasing number of lattice sites $N_\tau=\xi N_{\rm s}$ with anisotropies $\xi\in\{1,2,4,8\}$. Here, one can observe how the central prong of the Aoki phase separating the topological-insulating lobes is split into two peaks, each of which goes in a different direction as $\xi$ is increased. We note that this behavior differs markedly from the finite-temperature studies, which show that the lobe structure disappears completely as $N_\tau$ is varied~\cite{aoki_phases_finite_T}. Therefore, the anisotropy in the lattice constants gives rise to a different playground, which must be understood in terms of the symmetry-protected topological phases.

 Since $\kappa_0\to1/2$, while $\kappa_1\to 0$, as the anisotropy $\xi\to\infty$, one can identify the left-moving prong with the pole at $\boldsymbol{k}=(0,\pi)$ with mass  $-ma\to4\kappa_1\to 0$, and thus approaching the lower left corner. Similarly, the right-moving one can be identified  with the pole at $\boldsymbol{k}=(\pi,0)$ with mass  $-ma=4\kappa_0\to 2$ approaching the lower right corner. As a result of this movement, and considering the signs of the  corresponding Wilson masses, one finds that the region between these two poles correspond to a situation where both space- (time-) like doublers have a negative (positive) mass, such that the topological invariant vanishes $\varphi_{\textrm{Zak}}=0$, and one gets a trivial band insulator. Unfortunately, as the anisotropy increases, the two $\mathsf{BDI}$ topological lobes get smaller and smaller, such that the symmetry-protected topological phases vanish as we approach the time-continuum limit, and the central lobe corresponds to a trivial band insulator  (see Fig.~\ref{Fig:large_N_aoki_euclidean_dimensionless}{\bf (c)}). 
 
 This result seems to be in contradiction with our findings  for the Hamiltonian formalism  in  Fig.~\ref{Fig:large_N_aoki}, which predict that the central lobe should correspond to the correlated SPT phase with $\varphi_{\textrm{Zak}}=N\pi$. Moreover, since each of the two prongs now contain a pair of massless Dirac fermions, the continuum QFT that should emerge in the long-wavelength limit is no-longer that of the Gross-Neveu model for $N$ flavors, but rather that of the Gross-Neveu model for $2N$ flavors, which would indeed modify  the universal features of the  phase transition, and not only the non-universal shape of the critical line. As  mentioned at the beginning of this section, the Euclidean approach can lead to lattice artifacts that can modify qualitatively the phase diagram, and a detailed and careful account of the time-continuum limit is required to understand them. We address precisely this issue in the two following subsections.

\vspace{1ex}
{\it (ii)  Large-N phase diagrams with rescaled couplings:} We have found that one of the problems  leading to the apparent contradiction between the phase diagrams is the standard use of dimensionless quantities in the Euclidean lattice approach~\eqref{eq:dimensionless_action}. A detailed derivation of this action, which starts from the original action~\eqref{eq:euclidean_lattice_action_dimensionfull} rescaling the fields, shows that the dimensionless parameters are related to the original ones by the following expression 
\beq
\label{eq:rescaling}
\tilde{m}=\frac{1}{1+\xi}ma_1,\hspace{1ex}\tilde{g}^2=\frac{\xi}{(1+\xi)^2}g^2.
\eeq

\begin{figure}[t]
  \includegraphics[width=1\columnwidth]{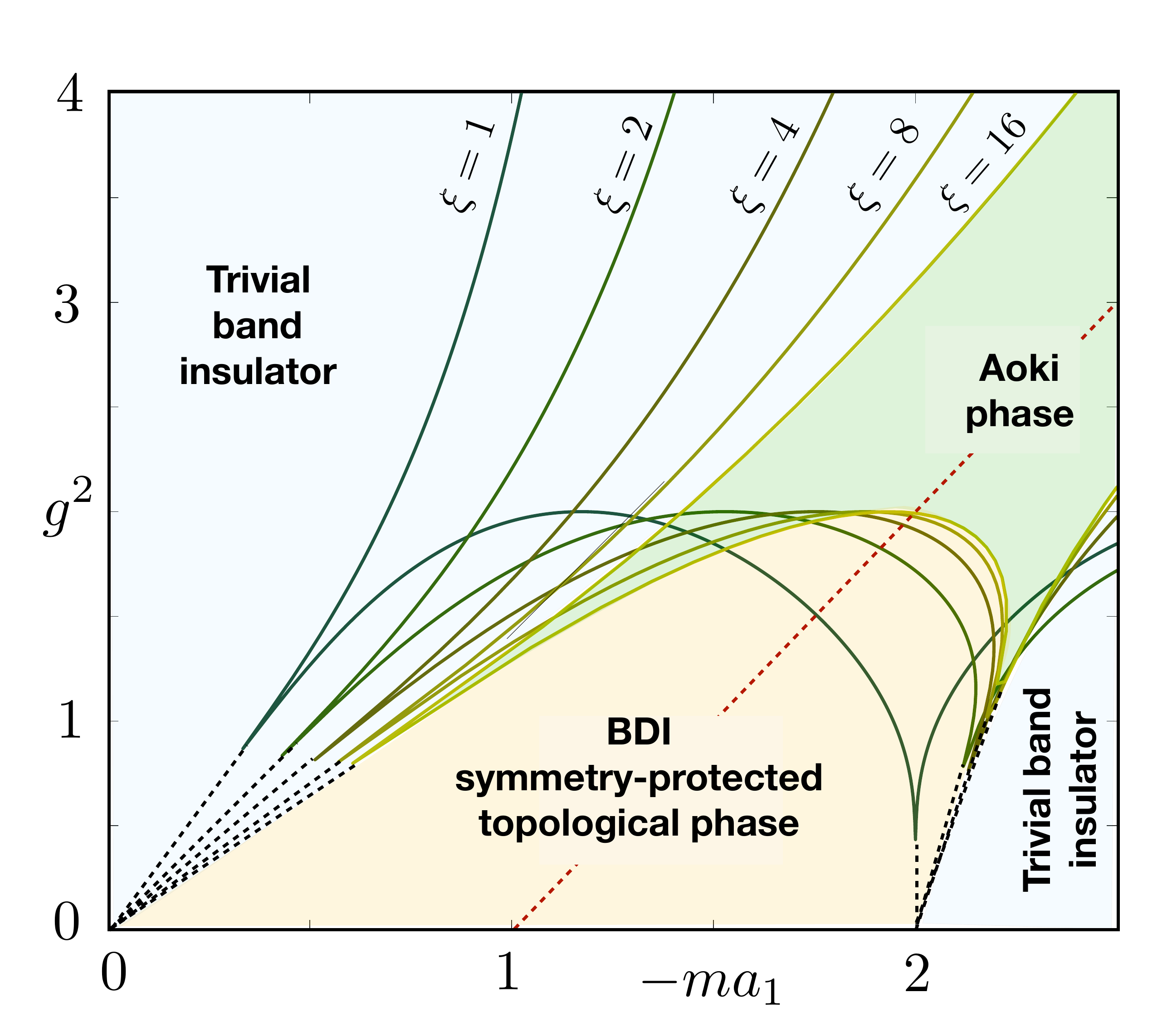}\\
  \caption{ {\bf Phase diagram  in the Euclidean lattice Gross-Neveu model with rescaled couplings:} The solid lines delimiting the Aoki phase with rescaled couplings~\eqref{eq:rescaling}  get distorted as the anisotropy parameter increases $\xi\in\{1,2,4,8,16\}$, and display a finite lobe hosting the symmetry-protected topological phase even in the continuum limit. As $\xi$ is increased, the  $\mathsf{BDI}$ topological insulator  and the Aoki phase display a mirror symmetry with respect to the red dashed line. }
  \label{Fig:large_N_aoki_euclidean_rescaled}
\end{figure}

Although apparently innocuous, this rescaling changes qualitatively the shape of the phase diagram (see Fig.~\ref{Fig:large_N_aoki_euclidean_rescaled}). In order to understand the main features of this phase diagram, the location of the non-interacting poles will be very useful again. For instance, at $g^2=0$, we note that the pole at $-\tilde{m}=4\kappa_0$ gets mapped into $-ma_1=4(1+\xi)\kappa_1$. Therefore, as the time-continuum limit is approached, this pole tends to $-ma_1\to2$ as $\xi\to\infty$, and  no longer  to the origin. Likewise, both time-like doublers at $-\tilde{m}\in\{4\kappa_0,4(\kappa_0+\kappa_1)\}$ are mapped  into $-ma_1\to\infty$ in the time-continuum limit. Accordingly,   in the region of interest displayed in Fig.~\ref{Fig:large_N_aoki_euclidean_rescaled}, these time doublers have a very large positive mass. Inspecting the sign of the corresponding Wilson masses, we can conclude that the region $-ma_1\in(0,2)$ will host an $\mathsf{BDI}$ topological insulator, while a trivial insulator will set in for $-ma_1>2$. 

Following a similar reasoning as in previous subsections, we know that these critical points surrounding the topological phase will flow as  the interactions are switched on and the $\sigma$ field acquires an non-zero vacuum expectation value. Accordingly, we identify the lobe of Fig.~\ref{Fig:large_N_aoki_euclidean_rescaled} as the $\mathsf{BDI}$ topological insulator that also appeared in the continuum-time Hamiltonian formalism of Fig.~\ref{Fig:large_N_aoki}. Moreover, the universal features are now in agreement as the critical lines are  controlled by a single pole, and the long-wavelength limit should now be controlled by the Gross-Neveu model for $N$ flavors.

Let us remark that, although the rescaled solution looks somewhat closer to the Hamiltonian results, there are still qualitative differences in the lattice approach that deserve a deeper understanding. For instance, the phase diagram does no longer display the mirror symmetry about $-ma_1=1$~\eqref{eq:symmetry_gap}. 

\vspace{1ex}
{\it (iii)  Continuum limit and connection to the Hamiltonian approach:} In order to understand these differences, and the connection to the gap equations
continuum limit~\eqref{eq:gap_eqs}, let us consider the original action with dimensional fields~\eqref{eq:euclidean_lattice_action_dimensionfull}. Following the same steps as before, 
one can  integrate the fermion fields,  $\mathsf{Z}=\int[{\rm d}{\sigma}{\rm d}{\Pi}]\ee^{-NS_{\rm eff}[\sigma,{\Pi}]}$, finding the following effective action 
\beq
S_{\rm eff}=\frac{L_{\rm s}L_{\tau}}{g^2}\!\left(\sigma^2+\Pi^2\right)\!-\!\sum_{\boldsymbol{k}\in{\rm BZ}_E}\!\!\log\!\left(\frac{\sin^2k_0a_0}{a_0^2}+s_{\boldsymbol{k}}^2(m+\sigma,\Pi)\right),
\eeq
which is the Euclidean lattice version of Eq.~\eqref{eq:eff_action}. Here, we have introduced the corresponding lengths $L_\tau=N_\tau a_0$, $L_{\rm s}=N_{\rm s}a_1$,  together with the  following function 
\beq
s_{\boldsymbol{k}}^2(x,y)=\left(x  +\sum_\nu\frac{1-\cos k_\nu a_\nu}{a_\nu}\right)^2+\frac{\sin^2k_1a_1}{a_1^2}+y^2.
\eeq

If we now take the limit of $N\to\infty$,  the saddle point conditions $\left.\partial_\sigma S_{\rm eff}\right|_{(\sigma_0,\Pi_0)}=\left.\partial_\Pi S_{\rm eff}\right|_{(\sigma_0,\Pi_0)}=0$ lead to the following pair of gap equations, which are equivalent to Eqs.~\eqref{eq:euclidean_gap_eqs_adim_1}-\eqref{eq:euclidean_gap_eqs_adim_2} but using dimensional couplings and dimensional fields, 
\beq
\label{eq:euclidean_gap_eqs}
\begin{split}
\frac{L_{\rm s}L_\tau}{g^2}&=\sum_{\boldsymbol{k}\in{\rm BZ}_E}\frac{1}{\sin^2(k_0a_0)/a_0^2+s_{\boldsymbol{k}}^2(m+\sigma,\Pi)},\\
-\frac{mL_{\rm s}L_\tau}{g^2}&=\sum_{\boldsymbol{k}\in{\rm BZ}_E}\frac{\sum_\nu(1-\cos k_\nu a_\nu)/a_\nu}{\sin^2(k_0a_0)/a_0^2+s_{\boldsymbol{k}}^2(m+\sigma,\Pi)}.
\end{split}
\eeq

In order to make a connection to the gap equations obtained with the Hamiltonian formalism~\eqref{eq:gap_eqs}, we should take the continuum limit in the imaginary time direction $N_{\tau}\to\infty$, and $a_0\to 0$, such that $L_\tau=L_{\rm s}$ remains constant imposing $\xi=a_1/a_0\to\infty$. To deal with the additional time doublers mentioned above, let us introduce a UV cutoff $\Lambda_\tau\ll 1/a_0$, and   make a long-wavelength approximation around $k_0\in\{0,\pi/a_0\}$. We find that  the  gap equation~\eqref{eq:euclidean_gap_eqs_adim_1} becomes
\beq
\begin{split}
\frac{L_{\rm s}L_\tau}{g^2}&\approx\sum_{|k_0|<\Lambda_\tau}\sum_{{k_1}\in{\rm BZ}_{\rm s}}\frac{1}{(k_0)^2+\epsilon_{k_1}^2(m+\sigma,\Pi)}+\\
&+\sum_{|k_0|<\Lambda_\tau}\sum_{{k_1}\in{\rm BZ}_{\rm s}}\frac{1}{(k_0)^2+\epsilon_{k_1}^2(m+2/a_0+\sigma,\Pi)},
\end{split}
\eeq
where we have used the single-particle energies of Eq.~\eqref{eq:sp_energy} and the spatial Brillouin zone, after identifying $a=a_1$. We note that the first line of this expression comes from the contribution around $k_0=0$, while the second line stems from the time doublers around $k_0=\pi/a_0$.

 We observe that the effective Wilson mass of these doublers becomes very large in the continuum limit  $m+2/a_0\to \infty$ if one keeps the bare mass $m$ non-zero. Hence, these doublers become very massive, and their contribution to above gap equation should become vanishingly small as described below Eq.~\eqref{eq:rescaling}. To prove that, let us  get rid of the cutoff $\Lambda_{\tau}\to\infty$,  and use $\sum_{|k_0|<\Lambda_\tau}\to L_\tau\int_{-\infty}^{\infty}{\rm d}k_0/2\pi$. After performing the integral using contour techniques, we directly  obtain 
\beq
\label{eq:euclidean_cont_gap_1}
\begin{split}
\frac{1}{g^2}&=\int_{-\pi}^\pi\frac{{\rm d}k_1}{4\pi}\frac{1}{\sqrt{(ma+\sigma a+(1-\cos k_1 a))^2+\sin^2k_1a}}\\
&+\int_{-\pi}^\pi\frac{{\rm d}k_1}{4\pi}\frac{1}{\sqrt{(ma+\sigma a+(1+2\xi-\cos k_1 a))^2+\sin^2k_1a}},
\end{split}
\eeq
where we have also taken the continuum limit in the space-like direction. Using the definition of the complete elliptic integrals~\eqref{eq:comp_ell_int}, this equation can be expressed
\beq
\label{eq:first_gap_equation}
\frac{1}{g^2}=\frac{\mathsf{K}(\theta_0)}{\pi(1+\eta_0)}+\frac{\mathsf{K}(\tilde{\theta}_0)}{\pi(1+\tilde{\eta}_0)}.
\eeq
Here, we have used the parameters of Eq.~\eqref{eq:param_gap_eqs}, together with
\beq
 \tilde{\eta}_0=1+ma+2\xi+\sigma_0a,\hspace{2ex}\tilde{\theta}_0=\frac{4\tilde{\xi}_0}{(1+\tilde{\xi}_0)^2},
\eeq
which determine the contribution of the time doublers to the gap equation (i.e. second term of  Eq.~\eqref{eq:euclidean_cont_gap_1}).
In the continuum limit, we take $\xi\to\infty$, such that $\tilde{\eta}_0\to\infty$, and $\tilde{\theta}_0\to 0$. This  makes $\mathsf{K}(\tilde{\theta}_0)\to \pi/2$, such that the time-doubler contribution  vanishes, and we recover exactly the gap equation of the Hamiltonian approach~\eqref{eq:gap_eqs}.

 \begin{figure}[t]
  \includegraphics[width=1\columnwidth]{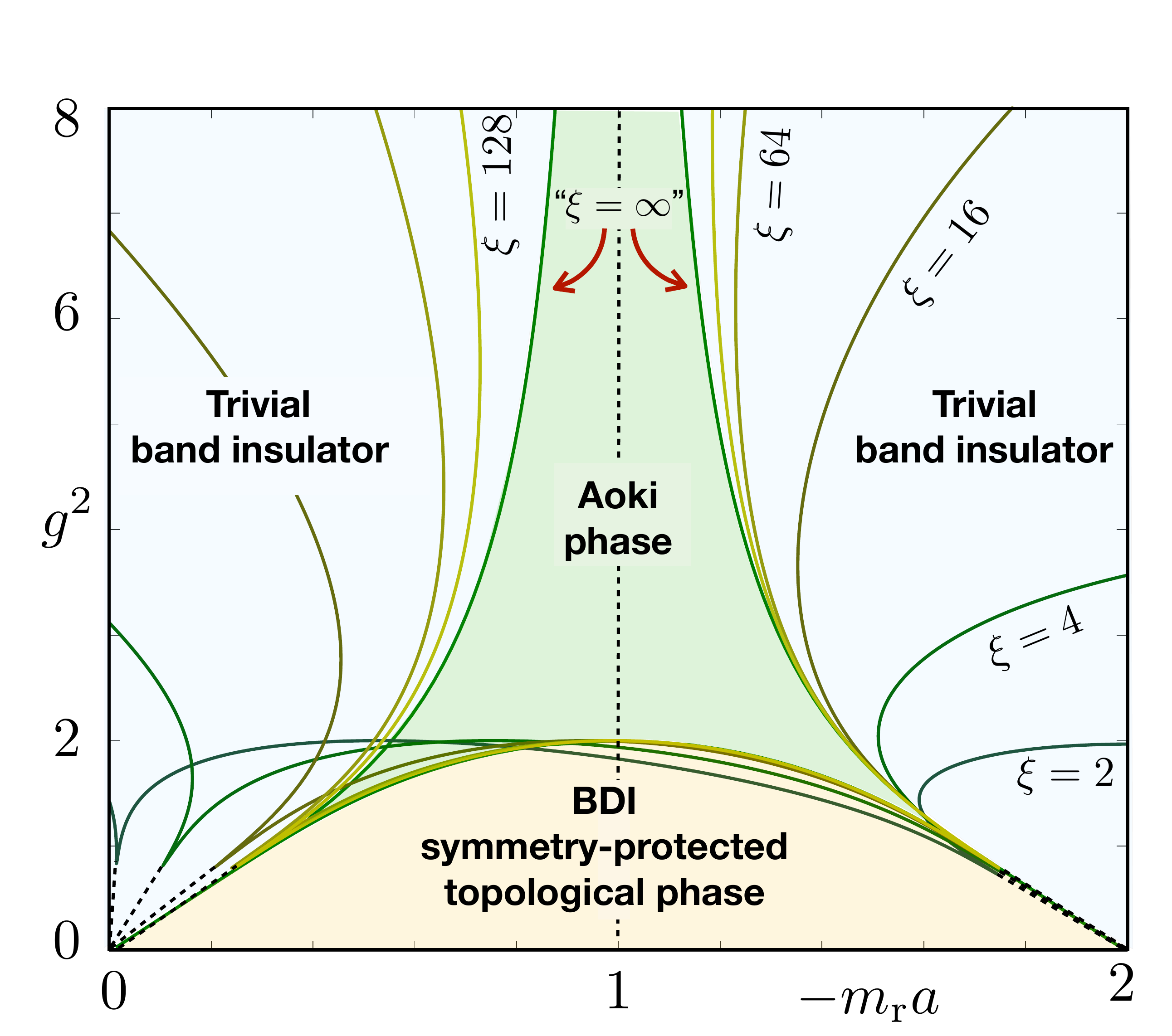}\\
  \caption{ {\bf Phase diagram  in the Euclidean lattice Gross-Neveu model with renormalized mass:} The solid lines delimiting the Aoki phase as a function of the renormalized mass for $\xi\in\{2,4,16,64,128\}$ show the tendency towards the mirror-symmetric  behavior about $-m_{\rm r}a=1$. The Hamiltonian prediction is also displayed, where $m_{\rm r}a=ma$, which is labelled by ``$\xi=\infty$''. }
  \label{Fig:large_N_aoki_euclidean_renormalised}
\end{figure}

The continuum limit of the remaining gap equation~\eqref{eq:euclidean_gap_eqs_adim_2} follows the same lines: we   perform a long-wavelength approximation around the time doublers,  let the cutoff  $\Lambda_{\tau}\to\infty$, and use contour integration to find
\beq
\begin{split}
-\frac{m a}{g^2}&=\int_{-\pi}^\pi\frac{{\rm d}k_1}{4\pi}\frac{1-\cos k_1 a}{\sqrt{(ma+\sigma a+(1-\cos k_1 a))^2+\sin^2k_1a}}\\
+&\int_{-\pi}^\pi\!\!\frac{{\rm d}k_1}{4\pi}\frac{1+2\xi-\cos k_1 a}{\sqrt{(ma+\sigma a+(1+2\xi-\cos k_1 a))^2+\sin^2k_1a}}.
\end{split}
\eeq
Using the definition of  the complete elliptic integrals~\eqref{eq:comp_ell_int}, this  equation becomes  
\beq
\begin{split}
\frac{-ma}{g^2}&=\frac{(1-\eta_0)^2\mathsf{K}(\theta_0)}{2\pi(1+\eta_0)}+\frac{(1+\eta_0)^2\mathsf{E}({\theta}_0)}{2\pi\eta_0(1+{\eta}_0)}\\
&+\frac{2\xi\mathsf{K}(\tilde{\theta}_0)}{\pi(1+\tilde{\eta}_0)}+\frac{(1+\tilde{\eta}_0)^2\mathsf{E}(\tilde{\theta}_0)-(1-\tilde{\eta}_0)^2\mathsf{K}(\tilde{\theta}_0)}{2\pi\tilde{\eta}_0(1+\tilde{\eta}_0)},
\end{split}
\eeq
where the contribution of the time doublers is  expressed in the second line. In this case, taking the time-continuum limit $\xi\to\infty$, such that $\mathsf{E}(\tilde{\theta}_0)\to \pi/2$,  leads to 
\beq
 ma=\frac{(1-\eta_0)^2}{2\eta_0}-\frac{g^2}{2\pi }\frac{(1+\eta_0)}{\eta_0}\mathsf{E}(\theta_0)-\frac{g^2}{2},
\eeq
which contains an additional $-g^2/2$ term with respect to the  gap equation of the Hamiltonian formalism~\eqref{eq:gap_eqs}. 

We thus  find that, in contrast to the first gap equation~\eqref{eq:first_gap_equation},  the contribution of the time doublers is no longer vanishing in this case, but can instead be understood as a finite renormalization of the bare mass 
\beq
\label{eq:ren_mass}
ma\to m_{\rm r}a=ma+g^2/2.
\eeq
 It is precisely this renormalization which is responsible for the lack of the mirror symmetry~\eqref{eq:symmetry_gap} in Fig.~\ref{Fig:large_N_aoki_euclidean_rescaled}, and its qualitative difference with respect to the Hamiltonian prediction of Fig.~\ref{Fig:large_N_aoki}. These results can thus help us to identify the corresponding mirror symmetry, which is no longer about the vertical line $-ma=1$, but instead about  $-m_{\rm r}a=1$, which corresponds to the  red dashed line $-ma=1+g^2/2$ of Fig.~\ref{Fig:large_N_aoki_euclidean_rescaled}. 

To study in more detail the onset of this symmetry in the continuum limit, and the quantitative agreement with the Hamiltonian prediction, we plot the phase diagram with the corresponding renormalized mass in Fig.~\ref{Fig:large_N_aoki_euclidean_renormalised}, and superimpose the continuum-time prediction of Fig.~\ref{Fig:large_N_aoki}. This figure shows the clear agreement between both approaches, and highlights the importance of performing a careful analysis of the continuum limit in order to avoid Euclidean lattice artifacts that can lead to qualitatively different predictions, even questioning the universal aspects of the emerging QFTs (see Fig.~\ref{Fig:large_N_aoki_euclidean_dimensionless}). It also highlights the fact that the time doublers, despite becoming infinitely heavy in the continuum limit, can leave an imprint in the non-universal properties of the low-energy  phase diagram, such as the particular value of the critical points (see the tilted phase diagram of Fig.~\ref{Fig:large_N_aoki_euclidean_renormalised}). From the perspective of the renormalization group, this effect does not come as a surprise, since the time doublers lie at the cutoff of the continuum-time limit of the lattice field theory, and their integration can thus renormalize the parameters of the long-wavelength light-fermion modes. In this case, a careful analysis of the gap equations has allowed us to extract  an additive  renormalization $\delta m=g^2/2a$ which, as usual in discretized QFTs, depends on the remaining UV cutoff and shows that the bare mass must be fine tuned to a cutoff-dependent value to yield the physical mass of the low-energy excitations.  

\subsubsection{Extent of the Aoki phase and tri-critical points}

Let us now focus on a question of interest that has not been discussed in detail in the previous sections, namely the extent of the Aoki phase. As already noted above, the solution of the gap equations, either in the Hamiltonian theory~\eqref{eq:gap_eqs} or in the time-continuum limit of the Euclidean approach~\eqref{eq:euclidean_gap_eqs_adim_1}-\eqref{eq:euclidean_gap_eqs_adim_2}, can only determine the critical lines that delimit the Aoki phase with $\Pi\neq 0$. The question we consider in this section is whether the Aoki phase extends all the way down to $(ma,g^2)=(0,0)$, and $(ma,g^2)=(-2,0)$ or, instead, it terminates at a non-zero value of the coupling strength. In this later case, there would be a direct transition between the $\mathsf{BDI}$ symmetry-protected phase and the trivial band insulator, not separated by an intermediate parity-broken Aoki phase.

 In order to address this point, we apply large-$N$ techniques away from half filling via
the introduction of a  chemical potential $\tilde\mu$ in the GNW model. Following the orthodox
prescription for Euclidean LFTs~\cite{Hasenfratz:1983ba}, the
hopping term $\kappa_\nu\left(1-s\hat{\gamma}_\nu\right)$ of the Euclidean action $S_{\rm W}^E$~\eqref{eq:dimensionless_action} is
modified to $e^{s\tilde\mu\delta_{\nu0}}\kappa_\nu\left(1-s\hat{\gamma}_\nu\right)$, such that time-like hopping is promoted
in the forwards direction by a factor $e^{\tilde\mu}$, and suppressed by $e^{-\tilde\mu}$ when hopping backwards. As a consequence, one can study the phase diagram of the GNW  model at finite densities by solving the  gap equations~\eqref{eq:euclidean_gap_eqs_adim_1}-\eqref{eq:euclidean_gap_eqs_adim_2}  with the sum over over  the time-like momenta now given by $k_0=2\pi(n_\tau+1/2)/N_\tau-{\rm i}\tilde\mu$. 

 Moreover, using  the Euclidean partition function, one finds that the
conserved  fermion charge density  $n_q=-{{\partial\ln\mathsf{Z}}\over{\partial\tilde\mu}}$   is
\begin{equation}
n_q=
\frac{\kappa_0}{\mathsf{Z}}\sum_{\boldsymbol{x},s}\int[{\rm d}\overline{\Psi}{\rm d}{\Psi}]\overline{\Psi}_{\boldsymbol{x}}(\hat\gamma_0-s)e^{s\tilde\mu}{\Psi}_{\boldsymbol{x}+s{\bf e}_0}\ee^{-S_{\rm W}^E[\overline{\Psi},{\Psi}]}.
\end{equation}
Setting $\tilde{\mu}\to 0$, this  quantity becomes proportional to the expectation value of the time-like component of the  vector current $J^E_\mu(\boldsymbol{x})=\overline{\Psi}_{\boldsymbol{x}}(1+\hat\gamma_\mu){\Psi}_{\boldsymbol{x}-{\bf e}_\mu}-\overline{\Psi}_{\boldsymbol{x}}(1-\hat\gamma_\mu){\Psi}_{\boldsymbol{x}+{\bf e}_\mu}$~\cite{detar_book}, which is the discretized version of the continuum  vector current $J^E_\mu(\boldsymbol{x})=:\!\overline{\Psi}(\boldsymbol{x})\hat\gamma_\mu{\Psi}(\boldsymbol{x})\!\!:$ for Wilson fermions.
Therefore, the time-like component is simply related to the fermion density in the continuum limit, and we can readily explore situations away from half-filling $n_q\neq 0$.
Interestingly, while the gap equations~\eqref{eq:euclidean_gap_eqs_adim_1}-\eqref{eq:euclidean_gap_eqs_adim_2} remain
symmetrical under the transformation~(\ref{eq:symmetry_gap}) using the renormalized mass~\eqref{eq:ren_mass}, the charge density $n_q(\tilde m, \tilde g^2, \tilde\mu)$ has  only
an approximate symmetry.
 
\begin{figure}[t]
  \includegraphics[width=1\columnwidth]{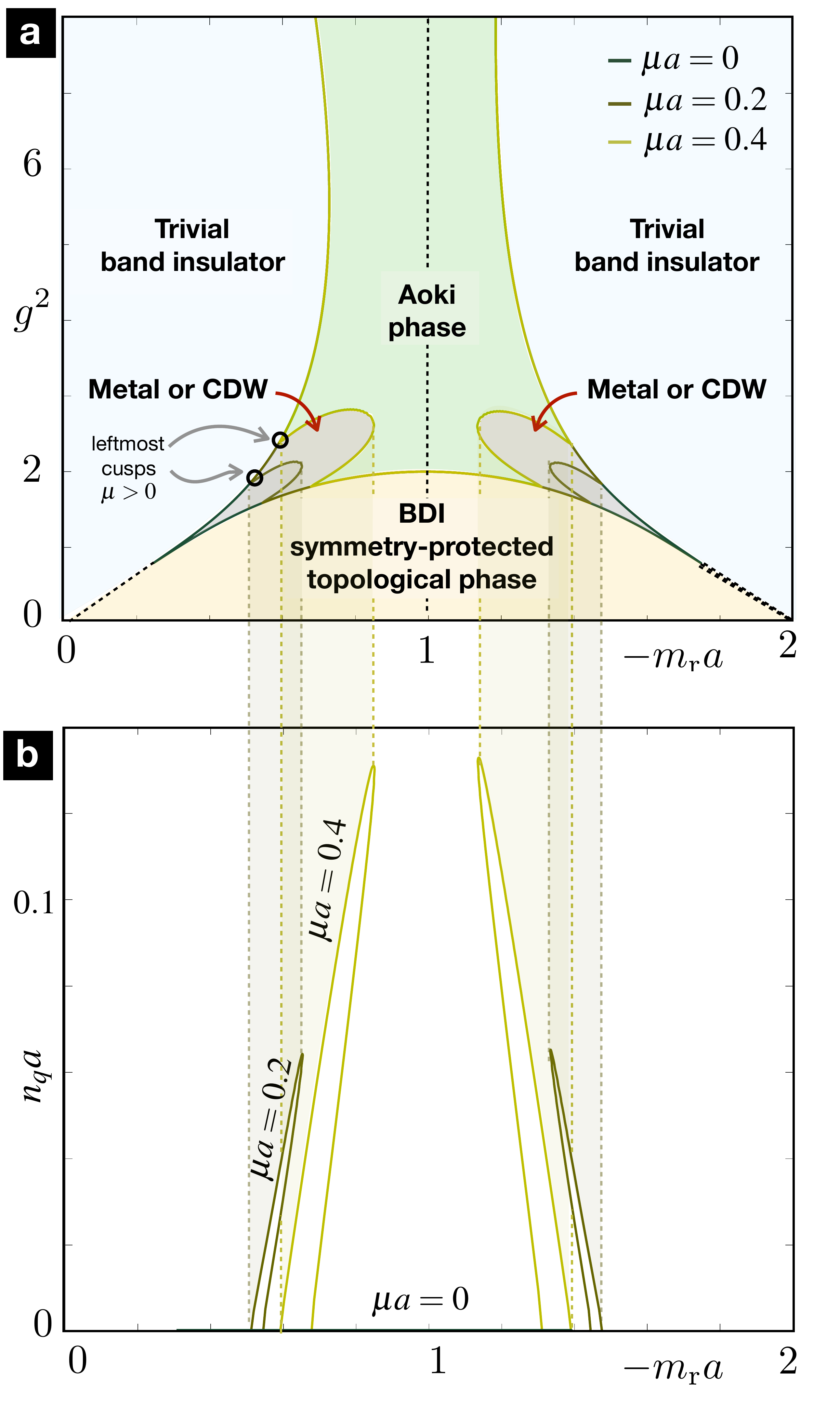}\\
  \caption{{\bf Large-$N$ phase diagram for a non-zero chemical potential:} {\bf (a)}
Solid lines delimit the Aoki phase as determined on a $512\times512\xi$ lattice
with $\xi=128$, for chemical potential $\mu a=0,0.2,0.4$. As the chemical potential increases, the region hosting a parity-breaking groundstate diminishes, leaving behind a droplet-shaped region between the trivial and topological phases of the half-filled case, which we conjecture corresponds to a new metallic phase. {\bf (b)} The solid lines lines
show the charge density $n_q(m_{\rm r})$ at the  phase boundary of each chemical potential, obtained by fixing the interaction strength $g^2=g^2_{\rm c}(m_{\rm r})$ to the corresponding critical line. We observe that the charge density  becomes non-zero only in the new line connecting the two cusps for a finite chemical potential.
}
  \label{Fig:chemo}
\end{figure}

We now solve the gap equations~\eqref{eq:euclidean_gap_eqs_adim_1}-\eqref{eq:euclidean_gap_eqs_adim_2} with a dimensional chemical potential
$\mu\equiv\xi\tilde{\mu}\not=0$, which yield the  phase diagram of Fig.~\ref{Fig:chemo}{\bf (a)}, where the axes  have been rescaled to match those of
Fig.~\ref{Fig:large_N_aoki_euclidean_renormalised}. We see that, as a consequence of the non-zero-chemical potential, the leftmost and rightmost cusps of the half-filled phase diagram of Fig.~\ref{Fig:large_N_aoki_euclidean_renormalised} split into a couple of cusps each, such that the region hosting the Aoki phase becomes smaller. The decrease of the Aoki phase can be qualitatively understood  as follows. For  $\mu>0$, one expects that the charge density 
$n_q$ will eventually rise from the value $n_q=0$ characterizing the half-filled regime.    As a consequence,  a 
Fermi surface will be formed in a certain parameter regime,  which consists of two disconnected Fermi points in 1+1$d$.
This  has the effect of disfavoring the particle--anti-particle pairing
required for the pseudoscalar condensate $\langle \Pi(x)\rangle \not=0$, since the excitation energy for such a
zero-momentum pair would be on the order of  $\Delta\epsilon\sim2\mu$. Accordingly, the Aoki phase should shrink as the chemical potential is increased. 
This is corroborated by the curves for $n_q(m_{\rm r})$ in Fig.~\ref{Fig:chemo}{\bf (b)}, which rise above zero only around
the borders of the droplet-shaped region between the two newly-formed cusps for $\mu>0$ (see the grey regions in Fig.~\ref{Fig:chemo}{\bf (a)}). These are precisely the regions where the half-filled Aoki phase has been expelled from. We also note that Fig.~\ref{Fig:chemo}{\bf (b)} shows an approximate
symmetry about $-m_{\rm r} a=1$ which we expect to become exact for
$\xi\to\infty$.

\begin{figure}[t]
  \includegraphics[width=1\columnwidth]{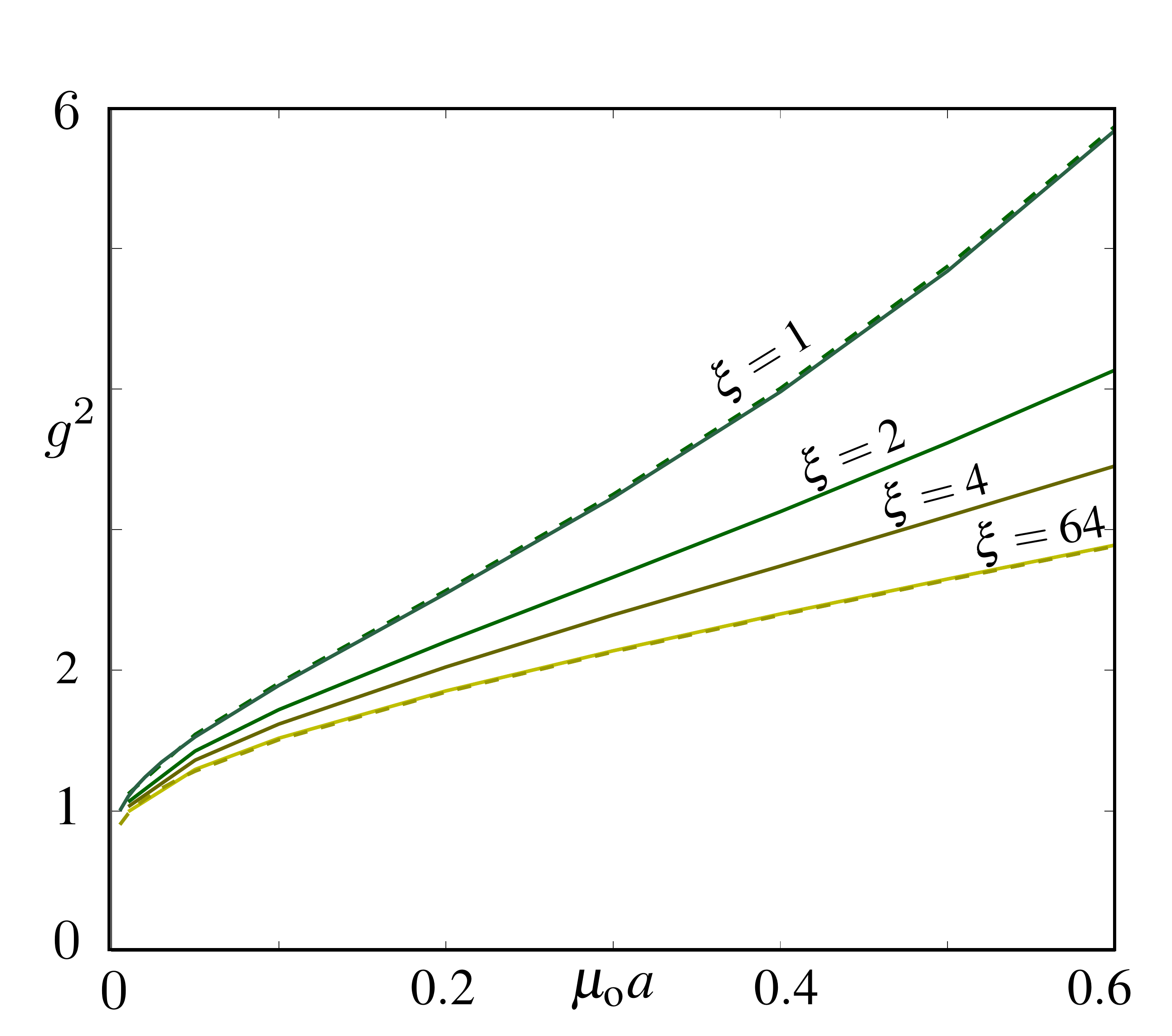}\\
  \caption{{\bf Location of the onset and tri-critical point of the phase diagram:}
the leftmost cusp of the phase boundary yields an estimate
for the onset chemical potential $\mu_{\rm o}(g^2)$. Results are obtained from solution of the gap 
equations (\ref{eq:euclidean_gap_eqs_adim_1})-(\ref{eq:euclidean_gap_eqs_adim_2}) on $512\times512\xi$, except for
`the dashed lines corresponding to $512\times1024\xi$ lattices. The limiting vale of $g^2$ as $\mu_{o}a\to 0$ marks the tri-critical point separating  the half-filled Aoki phase, from the trivial and topological insulators. 
}
  \label{Fig:tricritical}
\end{figure}

Let us now describe how these results can be used to determine, in a controlled way, the extent of the Aoki phase at half-filling.
The empty circles of Fig.~\ref{Fig:chemo}{\bf (a)} mark the so-called onset, beyond which the ground
state has a non-zero charge density (i.e.  $n_q>0$ for  $\mu>\mu_o(g^2)$). By numerically obtaining such onsets for a variety of parameters and time discretizations, we obtain 
Fig.~\ref{Fig:tricritical}. We note that the variation for finite $\xi$ is
probably due to non-universal effects since in the  sums over the Brillouin zone  of Eqs.~\eqref{eq:euclidean_gap_eqs_adim_1}-\eqref{eq:euclidean_gap_eqs_adim_1}, as the  chemical
potential enters as e.g. $\sinh(\xi\mu a)$. One observes from this figure that all curves come
closer in the limit $\mu\to0$, and seem to approach a limit as
$\xi\to\infty$. This limiting value   corresponds to the  point where half-filled Aoki phase terminates, proving that these phase does not extend all the way down to the weak coupling limit, but only survives down to $g^2(\mu_o=0)\approx0.8$ according to the results of Fig.~\ref{Fig:tricritical}.

Let us note that extracting  this  limiting value is numerically hard; for instance, the
curvature appears slightly sensitive to temperature, as revealed by calculations
on  $512\times1024\xi$ lattices. However, our  approximate prediction
$g^2(\mu_o=0)\approx0.8$ is consistent with the  cusps of Fig.~\ref{Fig:large_N_aoki_euclidean_renormalised}, where the Aoki phase
terminates. Fig.~\ref{Fig:tricritical} therefore strengthens our belief in the
existence of a {tricritical point} at non-zero $g^2(\mu_o=0)$; for couplings below
this value there is a direct transition between trivial and topological
insulating phases as $m$ is varied, and no parity breaking Aoki phase is encountered in the middle.

So far, we have used the large-$N$ results for a non-zero chemical potential to extract features of the half-filled phase diagram by taking the limit $\mu\to 0$ in a controlled manner. However, we note that another interesting question  would be to study the fate of the symmetry-protected topological phases, and the appearance of other new phases of matter, in the GNW model away from half filling.  In that respect, we note that our  large-$N$ results  point towards the appearance of a new phase (i.e. droplet-shape region of Fig.~\ref{Fig:chemo}). Since we have argued that the finite charge densities appear due to the formation of a Fermi surface, it is reasonable to expect that such densities will not drop abruptly to zero as we move away from the critical line. In that sense, the droplet-shaped region may either correspond to a metallic phase where the Fermi points occur at different momenta as the microscopic parameters are modified, or maybe to a kind of charge-density-wave where the fermionic density forms a regular periodic pattern. Understanding the nature of this phase lies outside of the scope of the present work, and  will be the subject of a future work. We advance at this point that the density-matrix renormalization group methods discussed in the following section could be adapted to study situations away from half filling, and are a potential tool to address the nature of this new phase. Moreover, we also note that  the sign problem for $\mu\not=0$ can be safely avoided for any discretized Gross-Neveu or Nambu-Jonal-Lasinio models, such that Monte Carlo techniques~\cite{Hands:1992ck} could also be applied to the present problem, and extensions thereof.

\subsection{Large-$N$ benchmark via matrix product states}
\label{sec:mps_benchmark}

 \begin{figure*}[t]
  \includegraphics[width=1.6\columnwidth]{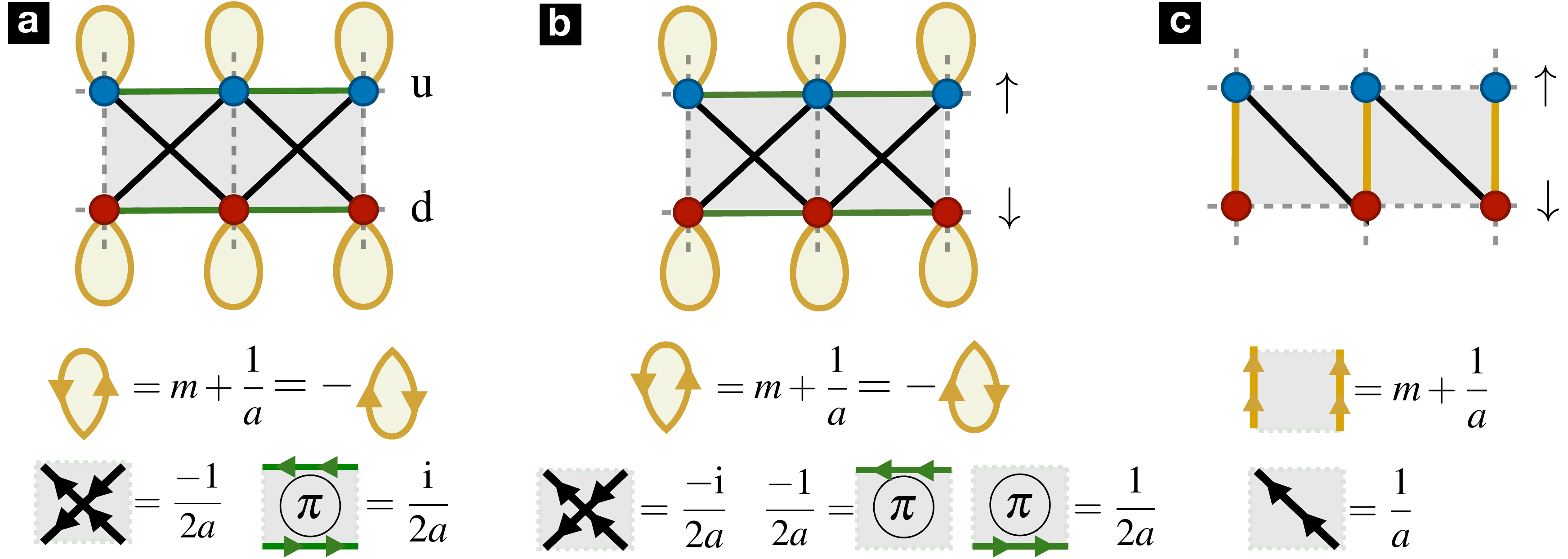}\\
  \caption{ {\bf Scheme for the Wilson-fermion kinetic energy:} {\bf (a)} The spinor index of the  fermionic field can be pictured as the legs of a virtual ladder $\ell\in\{\rm u,d\}$, which connects the GNW model to the imbalanced Creutz-Hubbard ladder~\eqref{eq:CH_ladder} after a gauge transformation.  Fermions can hop diagonally between the legs of the ladder with a tunneling strength $-t_{\times}=-1/2a$ (black crossed lines), and horizontally along the legs of the ladder with a  tunneling strength $-t_{\rm h}=\pm \ii /2a$ that is complex due to an external magnetic  $\pi$-flux  $-t_{\rm h}=-\ee^{\ii\int{\rm d}l A_{\rm ext}(x)} /2a$ (green horizontal lines). The yellow circles stand for an energy imbalance $\pm\Delta\epsilon/2=\pm(m+1/a)$  between the upper and lower legs. {\bf (b)} In the cold-atom implementation,  the legs of the synthetic ladder correspond to a couple of hyperfine states $\sigma\in\{\uparrow,\downarrow\}$, such that the original GNW model~\eqref{eq:naive_GN_lattice}-\eqref{eq:Wilson_GN_lattice} is similar to the Creutz-Hubbard ladder {\bf (a)} but with different tunneling strengths $-t_{\times}=\ii/2a$ (black crossed lines), and  $-t_{\rm h}=\pm 1/2a$ (green lines), which preserves the overall $\pi$-flux around square plaquettes. {\bf (c)} Using a different choice for the gamma matrices $\gamma^0=\sigma^x$, $\gamma^5=\sigma^y$, the discretization yields a much simpler tunneling for $r=1$. The kinetic energy consists only of diagonal $t_{\rm d}=1/a$ (black lines) and vertical  $t_{\rm v}=(m+1/a)$(yellow lines) hoppings.  }
  \label{Fig:wilson_scheme}
\end{figure*}

In this section, we test the above large-$N$ prediction for the   single-favor GNW lattice model using numerical routines based on matrix product states (MPS)~\cite{mps} (i.e. a variational version of  real-space  density-matrix renormalization group method~\cite{dmrg}).  On the one hand, this can be considered as the most stringent test of the validity of the  large-$N$ approach,   as  we are indeed  very far from the large-$N$ limit. On the other hand, the choice of $N=1$  is also motivated by the fact that the single-flavor GNW model can be realized in cold-atom experiments following the scheme of Sec.~\ref{sec:cold_atom}  described below. Note that the $N=1$ flavor of the continuum Gross-Neveu QFT~\eqref{eq:GN_cont_capital_psi} with an additional mass term corresponds to the so-called massive Thirring model~\cite{thirring_model}. The discretization of this QFT using the Wilson  approach allows us to discuss the occurrence of symmetry-protected topological phases in this LFT, and use it to benchmark the large-$N$ predictions for the phase diagram of the GNW model with a finite number of flavors.

\subsubsection{High-energy  physics to condensed matter  mapping} 
We consider the GNW lattice Hamiltonian~\eqref{eq:naive_GN_lattice}-\eqref{eq:Wilson_GN_lattice} for a single fermion flavor $N=1$. By performing a $U(1)$ gauge transformation to the   spinors $\Psi(x)\to\ee^{-\ii\frac{\pi}{2a} x}\Psi(x)$,  which can be understood as an instance of a Kawamoto-Smit phase rotation in  LFTs~\cite{kawa_smit}, and using the algebraic properties of the gamma matrices, we can rewrite $\tilde{H}_{\rm W}\to\tilde{H}_{\rm W}=a\!\sum_{{ x}\in\Lambda_\ell}\! :\!\tilde{\mathcal{H}}_{\rm W}\!:$, where 
\beq
\begin{split}
\tilde{\mathcal{H}}_{\rm W}=&\Psi^{\dagger}\!(x)\!\left(-\frac{\gamma^5}{2a}+\ii\frac{r \gamma^0}{2a}\right) \Psi\!\left(x+a\right)+{\rm H.c.}\\
+&\Psi^{\dagger}\!({ x})\!\left(m +\frac{r}{a}\right)\gamma^0\Psi(x)-\frac{g^2}{2N}\!\left(\Psi^\dagger\!(x)\gamma^0 \Psi(x)\right)^2\!.
\end{split}
\eeq
In this notation, the Hamiltonian looks similar to the Hubbard model~\cite{hubbard_model}, a paradigm of strongly-correlated electrons in condensed matter~\cite{facekas_book}, with an additional spin-orbit coupling. Note that this formulation only differs from Eqs.~\eqref{eq:naive_GN_lattice}-\eqref{eq:Wilson_GN_lattice} on the particular distribution of the complex tunnellings, which can be understood as a gauge transformation on a background magnetic field  maintaining an overall  $\pi$-flux. Indeed, the defining property of the above Kawamoto-Smit phases is that they yield a $\pi$-flux through an elementary plaquette.

In order to understand the origin of this magnetic flux, let us  introduce the following notation for the Dirac spinor $\Psi(x)=(c_{\rm u}(x), c_{\rm d}(x))^{\rm t}\to (c_{i,{\rm u}}, c_{i, {\rm d}})^{\rm t}/\sqrt{a}$ for $N=1$. Here, the dimensionless  fermion operators $c_{i,\ell}$ depend on a spinor index $\ell\in\{{\rm u,d}\}$ that can be interpreted  in terms of  the upper ($\ell={\rm u}$) and lower ($\ell\in {\rm d}$) legs of a synthetic ladder, and  $i\in\{1,\cdots ,N_{\rm s}\}$ labels the positions of the rungs of the ladder (see Fig.~\ref{Fig:wilson_scheme}{\bf (a)}). Considering our particular choice of gamma matrices $\gamma^5=\sigma^x$, $\gamma^0=\sigma^z$, the corresponding Hamiltonian $\tilde{H}_{\rm W}$ for the chosen Wilson parameter $r=1$ can be rewritten as
\beq
\label{eq:CH_ladder}
\begin{split}
H_{\rm W}=&\frac{-1}{2a}\sum_{i,\ell}\left(c^{\dagger}_{i,\ell}c^{\phantom{\dagger}}_{i+1,\bar{\ell}}-\ii s_\ell c^{\dagger}_{i,\ell}c^{\phantom{\dagger}}_{i+1,{\ell}}+{\rm H.c.}\right)\\
+&\sum_{i,\ell}\left(\left(m +\frac{1}{a}\right)s_\ell c^{\dagger}_{i,\ell}c^{\phantom{\dagger}}_{i,{\ell}}+\frac{g^2}{2a}c^{\dagger}_{i,\ell}c^{\dagger}_{i,\bar{\ell}}c^{\phantom{\dagger}}_{i,\bar{\ell}}c^{\phantom{\dagger}}_{i,{\ell}}\right),
\end{split}
\eeq
where we have introduced $s_{\ell}=+ 1$ ($s_{\ell}=-1$) for the upper   $\ell={\rm u}$ (lower $\ell={\rm d}$) leg of the ladder, and  $\bar{\ell}={\rm d,u}$ for $\ell={\rm u,d}$. As can be seen in Fig.~\ref{Fig:wilson_scheme}{\bf (a)}, there is a net $\pi$-flux due to an Aharonov-Bohm phase that the fermion would pick when tunneling around an elementary plaquette.

In particular, Eq.~\eqref{eq:CH_ladder} can be understood as a generalized Hubbard model on a ladder corresponding to the imbalanced Creutz-Hubbard model~\cite{creutz_hubbard}, which is an interacting version of the so-called Creutz ladder~\cite{creutz_ladder,creutz_ladder_refs}. The first line in  Eq.~\eqref{eq:CH_ladder} describes the horizontal and diagonal tunneling of fermions with strength $\tilde{t}=1/2a$, which are subjected to an external  magnetic $\pi$-flux threading the ladder (see Fig.~\ref{Fig:wilson_scheme}{\bf (a)}). One thus finds that the UV cutoff  of the GNW model $ \Lambda=1/a$ is provided by the maximum  energy within the band structure $\Lambda=2\tilde{t}$ of the Creutz-Hubbard model. Likewise, one understands that the first term in the second line of  Eq.~\eqref{eq:CH_ladder} corresponds to an energy imbalance between both legs of the ladder $\Delta\epsilon/2=(m+1/a)$, and yields a single-particle Hamiltonian in momentum space that is similar to Eq.~\eqref{eq:single-particle_h}, namely
 \beq
\label{eq:single-particle_h_CH}
\mathsf{h}^{\rm CH}_{k}=\left(\frac{\Delta\epsilon}{2}+2\tilde{t}\sin k\right)\gamma^0-2\tilde{t}\cos k\gamma^5.
\eeq
Finally,  the last term of Eq.~\eqref{eq:CH_ladder} amounts to a Hubbard-type density-density interaction $V_{\rm v}n_{n,{\rm u}}n_{n,{\rm d}}$ between fermions residing on the same rung of the ladder, which repel themselves with a strength  $V_{\rm v}=g^2/a$. 

According to this discussion, the high-energy-physics GNW lattice model is gauge equivalent to the condensed-matter  imbalanced Creutz-Hubbard model. Similarly to the high-energy physics convention of working with dimensionless parameters $m/\Lambda=ma$ and $g^2$,  the condensed-matter community normalizes the couplings to the tunneling strength $\tilde{t}$, such that the exact relation between the microscopic parameters of these two models is
\beq
\label{eq:mapping_parameters}
ma=\frac{\Delta\epsilon}{4\tilde{t}}-1, \hspace{1ex} g^2=\frac{V_{\rm v}}{2\tilde{t}}.
\eeq
  Let us also note that, in the condensed-matter context, the lattice constant $d$ of the model~\eqref{eq:CH_ladder} is fixed by the underlying Bravais lattice of the solid, which is typically set to $d=1$ in the calculations~\eqref{eq:single-particle_h_CH} (i.e. lattice units). Note, however, that this does not preclude us from taking the continuum limit. In this case, the continuum limit corresponds to the low-energy limit, where $\tilde{t}=1/2a$ (i.e. UV cutoff) is much larger than the energy scales of interest. By setting the model parameters in the vicinity of a second-order quantum phase transition, the relevant length scales fulfill $\xi_{\rm l}\gg d$, and one recovers universal features that are independent of  the microscopic lattice details, and can be described by a continuum  QFT.

  \begin{figure*}[t]
  \includegraphics[width=2.05\columnwidth]{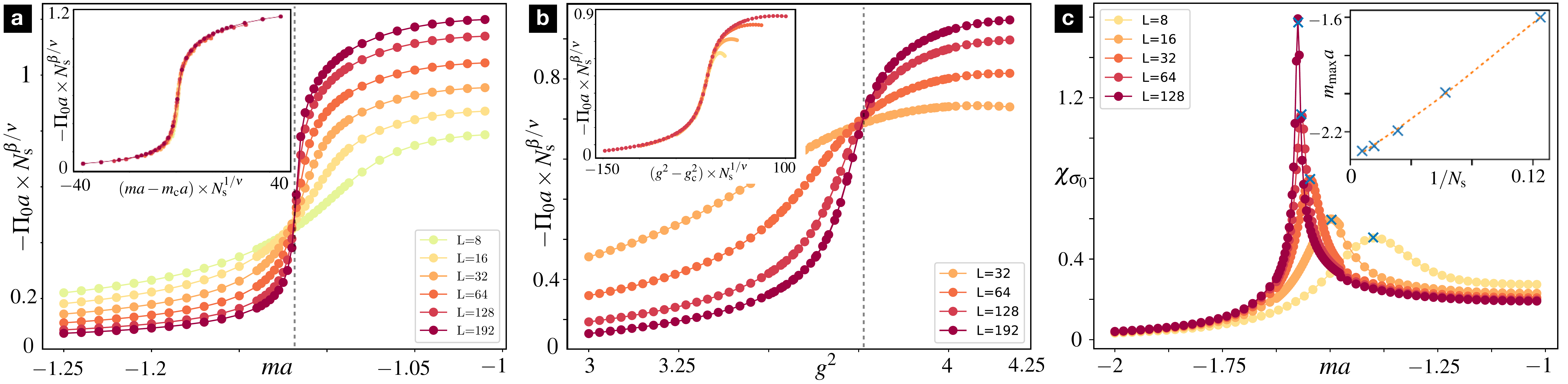}\\
  \caption{ {\bf Finite-size scaling for the quantum phase transitions of the Gross-Neveu-Wilson model:} {\bf (a)} (Main panel) Rescaling of the pseudoscalar condensate $\Pi_0a=\langle \bar{\Psi}\ii\gamma^5\Psi\rangle a$ with a power of the number of lattice sites $N_{\rm s}^{\beta/\nu}$ as a function of the bare mass $ma$. Assuming Ising critical exponents $\beta=1/8$ and $\nu=1$, we find a crossing for different lengths, signaling the critical point $m_{\rm c}a$ between the trivial band insulator and the Aoki phase. (Inset) Collapse of the data around the critical point, showing that the choice of Ising critical exponents leads to the correct universality. {\bf (b)} Same as {\bf (a)}, but studying the finite-size scaling of the pseudoscalar condensate as a function of the interactions $g^2$. The corresponding critical point $g^2_{\rm c}$ describes the transition between the SPT phase and the Aoki phase with Ising exponents  $\beta=1/8$ and $\nu=1$. {\bf (c)} The transition between the trivial band insulator and the SPT phase is captured by the divergence of the }
  \label{Fig:fss_mps}
\end{figure*}

\subsubsection{ Phase diagram of the $N=1$   Gross-Neveu-Wilson model}

 In this section, we exploit the above mapping~\eqref{eq:mapping_parameters} to explore the phase diagram of the $N=1$ lattice GNW model by importing some of the  condensed-matter and quantum-information techniques described in~\cite{creutz_hubbard}. In particular, we will use  the numerical matrix-product-state  results to benchmark the large-$N$ predictions.  We remark that this mapping also becomes very useful in the reverse direction, as certain aspects of the Creutz-Hubbard model become clarified from the high-energy perspective of the GNW model.

 In the  parameter regime $\Delta\epsilon/4\tilde{t}<1$, which corresponds to a bare mass  $-ma\in[0,1]$, we found that the  imbalanced Creutz-Hubbard model displays three distinct phases: an orbital paramagnet, an orbital ferromagnet, and an SPT phase~\cite{creutz_hubbard}.  The orbital paramagnet corresponds to a gapped phase of matter that is characterized by the absence of long-range order and any topological feature. Therefore, this phase should correspond to the trivial band insulator of the GNW model in Fig.~\ref{Fig:large_N_aoki}. 
 
 The orbital ferromagnet, on the other hand, is a phase displaying an Ising-type long-range  order due to the spontaneous breaking of a discrete orbital symmetry. Accordingly, it should correspond to the parity-broken Aoki phase of the Gross Neveu model in Fig.~\ref{Fig:large_N_aoki}. To show this correspondence in more detail, let us comment on   the orbital magnetization introduced for the Creutz-Hubbard ladder $T_0=\langle T_i^y\rangle=\half\langle \ii c_{i,{\rm u}}^\dagger c_{i,{\rm d}}^{\phantom{\dagger}}\rangle+{\rm c.c.}\neq 0$ $\forall i$, and show that it is related to an order parameter of the GNW model. The parity symmetry of the GNW model that is broken in the Aoki phase, namely $\Psi(x)\to\eta\mathbb{I}_N\otimes\gamma^0\Psi(-x)$ with $|\eta|^2=1$, corresponds to $c_{i,{\rm u}}\to\eta c_{N_{\rm s}-i,{\rm u}}$, and $ c_{i,{\rm d}}\to-\eta c_{N_{\rm s}-i,{\rm d}}$ in the Creutz-Hubbard ladder. Hence, one finds that  $T_0=\langle T_i^y\rangle\to -\langle T_{N_{\rm s}-i}^y\rangle=-T_0$ is spontaneously broken by the orbital ferromagnet. We thus see that, in the language of the synthetic Creutz-Hubbard ladder, the pseudoscalar condensate corresponds  to an Ising-type  ferromagnet with a non-zero orbital magnetization $T_0=-\Pi_0a/2=-\langle \overline{\Psi}\ii\gamma^5\Psi\rangle a/2$. This connection also teaches us that one can perform a rigorous finite-size scaling  of the pseudoscalar condensate to obtain  accurate predictions of  the critical lines enclosing the whole Aoki phase, instead of using the various mappings discussed in~\cite{creutz_hubbard}. 

  Finally, as shown explicitly  in~\cite{creutz_hubbard}, the Creutz-Hubbard ladder also hosts a correlated SPT phase, which displays a double-degenerate entanglement spectrum~\cite{ent_spectrum} due to a couple of zero-energy edge modes. This phase should thus corresponds to the $\mathsf{BDI}$ symmetry-protected topological phase of the GNW model discussed throughout this work (see Fig.~\ref{Fig:large_N_aoki}). Let us remark, however, that  the topological insulator of the Creutz-Hubbard model lies in the symmetry class $\mathsf{AIII}$, breaking explicitly the time-reversal $\mathsf{T}$ and charge-conjugation $\mathsf{C}$ symmetries, yet  maintaining the sublattice symmetry. According to our discussion below Eq.~\eqref{eq:zak_masses}, we see that the Creutz-Hubbard single-particle Hamiltonian breaks  $\mathsf{T}$: $\gamma^0 (h^{\rm CH}_{-k})^*\gamma^0\neq h^{\rm CH}_{k}$,  and $\mathsf{C}$: $\gamma^5 (h_{-k}^{\rm CH})^*\gamma^5\neq-h_{k}^{\rm CH}$, explicitly. On the other hand, the combination  $\mathsf{S}=\mathsf{T}\mathsf{C}$ yields $(\gamma^1)^\dagger h^{\rm CH}_{k}\gamma^1=-h^{\rm CH}_{k}$, such that the topological insulator of the Creutz-Hubbard ladder is in the $\mathsf{AIII}$ symmetry class. Therefore, the last element of our high-energy physics to condensed-matter dictionary is the mapping between the symmetry classes $\mathsf{BDI}\leftrightarrow\mathsf{AIII}$, which is  a direct consequence of  the above local gauge transformation/Kawamoto-Smit phase rotation. Although differences will arise regarding perturbations that explicitly  break/preserve the corresponding symmetries (e.g. disorder), the phase diagram of the translationally-invariant GNW model should coincide exactly with that of the Creutz-Hubbard ladder provided that one uses the relation between microscopic parameters in Eq.~\eqref{eq:mapping_parameters}.

  With this interesting dictionary for the correspondence of phases, and the microscopic parameter mapping in Eq.~\eqref{eq:mapping_parameters}, we can use  numerical matrix-product-state  simulations, extending the parameter regime studied in~\cite{creutz_hubbard}  from $\Delta\epsilon/4\tilde{t}<1$ to $-1<\Delta\epsilon/4\tilde{t}<1$. In this way,   we can explore  the full phase diagram diagram of the $N=1$    GNW model, and compare it to our previous large-$N$ predictions for $-ma\in[0,2]$. Let us recall that the large-$N$ approach fulfills~\eqref{eq:symmetry_gap}, such that the obtained phase diagrams have a mirror symmetry about $-ma=1$. However, it is not  clear a priori if this symmetry is a property of the model, or if it is instead rooted in the approximations of the large-$N$ prediction.  We will be able to address this question with our new matrix-product-state simulations.

  In Figs.~\ref{Fig:fss_mps} {\bf (a)}-{\bf (b)}, we discuss a representative example of the finite-size scaling of the pseudoscalar condensate $\Pi_0=\langle\overline{\Psi}\ii\gamma^5\Psi\rangle$ for the transition between the trivial, or topological, band insulators and the Aoki phase. One clearly sees that the matrix-product-state numerical simulations for different lengths display a crossing that gives access to the critical point (main panel of Figs.~\ref{Fig:fss_mps} {\bf (a)}-{\bf (b)}), and that the data collapse of (inset of Figs.~\ref{Fig:fss_mps} {\bf (a)}-{\bf (b)}) corroborates that this critical point lies in the Ising universality class.  Note that the pseudoscalar condensate gives no information about the phase transition between the trivial and topological insulators. In order to access this information, the mapping to the Creutz-Hubbard ladder becomes very useful, as it points to the possibility of using   a generalized susceptibility   associated to the variation of the scalar  condensate  $\sigma_0=\langle\overline{\Psi}\Psi\rangle$ with the bare mass $\chi_{\sigma_0}=\partial\sigma_0/\partial m$. As shown in the main panel of Fig.~\ref{Fig:fss_mps} {\bf (c)}, this susceptibility diverges at the critical point of the thermodynamic limit, and can be used to perform a finite-size scaling.
  
\begin{figure}[t]
  \includegraphics[width=1\columnwidth]{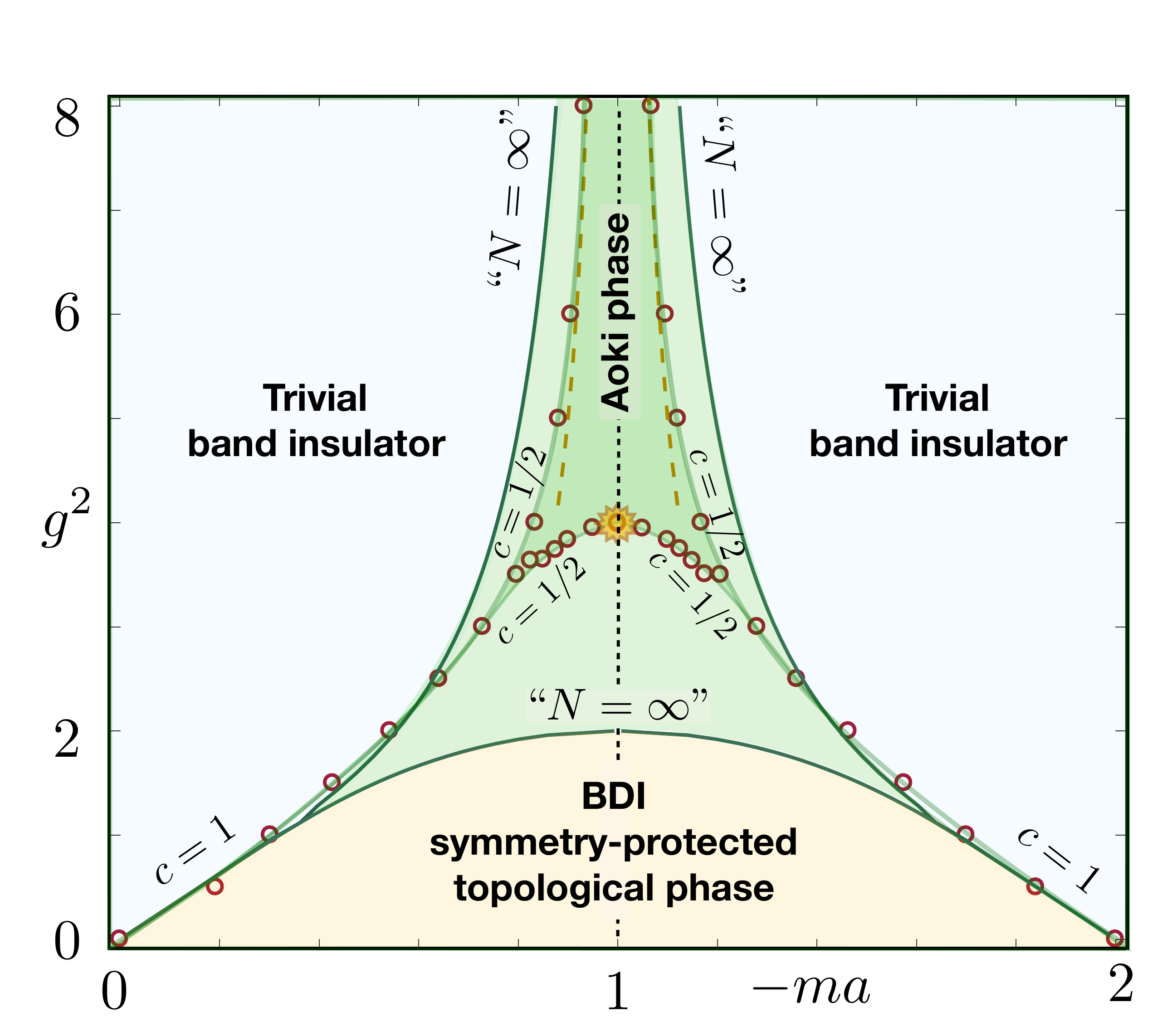}\\
  \caption{ {\bf Phase diagram from matrix-product-state methods:} The red circles represent the critical points of the $N=1$ Gross Neveu lattice model obtained by with matrix product states. The semi-transparent green lines joining these points delimit the trivial band insulator, Aoki phase, and the $\mathsf{BDI}$ symmetry-protected topological phase. Note again that this SPT phase corresponds to the $\mathsf{AIII}$ topological insulator of the Creutz-Hubbard model .These lines are labelled by $N=1$, and by the central charge $c\in\{1/2,1\}$ of the conformal field theory that controls the continuum QFT at criticality. These results serve to benchmark the large-$N$ predictions for the critical lines, which are represented by dark solid lines, and labelled with $``N=\infty"$. We also include the exact critical point at $(-ma,g^2)=(1,4)$, which is depicted by an orange star, and the strong-coupling critical lines that become exact in the limit of $g^2\to\infty$, which are depicted by dashed orange lines.  The matrix-product-states  predictions match remarkably well these exact results. }
  \label{Fig:MPS_aoki}
\end{figure}

   Repeating this procedure for various critical points, we obtain the red empty circles  displayed in Fig.~\ref{Fig:MPS_aoki}, which are compared with the large-$N$ results of Fig.~\ref{Fig:large_N_aoki} represented as solid lines.   
  We can thus conclude that the large-$N$ predictions are qualitatively correct,  as they predict the same three possible phases, and the shape of the critical lines is qualitatively similar to the matrix-product-state prediction. Moreover, the agreement between the critical lines becomes quantitatively correct in the weak-coupling limit $g^2,|ma|\ll 1$, which is the  regime where the asymptotically-free  Gross-Neveu QFT~\eqref{eq:GN_continuum} is expected to emerge from the lattice model. Since both the mass and the interaction strengths are relevant perturbations  growing with the renormalization-group transformations, one expects that a continuum limit with physical parameters well below  the UV cutoff can be recovered provided that $g^2,|ma|\ll 1$. Let us also remark that the matrix-product-state simulations are consistent with the mirror symmetry about $-ma=1$ of the large-$N$ gap equations. Therefore, it seems that this symmetry is an intrinsic property of the GNW model, which is easy to understand in the non-interacting limit, but not so obvious in the interacting case. On  general grounds, Fig.~\ref{Fig:MPS_aoki} shows that the large-$N$ prediction tends to overestimate the extent of the Aoki phase, predicting that the spontaneous breaking of the parity symmetry occurs for weaker interactions and  smaller  masses. This trend could be improved by considering next-to-leading-order (NLO) corrections to the saddle-point solution, and will be the subject of a future study. In this sense, our results suggest that  large-$N$ methods from a high-energy context can be a useful and systematic tool to study problems of correlated symmetry-protected topological phases in condensed matter. 
  
\begin{figure}[t]
  \includegraphics[width=0.9\columnwidth]{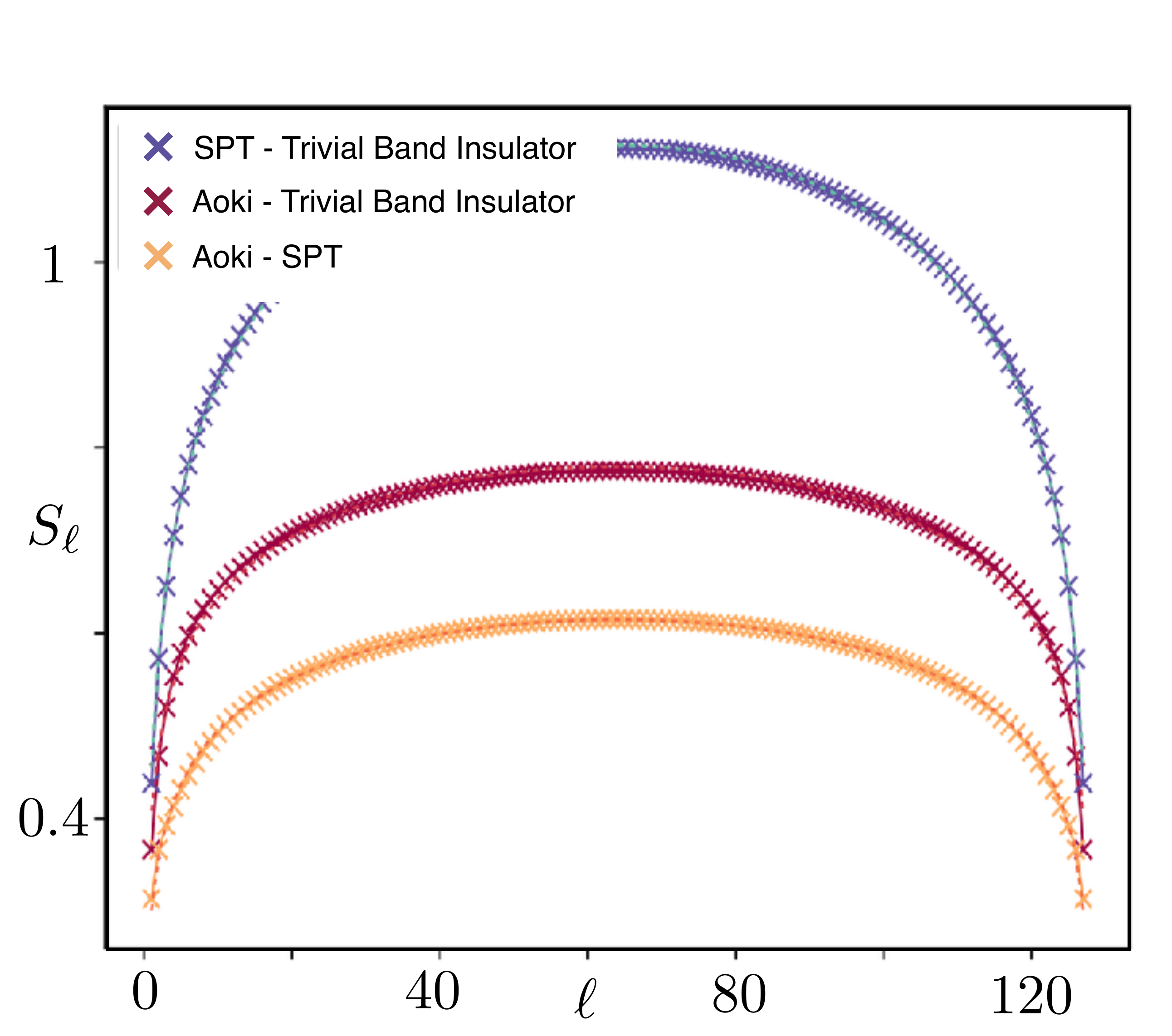}\\
  \caption{ {\bf Scaling of the block entanglement entropy :} Entanglement entropy $S_\ell=-{\rm Tr}\{\rho_\ell\log\rho_\ell\}$ for a $\ell$-sites block reduced density matrix $\rho_\ell={\rm Tr}_{N_{\rm s}-\ell}\{\ket{\rm gs}\bra{\rm gs}\}$ obtained from a chain of $N_{\rm s}=128$ sites. The blue, red, and yellow crosses correspond to the data for the critical points $(-ma,g^2)=(1.5787,1.5)$, $(-ma,g^2)=(1.1182,5)$, and $(-ma,g^2)=(1.125,3.775)$, respectively. The solid lines correspond to the fittings with the conformal field theory predictions $S_\ell=\frac{c}{6}\log(D_\ell)+a$, where $c$ is the central chrage, $a$ a non-universal constant, and $D_\ell=2N_{\rm s}\sin(\pi\ell/N_{\rm s})/\pi$ is the chord distance. The fitted central charges are $c=1.02$, $c=0.503$, and $c=0.49$.}
  \label{Fig:central_charges}
\end{figure}

We now comment on further interesting features that can be learned from this dictionary, and imported from  condensed matter  into the high-energy physics context.  In Fig.~\ref{Fig:MPS_aoki}, we have highlighted with a semi-transparent orange star the critical point separating the topological and Aoki phases at $(-ma,g^2)=(1,4)$. This point corresponds to a  Creutz-Hubbard model with  vanishing imbalance $\Delta\epsilon=0$, and  strong  repulsion $V_{\rm v}=8\tilde{t}$. Interestingly, it is precisely at this point that  an exact quantum phase transition is found by  mapping the Creutz-Hubbard ladder onto an exactly-solvable quantum impurity Ising-type model via the so-called maximally-localized Wannier functions~\cite{creutz_hubbard}. In this way, one learns that the lattice GNW model can be solved exactly for a particular limit with relatively strong couplings, and that the corresponding quantum phase transition  must lie in the Ising  class. From a high-energy  perspective, the whole critical line separating the topological and Aoki phases should be controlled by the continuum QFT of a Majorana fermion, and not by the standard Dirac-fermion QFT expected at weak couplings (i.e. along the critical line separating the topological and trivial insulators). We have  proved this rigorously using the numerical scaling of the entanglement entropy~\cite{ent_entropy}, which shows that this critical line is characterized by a central charge $c\approx1/2$ in agreement with the conformal field theory of a massless Majorana fermion. Conversely, at weak couplings, the scaling of the entanglement entropy yields a central charge $c\approx1$ in agreement with the massless Dirac fermion (see  Fig.~\ref{Fig:central_charges}). 
  
  Let us now discuss the  orange dashed line of Fig.~\ref{Fig:MPS_aoki}, which describes an exact solution that becomes valid in the strong-coupling limit $g^2\gg1$.  From the parameter correspondence~\eqref{eq:mapping_parameters}, this regime corresponds to the strongly-interacting Hubbard model, where one expects to find super-exchange interactions between the fermions~\cite{superexchange}. In this case, these super-exchange can be described in terms of an orbital Ising model with ferromagnetic coupling $J=-2/g^2a$, and subjected to a transverse magnetic field $B=2(m+1/a)$. According to the exact solution of the transverse Ising model~\cite{qim}, the strong-coupling critical line $J=2B$ corresponds to $g^2=1/2(ma+1)$. This line, and its mirror image,  have been depicted by the orange dashed lines of Fig.~\ref{Fig:MPS_aoki}, and shows a very good agreement with the numerical critical points of the GNW model at strong-couplings $g^2\gg1$. Since the strong-coupling mapping yields a transverse Ising model, we learn again that the corresponding continuum QFT at criticality is that of a Majorana fermion, which is corroborated again by the matrix-product-state scaling of the entanglement entropy yielding a central charge of central charge $c\approx1/2$ (see  Fig.~\ref{Fig:central_charges}). Therefore, the condensed-matter mapping teaches us that the GNW lattice model has an exact solution in the strong-coupling limit, and both critical lines delimiting the Aoki phase lead to a continuum limit controlled by a Majorana-fermion QFT.  These results show that condensed-matter methods can offer a useful and systematic tool to benchmark large-$N$ methods applied to problems of asymptotically-free LFTs  in a high-energy context. In future works, we will study  leading order $1/N$ corrections to the present large-$N$ approach, and see how fast they approach the exact and quasi-exact results for the phase diagram discussed in this section. 
    
  \subsection{Cold-atom Gross-Neveu-Wilson model}
  
 In this section, we  describe  possible routes for the cold-atom realization of the lattice GNW model starting from Eqs.~\eqref{eq:cold_atoms_ho}-\eqref{eq:cold_atoms_v}. At this stage, we could simply build on the mapping to the Creutz-Hubbard ladder~\eqref{eq:CH_ladder} to adopt the quantum simulation scheme recently proposed in~\cite{creutz_hubbard}. However, this would lead to an SPT phase in a different symmetry class, so we will now focus on the cold-atom quantum simulation of the original GNW Hamiltonian~\eqref{eq:naive_GN_lattice}-\eqref{eq:Wilson_GN_lattice}. Moreover, as discussed below,  the original arrangement of tunnellings can  simplify  the experimental requirements. 
 
 Therefore, in this section, we describe in detail the scheme based on a two-component $\sigma=\{\uparrow,\downarrow\}$ single-species $N_{\rm sp}=1$ Fermi gas in a state-independent optical lattice~\eqref{eq:cold_atoms_ho}-\eqref{eq:cold_atoms_v}. The internal states can be interpreted as a synthetic dimension~\cite{synthetic_dimension}, such that the  target GNW Hamiltonian density 
 \beq
 \label{eq:GNLH}
\begin{split}
\tilde{\mathcal{H}}_{\rm W}=&\,\Psi^{\dagger}\!(x)\!\left(-\ii\frac{\gamma^5}{2a}-\frac{r \gamma^0}{2a}\right) \Psi\!\left(x+a\right)+{\rm H.c.}\\
+&\,\Psi^{\dagger}\!({ x})\!\left(m +\frac{r}{a}\,\right)\gamma^0\Psi(x)-\frac{g^2}{2N}\!\left(\Psi^\dagger\!(x)\gamma^0 \Psi(x)\right)^2\!.
\end{split}
\eeq
  can be depicted using the scheme of Fig.~\ref{Fig:wilson_scheme}{\bf (b)}.
 
 As usual~\cite{cold_atoms_review}, one  makes use of the Wannier functions localized around  $\boldsymbol{x}^0_{\bf i}=\sum_\nu ({\pi}/k_{{\rm L},\nu}){i}_\nu{\bf e}_\nu$, where ${\bf i}$ is a vector of integers labelling the optical lattice minima,  to transform the  Fermi fields as $\Psi_\sigma({\boldsymbol{x}})=\sum_{\bf i}w(\boldsymbol{x}-\boldsymbol{x}^0_{\bf i})f_{{\bf i},\sigma}$, where we have eliminated the species index from the lattice annihilation operators $f_{{\bf i},\sigma}$. Assuming that the optical lattice is much deeper along two axes $V_{0,y},V_{0,z}\gg V_{0,x}\gg E_{\rm R}$, the dynamics for the time-scale of interest occurs along the $x$-axis, and can be  described by a Hubbard-type model
 \beq
 \label{eq:hubbard_atoms}
 \begin{split}
 H=-t &\sum_{i,\sigma}  \left(f^{\dagger}_{i,\sigma}f^{\phantom{\dagger}}_{i+1,\sigma}+ {\rm H.c.}\right)+\frac{U}{2}\sum_i\sum_{\sigma,\sigma'}f^{\dagger}_{i,\sigma}f^{{\dagger}}_{i,\sigma'}f^{\phantom{\dagger}}_{i,\sigma'}f^{\phantom{\dagger}}_{i,\sigma}\\
+&\sum_{i,\sigma}(\epsilon_\sigma+\Delta i)f^{{\dagger}}_{i,\sigma}f^{\phantom{\dagger}}_{i,\sigma}+\sum_{i,j}\sum_{\sigma,\sigma'}v_{\sigma,\sigma'}^{i,j}(t)f^{\dagger}_{
j,\sigma'}f^{\phantom{\dagger}}_{i,\sigma}.
\end{split}
 \eeq
 Here, the terms of the first line corresponds to the standard tunneling with $t=4 E_{\rm R}(V_{0,x}/E_{\rm R})^{3/4}\ee^{-2(V_{0,x}/E_{\rm R})^{1/2}}$, and on-site interactions $U=\sqrt{8/\pi}k_{{\rm L},x}a_{\uparrow\downarrow}E_{\rm R}(V_{0,x}V_{0,y}V_{0,z}/E_{\rm R}^3)^{1/4}$ of the cold-atom Hubbard model~\cite{cold_atoms_review}. In addition to these terms, the second line contains the internal energies $\epsilon_{\sigma}$, and a static gradient $\Delta$ that comes from the so-called lattice acceleration~\cite{lattice_acceleration}, i.e. the detuning of the optical-lattice beams is modified linearly   in time, yielding  a linear gradient (i.e. constant force) in the lattice reference frame. The remaining terms $v_{\sigma,\sigma'}^{i,j}(t)$ in Eq.~\eqref{eq:hubbard_atoms} stem from the  pairs of laser beams in a Raman configuration~\eqref{eq:on-site_potentials}, which need to be exploited such that the tunneling dynamics of the atoms corresponds to Fig.~\ref{Fig:wilson_scheme}{\bf (b)}.
 
 First of all, the bare tunneling must be inhibited by the gradient $t\ll\Delta$. Then, the inter-leg tunnellings of Fig.~\ref{Fig:wilson_scheme}{\bf (b)} (crossed black lines) can be laser-assisted by a  Raman pair~\cite{raman_assisted_scheme}, which  also leads to the energy imbalance terms (yellow loops). We set {\it (i)} the Raman frequencies to $\Delta\omega_1=(\epsilon_{\uparrow}-\epsilon_{\downarrow})+\Delta+\Delta\epsilon/2$, and $\Delta\omega_2=(\epsilon_{\downarrow}-\epsilon_{\uparrow})+\Delta-\Delta\epsilon/2$, where $\Delta\epsilon$ is small detuning, {\it (ii)} the two-photon Rabi frequencies (phases) to $\Omega_1=\Omega_2=:\Omega$ ($\varphi_1=\varphi_2=:\varphi$),  and {\it (iii)}  the corresponding  Raman wave-vectors to  $\Delta \boldsymbol{k}_1\cdot{\bf e}_x=\Delta \boldsymbol{k}_2\cdot{\bf e}_x=0$.
In a rotating frame, the Raman-assisted tunneling arising from the corresponding $v_{\sigma,\sigma'}^{i,j}(t)$ term contributes with
\beq
\label{eq:cross_tunneling}
\begin{split}
H_{\rm R,1}&=\frac{\Delta\epsilon}{2}\sum_{i}\left(f^{\dagger}_{i,\uparrow}f^{\phantom{\dagger}}_{i,\uparrow}-f^{\dagger}_{i,\downarrow}f^{\phantom{\dagger}}_{i,\downarrow}\right)\\
&+\sum_i\Omega\ee^{-\sqrt{\frac{V_{0,x}}{E_{\rm R}}}}\left(\ee^{\ii \varphi} f^{\dagger}_{i,\uparrow}f^{\phantom{\dagger}}_{i+1,\downarrow}+\ee^{\ii \varphi} f^{\dagger}_{i,\downarrow}f^{\phantom{\dagger}}_{i+1,\uparrow}+{\rm H.c.}\right),
\end{split}
\eeq
which contains precisely  the desired crossed tunnellings for $\varphi=\pi/2$, and the energy imbalance of Fig.~\ref{Fig:wilson_scheme}{\bf (b)}. 

In order to engineer the horizontal tunneling of  Fig.~\ref{Fig:wilson_scheme}{\bf (b)} (green  lines), we shall make use of  a third  Raman pair, but this time  far detuned  from the atomic transition $\Delta\omega_3\ll(\epsilon_{\uparrow}-\epsilon_{\downarrow})$. In this situation, when the corresponding laser intensities  are weak, the Raman term leads to  a crossed-beam ac-Stark shift that can be interpreted   as slowly-moving shallow optical lattice that acts as a periodic modulation of the on-site energies
\beq
\label{eq:periodic_modulation}
H_{\rm m}(t)=\sum_{i}\Omega_\sigma\cos(\Delta {\bf k}_3\cdot\boldsymbol{x}_i^0-\Delta\omega_3t+\varphi_3)f^{\dagger}_{i,\sigma}f^{\phantom{\dagger}}_{i,\sigma},
\eeq 
where $\Omega_{\sigma}$ is the two-photon ac-Stark shift for each of the hyperfine levels, which can be controlled by tuning the intensity and polarization of the lasers. We set {\it (i)} the Raman frequency in resonance with the gradient $\Delta\omega_3=\Delta\ll (\epsilon_\uparrow-\epsilon_\downarrow)$;  {\it (ii)} the Raman wave-vector  to   $\Delta \boldsymbol{k}_3\cdot{\bf e}_x=k_{{\rm L},x}$ with respect to the static optical lattice; and {\it (iii)} the Raman phase $\varphi_3=0$. In a rotating frame,  the atoms can absorb energy from this shallow moving lattice, such that the  horizontal tunneling gets reactivated~\cite{pat_gauge,pat_gauge_ol}, according to
\beq
H_{\rm R,2}=H=-t \sum_{i,\sigma}J_{1}\left(\frac{\Omega_{\sigma}}{\Delta}\right)  \left(f^{\dagger}_{i,\sigma}f^{\phantom{\dagger}}_{i+1,\sigma}+{\rm H.c.}\right),
\eeq
where we have introduced the $n$-th order Bessel function of the first class $J_n(x)$. According to this expression, we can laser-assist the  horizontal hopping with the desired  signs of Fig.~\ref{Fig:wilson_scheme}{\bf (b)} by exploiting the state-dependence of the dressed tunneling rates, and setting 
\beq
\label{eq:constraint_1}
J_1\left(\frac{\Omega_\uparrow}{\Delta}\right)=-J_1\left(\frac{\Omega_\downarrow}{\Delta}\right).
\eeq
 This can be achieved, while simultaneously maximizing the dressed tunneling, by setting $\Omega_\uparrow=3\Delta\approx0.6\Omega_\downarrow$.

 Let us note that the cross-tunneling~\eqref{eq:cross_tunneling} will also get a multiplicative renormalization due to this periodic modulation~\eqref{eq:periodic_modulation}, which will be  proportional to $J_0((\Omega_\uparrow+\Omega_\downarrow)/\Delta)$. This dressing is similar to the effect exploited for the so-called coherent destruction of tunneling~\cite{cdt}. To achieve  the relation of the tunnellings of Fig.~\ref{Fig:wilson_scheme}{\bf (b)}, one should  modify the Rabi frequency of the Raman beams $\Omega$, such that 
 \beq
 \label{eq:constraint_2}
 \Omega\ee^{-\sqrt{\frac{V_{0,x}}{E_{\rm R}}}}J_0\left(\frac{\Omega_\uparrow+\Omega_\downarrow}{\Delta}\right)=tJ_1(\Omega_\uparrow),
 \eeq
although we note that there might be other strategies to fulfill both constraints~\eqref{eq:constraint_1}-\eqref{eq:constraint_2} simultaneously. Altogether, considering also Hubbard interactions, the correspondence between the cold-atom and the Gross-Neveu parameters is 
 \beq
 \frac{1}{a}=2tJ_1\left(\frac{\Omega_\uparrow}{\Delta}\right), \hspace{1.5ex} m=\Delta\epsilon-2tJ_1\left(\frac{\Omega_\uparrow}{\Delta}\right), \hspace{1.5ex}\frac{g^2}{a}=U_{\uparrow\downarrow}.
 \eeq
 
 As announced at the beginning of this section, this scheme provides a slight simplification over the proposal for the Creutz-Hubbard model~\cite{creutz_hubbard}, which required the use of an intensity-modulated superlattice, instead of the shallow moving lattice~\eqref{eq:periodic_modulation}  already  implemented in experiments~\cite{pat_gauge_ol}. At this point, we comment on an interesting alternative that would simplify considerably the cold-atom scheme. As realized recently~\cite{wilson_hauke}, a different choice of the gamma matrices $\gamma^0=\sigma^x$, $\gamma^5=\sigma^y$, simplifies considerably the tunneling of Eq.~\eqref{eq:GNLH}, since $\ii\gamma^5+r\gamma^0=2\sigma^+$ for a Wilson parameter $r=1$, where we have introduced the raising operator $\sigma^+=\ket{{\uparrow}}\bra{{\downarrow}}$. Accordingly, the kinetic energy of the Wilson fermions can be depicted by the scheme of Fig.~\ref{Fig:wilson_scheme}{\bf (c)}. Let us  note that the  $\mathsf{BDI}$ symmetry class   can be readily understood by realizing that the synthetic ladder of this figure can be deformed into a single chain with dimerized tunnellings, and thus corresponds to the Su-Schrieffer-Hegger $\mathsf{BDI}$ topological insulator~\cite{ssh}. 

This representation was exploited in~\cite{wilson_hauke} to propose a cold-atom realization of quantum electrodynamics with Wilson fermions  in $(1+1)$ dimensions (i.e. Schwinger model). In that case,  one should introduce a bosonic species to simulate the gauge field, and exploit the spin-changing boson-fermion atomic scattering to obtain the gauge-invariant tunneling of the lattice gauge theory. In our case, the required experimental tools are already contained in our previous  description and, more importantly, can be considerably simplified with respect to the above discussion. The vertical tunnellings of Fig.~\ref{Fig:wilson_scheme}{\bf (c)} can be obtained from a Raman pair with $\Delta\omega_1=(\epsilon_\uparrow-\epsilon_\downarrow)$, whereas the diagonal tunnellings require another pair of Raman beams with $\Delta\omega_2=(\epsilon_\downarrow-\epsilon_\uparrow)+\Delta$, but no additional periodic modulations would be required. Therefore, if no additional disorder is to be considered, which could depend on the particular symmetry class and choice of gamma matrices,  this later approach should be followed  for  the cold-atom experiment, as it simplifies the experimental requirements for the quantum simulation of the GNW model.

 Let us finally comment on another interesting alternative. The non-interacting Creutz ladder has been recently realized in multi-orbital optical-lattice experiments~\cite{sp_creutz_atoms} that exploit two orbital states of the optical lattice to encode the legs of the ladder, and orbital-changing Raman transitions to implement  the  inter-leg tunnelings. It would be interesting to study the type of multi-orbital interactions~\cite{ghm_review} that can be generated in this setup, and the possibility of simulating directly the GNW model studied in this work.

\section{Conclusions and Outlook}

In this work, we have described the existence of correlated symmetry-protected topological phases in a discretized version of the Gross-Neveu model. We have applied large-$N$ techniques borrowed from high-energy physics, complemented with the study of topological invariants from condensed matter, to unveil a rich phase diagram that contains a wide region hosting a $\mathsf{BDI}$ topological insulator. This region extends  to considerably strong interactions, and must thus correspond to a strongly-correlated symmetry-protected topological phase. We have shown that this phase, and the underlying topological invariant, can be understood in terms of the renormalization of Wilson masses due to interactions (i.e. dynamic mass generation due to a scalar fermionic condensate). This renormalization has been used to find a critical line at weak couplings that separates the topological insulator from a  gapped phase that can be adiabatically deformed into a trivial product state (i.e. trivial band insulator). Moreover, we have shown that for sufficiently-strong interactions,  a gapped phase where parity symmetry is spontaneously broken (i.e. Aoki phase) is formed due to the appearance of a pseudoscalar fermion condensate. The large-$N$ prediction has allowed us to find the    critical line separating the topological insulator from the Aoki phase by studying the onset of the  pseudoscalar  condensate, and show that it terminates at a tri-critical point where all these three phases of matter coexist.

By using both  Hamiltonian and Euclidean lattice approaches, we have been able to pinpoint important details that must be carefully considered when taking the time-continuum limit of the lattice approaches, such that standard methods of lattice field theories can be used to describe quantitatively the phase diagram of the Gross-Neveu-Wilson Hamiltonian. In particular, we have described how  lattice artifacts can appear in the standard dimensionless formulation  of the Euclidean field theory, and how the spurious time doublers, even when residing at the cutoff of the theory, can renormalize the bare parameters and introduce qualitative modifications to the layout of the phases. The results hereby presented will serve as the starting point for the application of other well-established Euclidean lattice techniques to explore the phenomenology of leading-order corrections that appear for finite $N$.

Motivated by the possibility of implementing a cold-atom quantum simulator of the Gross-Neveu-Wilson model for a single flavor $N=1$, which has also been described in this work, we have benchmarked these large-$N$ predictions by means of quasi-exact numerical methods based on matrix product states. In particular, we have shown that the single-flavor model, corresponding to a discretized version of the massive Thirring model, can also be mapped into a condensed-matter Hamiltonian of spinless fermions hopping on a two-leg ladder, and interacting via Hubbard-type couplings. This connection has allowed us to identify the phases of the Gross-Neveu-Wilson model, discussed above, with condensed-matter counterparts that include orbital paramagnets and ferromagnets, as well as a chiral-unitary topological phase. In this way, the matrix-product-state simulations can readily access a variety of observables to determine the position of the critical lines, which show a remarkable qualitative agreement with the large-$N$ predictions that becomes even quantitative in the region where the continuum Gross-Neveu QFT is expected to emerge (i.e. weak couplings). These numerical simulations also prove that the symmetry of the large-$N$ phase diagram holds for $N=1$, and should then be maintained at all orders  $\mathcal{O}(1/N^\alpha)$. Beyond the matrix-product-state simulations, the aforementioned mapping has allowed us to import exact results  for the Gross-Neveu-Wilson model in the regime of intermediate and strong couplings, which originate from quantum-impurity and quantum magnetism techniques in condensed matter. 

 Therefore, we believe that our work constitutes an  example of the useful dialogue and exchange of ideas between the high-energy physics, condensed-matter, quantum-information, and quantum optics communities, stimulating further cross-disciplinary efforts in the future. As an outlook, one can easily foresee that lattice field-theory techniques to study leading-order corrections to the large-$N$ behavior will be very useful to elucidate the mechanism that induces strong correlations in the symmetry-protected topological phase of the Gross-Neveu-Wilson model. Likewise, quantum-information approaches might be useful to understand the entanglement content of those phases, making a connection to the lattice field-theory techniques. As already pointed out by the Euclidean lattice results, new phases of the Gross-Neveu-Wilson model can arise as one moves  away from half-filling. It will be very interesting to explore the nature of these phases using some of the high-energy and condensed-matter techniques hereby discussed. We also note that the techniques hereby presented can be generalized to other lattice Hubbard-type models, not necessarily connected to well-known relativistic QFTs. In particular, it will be very interesting to apply them to the study of higher-dimensional models  hosting topological phases of matter.  In this context, Aoki phases have been identified in the limit of very-strong Coulomb interactions via strong-coupling techniques  of lattice gauge theories~\cite{strong_coupling_wilson_2d_U(1), strong_coupling_wilson_3d_U(1)}. These results have been used to conjecture the qualitative shape of the phase diagram in the regime of weak to intermediate inetractions~\cite{strong_coupling_wilson_2d_U(1)}, which is expected to be more relevant for the understanding of correlation effects in topological insulating materials.
 
\section*{Acknowledgments}

A.B. acknowledges support from the Ram\'on y Cajal program under RYC-2016-20066, Spanish MINECO project FIS2015-70856-P,  and CAM PRICYT project  QUITEMAD+ S2013/ICE-2801. E.T. and M.L. acknowledge the Spanish Ministry MINECO (National Plan 15 Grant: FISICATEAMO No. FIS2016-79508-P, SEVERO OCHOA No. SEV-2015-0522, FPI), European Social Fund, Fundaci\'o Cellex, Generalitat de Catalunya (AGAUR Grant No. 2017 SGR 1341 and CERCA/Program), ERC AdG OSYRIS, EU FETPRO QUIC, and the National Science Centre, Poland-Symfonia Grant No. 2016/20/W/ST4/00314. S.H. is supported in part by STFC grant ST/P00055X/1. Some of the MPS simulations were run on the Mogon cluster of the JGU (made available by the CSM and AHRP), with a code based on a flexible Abelian Symmetric Tensor Networks Library, developed in collaboration with the group of S. Montangero at the University of Ulm. We thank Gert Aarts for valuable discussions.



\begin{thebibliography}{100}

\bibitem{sachdev}
S. Sachdev, {\it Quantum Phase Transitions} (Cambridge University
Press, Cambridge, 1999).

\bibitem{hands_phases_qcd}
S. Hands, \href{https://www.tandfonline.com/doi/abs/10.1080/00107510110063843}{Contemp. Phys. {\bf 42,} 209 (2001).}

\bibitem{yoshida}
B. Yoshida, \href{https://www.sciencedirect.com/science/article/pii/S0003491610001867#%21}{Ann. Phys. {\bf 326,}  15 (2011)}.



\bibitem{landau_sb}
L. Landau, Zh. Eksp. Teor. Fiz. 7, 19 (1937) [Phys. Z. Sowjetunion 11, 26 (1937)]

\bibitem{wilson_kogut}
K. G. Wilson, and J. Kogut, \href{https://www.sciencedirect.com/science/article/pii/0370157374900234}{Phys. Rep. {\bf 12,}  75 (1974)}.

\bibitem{wen_book}
X. Wen , {\it Quantum Field Theory of Many-Body
Systems: From the Origin of Sound to an Origin of Light and Electrons} ( Oxford University Press, Oxford, 2004).  

\bibitem{chaos_book}
F. Haake, {\it Quantum Signatures of Chaos,}, (Springer Verlag, Berlin, 2001).

\bibitem{az_ten_fold}
A. Altland, and M. R. Zirnbauer, \href{https://journals.aps.org/prb/abstract/10.1103/PhysRevB.55.1142}{Phys. Rev. B {\bf 55,} 1142 (1997)}.



\bibitem{table_top_insulators}
A. P. Schnyder, S. Ryu, A. Furusaki, and  A. W. W. Ludwig, \href{https://journals.aps.org/prb/abstract/10.1103/PhysRevB.78.195125}{  Phys. Rev. B {\bf 78,} 195125 (2008)};  A. Y. Kitaev, \href{http://aip.scitation.org/doi/abs/10.1063/1.3149495}{
 AIP Conf. Proc. {\bf 1134,} 22 (2009)}.
 
\bibitem{chern_tknn}
D. J. Thouless, M. Kohmoto, M. P. Nightingale, and M. den Nijs, \href{http://journals.aps.org/prl/abstract/10.1103/PhysRevLett.49.405}{Phys. Rev. Lett. {\bf 49,} 405 (1982)}.

\bibitem{haldane}
F. D. M. Haldane,  \href{https://journals.aps.org/prl/abstract/10.1103/PhysRevLett.61.2015}{Phys. Rev. Lett. {\bf 61,} 2015 (1988)}.

\bibitem{kane_mele}
C. L. Kane and E. J. Mele, \href{https://journals.aps.org/prl/abstract/10.1103/PhysRevLett.95.146802}{Phys. Rev. Lett. {\bf 95,} 146802 (2005)}.

 \bibitem{spt_review}
{\it See } T. Senthil, \href{https://doi.org/10.1146%2Fannurev-conmatphys-031214-014740.}{Ann. Rev.  Cond. Matt. Phys. {\bf 6,} 299 (2015)}, {\it and references therein}.
 
 \bibitem{mps}
{\it See }U. Schollwoeck, \href{https://linkinghub.elsevier.com/retrieve/pii/S0003491610001752}{ Ann. Phys. {\bf 326,} 96 (2011)}; R. Orus, \href{https://arxiv.org/ct?url=http%3A%2F%2Fdx.doi.org%2F10%252E1016%2Fj%252Eaop%252E2014%252E06%252E013&v=86bed539}{Ann.Phys. {\bf 349,}  117 (2014)}; P. Silvi, F. Tschirsich, M. Gerster, J. J\"{u}nemann, D. Jaschke, M. Rizzi, and S. Montangero, \href{https://arxiv.org/abs/1710.03733}{arXiv:1710.03733 (2017)}; {\it and references therein}.

\bibitem{classification_ti}
 {\it See } C.-K Chiu, J. C. Y. Teo, A. P. Schnyder, and S. Ryu, \href{https://journals.aps.org/rmp/abstract/10.1103/RevModPhys.88.035005}{Rev. Mod. Phys. {\bf 88,} 035005 (2016)}, {\it and references therein}.


\bibitem{zhang}
L. Zhang, L. Zhang, S. Niu, and X.-J. Liu,  \href{https://arxiv.org/abs/1802.10061}{arXiv:1802.10061 (2018)}.

\bibitem{classification_1d}
N. Schuch, D. Perez-Garcia, J. I. Cirac, \href{https://journals.aps.org/prb/abstract/10.1103/PhysRevB.84.165139}{Phys. Rev. B {\bf 84,} 165139 (2011)}; X. Chen, Z.-C. Gu, and X.-G. Wen, \href{https://journals.aps.org/prb/abstract/10.1103/PhysRevB.84.235128}{Phys. Rev. B {\bf 84,} 235128 (2011)}.


\bibitem{GN_model}
D. J. Gross and A. Neveu, \href{https://journals.aps.org/prd/abstract/10.1103/PhysRevD.10.3235}{Phys. Rev. D {\bf 10,} 3235 (1974)}.

\bibitem{wilson_fermions}
K. Wilson, New Phenomena in Subnuclear Physics. (ed. A.
Zichichi, Plenum, New York, 1977).


\bibitem{NJL_model}
Y. Nambu and G. Jona-Lasinio, \href{https://journals.aps.org/pr/abstract/10.1103/PhysRev.122.345}{Phys. Rev. {\bf 122,} 345 (1961)}.

\bibitem{NJL_review}
{\it See} S. P. Klevansky, \href{https://journals.aps.org/rmp/abstract/10.1103/RevModPhys.64.649}{Rev. Mod. Phys. {\bf 64,} 649 (1992)}, {\it and references therein}.

\bibitem{yang_mills}
 C. Yang, and R.  Mills,  \href{https://journals.aps.org/pr/abstract/10.1103/PhysRev.96.191}{Phys. Rev. {\bf 96,}  191 (1954)}.

\bibitem{qcd}
D.J. Gross and F. Wilczek, \href{https://journals.aps.org/prl/abstract/10.1103/PhysRevLett.30.1343}{Phys. Rev. Lett. {\bf 30,}1343 (1973)}; H.D. Politzer \href{https://journals.aps.org/prl/abstract/10.1103/PhysRevLett.30.1346}{Phys. Rev. Lett. {\bf 30,} 1346 (1973).}

\bibitem{gawedzki}
C. Kopper, J. Magnen, and V. Rivasseau, \href{https://link.springer.com/article/10.1007/BF02101599}{Commun.  Math. Phys. {\bf 169,} 121 (1995)}; E. Pereira, and A. Procacci, \href{https://www.sciencedirect.com/science/article/pii/S0003491696956511}{Ann.  Phys. {\bf 255,} 19 (1997)};  K. Gaw\c edzki and A. Kupiainen, \href{https://link.springer.com/article/10.1007/BF01208817#citeas}{Commun. Math. Phys. {\bf 102,}  1  (1985).}

\bibitem{note_gamma}
In the Hamiltonian formulation,  one also uses the Dirac matrices $\boldsymbol{\alpha},\beta$, such that  $\mathcal{H}_{\rm D}(x)=\sum_n\psi_n^\dagger(x)(\ii\boldsymbol{\alpha}\cdot\boldsymbol{\nabla}+\beta m)  \psi_n(x)$~\cite{rqm_book}. In the one-dimensional case, one can chose $\alpha_x=\sigma^x$  and $\beta=\sigma^z$, such that the gamma matrices are  defined as $\gamma^0=\beta$ and $\gamma^1=\beta\alpha_x=\ii\sigma^y$. Accordingly, the chiral gamma matrix $\gamma^5=\gamma^0\gamma^1=\sigma^x$ coincides with the $\alpha_x$ Dirac matrix. The time-reversal $T$ and charge-conjugation $C$ matrices must fulfill $T\boldsymbol{\alpha}^*T^{\dagger}=-\boldsymbol{\alpha}$, $T\beta^*T^{\dagger}=\beta$, and $C\boldsymbol{\alpha}^*C^{\dagger}=\boldsymbol{\alpha}$, $C\beta^*C^{\dagger}=-\beta$~\cite{rqm_book}. In  $(3+1)$-dimensions, these equations are fulfilled by choosing $T=-\ii\alpha_x\alpha_z$, and $C=\ii\beta\alpha_y$. In the one-dimensional case, there is only one remaining Pauli matrix $\alpha_y=\sigma^y$, such that $C=\ii\sigma^z\sigma^y=\gamma^5$. The time-reversal condition can be fulfilled by taking $\alpha_z=\sigma^y$, such that $T=-\ii\sigma^x\sigma^y=\gamma^0$. The complete time-reversal, and charge-conjugation, anti-unitary operators lead to  $\psi_n(x^0,\boldsymbol{x})\to T\psi_n(-x^0,\boldsymbol{x})$, and $\psi_n(x^0,\boldsymbol{x})\to C\psi_n^*(x^0,\boldsymbol{x})$, such that $\overline{\psi}_n\psi_n\to\overline{\psi}_n\psi_n$, and $\overline{\psi}_n\gamma^5\psi_n\to\mp\overline{\psi}_n\gamma^5\psi_n$~\cite{qft_book}.



\bibitem{rqm_book}
P. Strange, {\it Relativistic quantum mechanics}, (Cambridge University Press, Cambridge, 2005).

\bibitem{qft_book}
M. E. Peskin and D. V. Schroeder, {\it An introduction to quantum field theory} , (Adison Wesley, Reading, 1995).

\bibitem{lattice_book}
J. Smit, {\it Introduction to Quantum Fields on a Lattice} (Cambridge Lecture Notes in Physics, Cambridge, 2003).

\bibitem{nn_theorem}
H. B. Nielsen and M. Ninomiya, \href{http://www.sciencedirect.com/science/article/pii/0550321381903618}{Nuc. Phys. B {\bf 185,} 20 (1981)}; ibid,  \href{http://www.sciencedirect.com/science/article/pii/0550321381905241}{Nuc. Phys. B {\bf 193,} 173 (1981)}.


\bibitem{susskind_1d}
L. Susskind, \href{https://journals.aps.org/prd/abstract/10.1103/PhysRevD.16.3031}{Phys. Rev. D, {\bf 16,} 3031 (1977)}.





\bibitem{review_top_insulators}
{\it See} M. Z. Hasan and C. L. Kane, \href{https://journals.aps.org/rmp/abstract/10.1103/RevModPhys.82.3045}{Rev. Mod. Phys. {\bf 82,} 3045 (2010)}; X.-L. Qi and S.-C. Zhang, \href{https://journals.aps.org/rmp/abstract/10.1103/RevModPhys.83.1057}{Rev. Mod. Phys. {\bf 83,} 1057 (2011)}; {\it and references therein}.

\bibitem{Bernevig}
B. A.~Bernevig and T. L.~Hughes, {\it Topological Insulators and Topological
Superconductors}, (Princeton, 2013).




\bibitem{10_fold_ryu}
S. Ryu, A. P. Schnyder, A.
Furusaki,  and A. W. W. Ludwig, \href{http://iopscience.iop.org/article/10.1088/1367-2630/12/6/065010/meta}{New J. Phys. {\bf 12,} 065010 (2010)}.

\bibitem{iqhe_exp}
K. Klitzing, G. Dorda, M. Pepper, \href{http://journals.aps.org/prl/abstract/10.1103/PhysRevLett.45.494}{Phys. Rev. Lett. {\bf 45,} 494 (1980)}.




\bibitem{edge_iqhe}
B. I. Halperin, \href{http://journals.aps.org/prb/abstract/10.1103/PhysRevB.25.2185}{Phys. Rev. B {\bf 25,} 2185 (1982)}.



\bibitem{wilson_atoms}
A. Bermudez, L. Mazza, M. Rizzi, N. Goldman, M. Lewenstein, and M.A. Martin-Delgado, \href{https://journals.aps.org/prl/abstract/10.1103/PhysRevLett.105.190404}{Phys. Rev. Lett. {\bf 105,} 190404 (2010)}; L. Mazza, A. Bermudez, N. Goldman, M. Rizzi, M.A.
Martin-Delgado, and M. Lewenstein, \href{http://iopscience.iop.org/article/10.1088/1367-2630/14/1/015007}{New J. Phys. {\bf 14,} 015007 (2012)}.


\bibitem{strong_coupling_wilson_3d_U(1)}
A. Sekine, T. Z. Nakano, Y. Araki, and K. Nomura, \href{https://journals.aps.org/prb/abstract/10.1103/PhysRevB.87.165142}{Phys. Rev. B {\bf 87,} 165142 (2013)}.

\bibitem{domain_wall}
D. B. Kaplan, \href{http://www.sciencedirect.com/science/article/pii/037026939291112M?via%3Dihub}{Phys. Lett. B {\bf 288,} 342 (1992)}.

\bibitem{zak_phase}
J. Zak, \href{https://journals.aps.org/prl/abstract/10.1103/PhysRevLett.62.2747}{Phys. Rev. Lett. {\bf 62,} 2747 (1989)}.

\bibitem{polarization}
R. Resta, \href{https://www.tandfonline.com/doi/abs/10.1080/00150199208016065}{Ferroelectrics {\bf 136,} 51 (1992)}; R. D. King-Smith and D. Vanderbilt,  \href{https://journals.aps.org/prb/abstract/10.1103/PhysRevB.47.1651}{Phys. Rev. B {\bf 47,} 1651 (1993).}

\bibitem{polarization_zak}
{\it See} D. Xiao, M.-C. Chang, and Q. Niu, \href{https://journals.aps.org/rmp/abstract/10.1103/RevModPhys.82.1959}{Rev. Mod. Phys. {\bf 82,} 1959 (2010)}, {\it 
and references therein.}

\bibitem{TI_field_theories}
X.-L- Qi, T.L. Hughes, and S.-C. Zhang, \href{https://journals.aps.org/prb/abstract/10.1103/PhysRevB.78.195424}{Phys. Rev. B {\bf 78,} 195424 (2008)}.

 

\bibitem{ti_interactions_reviews} {\it See,} M. Hohenadler, and F. F. Assaad,  \href{http://iopscience.iop.org/article/10.1088/0953-8984/25/14/143201/meta}{J. Phys.: Condens. Matter {\bf 25,} 143201 (2013)}; S.A. Parameswaran, R. Roy, and S.L. Sondhi, \href{https://www.sciencedirect.com/science/article/pii/S163107051300073X}{Compt. Rend. Phys. {\bf 14,} 816 (2013)};
S. Rachel, \href{https://arxiv.org/abs/1804.10656}{arXiv:1804.10656(2018)}, {\it and references therein}.

\bibitem{hamiltonian_lgt}
J. Kogut and L. Susskind, \href{https://journals.aps.org/prd/abstract/10.1103/PhysRevD.11.395}{Phys. Rev. D {\bf 11,} 395 (1975)}.

\bibitem{mps_lft}
 A. Milsted, J. Haegeman, and T. J. Osborne, \href{https://journals.aps.org/prd/abstract/10.1103/PhysRevD.88.085030}{Phys. Rev. D {\bf 88,} 085030 (2013)}; M. Ba\~nuls, K. Cichy, J. I, Cirac, and K. Jansen, \href{https://link.springer.com/article/10.1007/JHEP11(2013)158}{J. High Energ. Phys.  {\bf 158,} (2013)}; B. Buyens, J. Haegeman, K. Van Acoleyen, H. Verschelde, and F. Verstraete, \href{https://journals.aps.org/prl/abstract/10.1103/PhysRevLett.113.091601}{Phys. Rev. Lett. {\bf 113,} 091601 (2014)}; E. Rico, T. Pichler, M. Dalmonte, P. Zoller, and S. Montangero, \href{https://journals.aps.org/prl/abstract/10.1103/PhysRevLett.112.201601}{
Phys. Rev. Lett. {\bf 112,} 201601  (2014)}; L. Tagliacozzo, A. Celi, and M. Lewenstein, \href{https://journals.aps.org/prx/abstract/10.1103/PhysRevX.4.041024}{
Phys. Rev. X {\bf 4,} 041024 (2014)}.



\bibitem{qs_feynman}
R. P. Feynman, \href{https://link.springer.com/article/10.1007%2FBF02650179}{Int. J. Theor. Phys. {\bf 21, }467 (1982)}.

\bibitem{qs_lloyd}
S. Lloyd, \href{http://science.sciencemag.org/content/273/5278/1073}{Science {\bf 273,} 1073 (1996)}.



\bibitem{qs_cold_atoms_1}
{\it See} M. Lewenstein, A. Sanpera, V. Ahufinger, B. Damski, A.
Sen, and U. Sen, Adv. in Phys. {\bf 56,} 243 (2007), {\it and references
therein}.

\bibitem{qs_cold_atoms_2}
{\it See} I. Bloch, J. Dalibard, and S. Nascimbene, \href{http://www.nature.com/nphys/journal/v8/n4/full/nphys2259.html}{Nat. Phys. {\bf 8,} 267
(2012)}, {\it and references therein}.

\bibitem{cold_atoms_review}
{\it See} I. Bloch, J. Dalibard, and W. Zwerger, \href{https://journals.aps.org/rmp/abstract/10.1103/RevModPhys.80.885}{Rev. Mod. Phys. {\bf 80,}
885 (2008)}, {\it and references therein}.

\bibitem{su_N_review}
{\it See} M. A. Cazalilla and A. M. Rey, \href{http://iopscience.iop.org/article/10.1088/0034-4885/77/12/124401/meta}{Rep. Prog. Phys. {\bf 77,} 124401 (2014)}, {\it and references therein}.

\bibitem{hubbard_atoms}
D. Jaksch, C. Bruder, J. I. Cirac, C. W. Gardiner, and P. Zoller, \href{https://journals.aps.org/prl/abstract/10.1103/PhysRevLett.81.3108}{
Phys. Rev. Lett. {\bf 81,} 3108 (1998)}; W. Hofstetter, J. I. Cirac, P. Zoller, E. Demler, and M. D. Lukin,
\href{https://journals.aps.org/prl/abstract/10.1103/PhysRevLett.89.220407}{Phys. Rev. Lett. {\bf 89,} 220407 (2002)}; A. Albus, F. Illuminati, and J. Eisert, \href{https://arxiv.org/ct?url=http%3A%2F%2Fdx.doi.org%2F10%252E1103%2FPhysRevA%252E68%252E023606&v=ffed227b}{Phys. Rev. A {\bf 68,} 023606 (2003)}.

\bibitem{hubbard_exp}
M. Greiner, O. Mandel, T. Esslinger, T. W. H\"{a}nsch, and I.
Bloch, \href{http://www.nature.com/nature/journal/v415/n6867/abs/415039a.html}{Nature {\bf 415,} 39 (2002)}; R. J\"{o}rdens, N. Strohmaier, K. G\"{u}nter, H. Moritz, and T.
Esslinger, \href{http://www.nature.com/nature/journal/v455/n7210/full/nature07244.html}{Nature {\bf 455,} 204 (2008)}; U. Schneider, L. Hackerm\"{u}ller, S. Will, Th. Best, I. Bloch, T. A. Costi, R. W. Helmes, D. Rasch, and A. Rosch, \href{http://science.sciencemag.org/content/322/5907/1520}{Science {\bf 322,} 1520 (2008)}.

\bibitem{hubbard_toolbox}
D. Jaksch, and  P. Zoller, \href{https://www.sciencedirect.com/science/article/pii/S0003491604001782}{Ann.  Phys. {\bf 315,} 52 (2005)}.




\bibitem{aqs_qft_fermions_2d} 
N. Goldman, A. Kubasiak, A. Bermudez, P. Gaspard, M. Lewenstein, and M. A. Martin-Delgado, \href{https://journals.aps.org/prl/abstract/10.1103/PhysRevLett.103.035301}{Phys. Rev. Lett. {\bf 103,} 035301 (2009)}; 
A. Bermudez, N. Goldman, A. Kubasiak, M. Lewenstein, and M. A. Martin-Delgado, \href{http://iopscience.iop.org/article/10.1088/1367-2630/12/3/033041/meta}{New J. Phys. {\bf 12,} 033041 (2010)}; Z. Lan, N. Goldman, A. Bermudez, W. Lu, and P. \"{O}hberg, \href{https://journals.aps.org/prb/abstract/10.1103/PhysRevB.84.165115}{Phys. Rev. B {\bf 84,}  165115 (2011).}



\bibitem{aqs_qft_fermions_interactions} 
J. I. Cirac, P. Maraner, and J. K. Pachos, \href{http://journals.aps.org/prl/abstract/10.1103/PhysRevLett.105.190403}{Phys. Rev. Lett. {\bf 105,} 190403 (2010)}.



\bibitem{dqs_qft_phi4}
S. P. Jordan, K. S. Lee, and J. Preskill, \href{http://science.sciencemag.org/content/336/6085/1130}{Science {\bf 336,} 1130 (2012)}; {\it ibid.}, \href{http://dl.acm.org/citation.cfm?id=2685163&CFID=683947881&CFTOKEN=27608053}{Quant. Inf. and Comp. {\bf 14,} 1014 (2014)}.

\bibitem{O_N_models}
H. Zou, Y. Liu, C.-Y. Lai, J. Unmuth-Yockey, L.-P. Yang, A. Bazavov, Z. Y. Xie, T. Xiang, S. Chandrasekharan, S. -W. Tsai,  and Y. Meurice, \href{https://journals.aps.org/pra/abstract/10.1103/PhysRevA.90.063603}{Phys. Rev. A {\bf 90,} 063603 (2014)}.

\bibitem{aqs_phi_4}
A. Bermudez, G. Aarts, and M. M\"{u}ller, \href{https://journals.aps.org/prx/abstract/10.1103/PhysRevX.7.041012}{Phys. Rev. X {\bf 7,} 041012 (2017)}.


 \bibitem{dqs_lattice_gauge_theories}
 T. Byrnes and Y. Yamamoto, \href{http://journals.aps.org/pra/abstract/10.1103/PhysRevA.73.022328}{Phys. Rev. A {\bf 73,} 022328 (2006)}.
 


\bibitem{aqs_qft_gauge_U(1)_qlinks} 
H. P. B\"{u}chler, M. Hermele, S. D. Huber, M. P. A. Fisher, and P. Zoller, \href{http://journals.aps.org/prl/abstract/10.1103/PhysRevLett.95.040402}{Phys. Rev. Lett. {\bf 95,} 040402 (2005)}.

\bibitem{aqs_qft_gauge_abelian_qlinks}
H. Weimer, M. M\"{u}ller, I. Lesanovsky, P. Zoller, and H. P. B\"{u}chler, \href{http://www.nature.com/nphys/journal/v6/n5/abs/nphys1614.html}{Nat. Phys. {\bf 6,} 382 (2010)}; L. Tagliacozzo, A. Celi, A. Zamora, and M. Lewenstein,  \href{http://www.sciencedirect.com/science/article/pii/S0003491612001819}{Ann. Phys. {\bf 330,} 160 (2013)}.

\bibitem{dqs_qft_SU(2)_qlinks}
L.Tagliacozzo, A. Celi, P. Orland, M. W. Mitchell, and M. Lewenstein, \href{http://www.nature.com/articles/ncomms3615?message-global=remove}{Nat. Comm. {\bf 4,} 2615 (2013)}.

\bibitem{compresed_qs_gauge}
D. B. Kaplan and J. R. Stryker, \href{https://arxiv.org/abs/1806.08797}{arXiv:1806.08797 (2018)}.

\bibitem{gauge_higgs_atoms}
K. Kasamatsu, I. Ichinose, and T. Matsui, \href{https://journals.aps.org/prl/abstract/10.1103/PhysRevLett.111.115303}{
Phys. Rev. Lett. {\bf 111,} 115303 (2013)};  
Y. Kuno, K. Kasamatsu, Y. Takahashi, I. Ichinose, and T. Matsui, \href{https://www.google.com/search?client=safari&rls=en&q=Real+time+dynamics+and+proposal+for+feasible+experiments+of+lattice+gauge-Higgs+model+simulated+by+cold+atoms&ie=UTF-8&oe=UTF-8}{New J. Phys. {\bf 17,} 063005 (2015)}; Y. Kuno, S. Sakane, K. Kasamatsu, I. Ichinose, and T. Matsui, \href{https://journals.aps.org/prd/abstract/10.1103/PhysRevD.95.094507}{Phys. Rev. D {\bf 95,} 094507 (2017).}

\bibitem{gauge_higgs_meurice}
A. Bazavov, Y. Meurice, S.-W. Tsai, J. Unmuth-Yockey, and J. Zhang, \href{https://doi.org/10.1103/PhysRevD.92.076003}{ Phys. Rev. D {\bf 92,} 076003 (2015)}; J. Zhang, J. Unmuth-Yockey, J. Zeiher, A. Bazavov, S. -W. Tsai, and Y. Meurice, \href{https://arxiv.org/abs/1803.11166}{arXiv:1803.11166 (2018)}.

\bibitem{aqs_QED}
 E. Kapit, and E. Mueller, \href{http://journals.aps.org/pra/abstract/10.1103/PhysRevA.83.033625}{Phys. Rev. A {\bf 83,} 033625 (2011)};  E. Zohar, and B. Reznik, \href{http://journals.aps.org/prl/abstract/10.1103/PhysRevLett.107.275301}{Phys. Rev. Lett. {\bf 107,} 275301 (2011)}; E. Zohar, J. Cirac, and B. Reznik, \href{http://journals.aps.org/prl/abstract/10.1103/PhysRevLett.109.125302}{Phys. Rev. Lett. {\bf 109,} 125302 (2012)}.

\bibitem{aqs_qft_fermions_U(1)_qlinks}
D. Banerjee, M. Dalmonte, M. M\"{u}ller, E. Rico, P. Stebler, U.-J. Wiese, and P. Zoller, \href{http://journals.aps.org/prl/abstract/10.1103/PhysRevLett.109.175302}{Phys. Rev. Lett. {\bf 109,}175302  (2012)}; E. Zohar, J. Cirac, and B. Reznik, \href{http://journals.aps.org/prl/abstract/10.1103/PhysRevLett.110.055302}{Phys. Rev. Lett. {\bf 110,} 055302 (2013)}; G. Magnifico, D. Vodola, E. Ercolessi, S. P. Kumar, M. M\"{u}ller, and A. Bermudez,  \href{https://arxiv.org/abs/1804.10568}{arXiv:1804.10568 (2018)}.




\bibitem{aqs_qft_fermions_SU(2)_qlinks}
 D. Banerjee, M. B\"{o}gli, M. Dalmonte, E. Rico, P. Stebler, U.-J. Wiese, and P. Zoller, \href{http://journals.aps.org/prl/abstract/10.1103/PhysRevLett.110.125303}{Phys. Rev. Lett. {\bf 110,} 125303 (2013)}; E. Zohar, J. Cirac, and B. Reznik, \href{http://journals.aps.org/prl/abstract/10.1103/PhysRevLett.110.125304}{Phys. Rev. Lett. {\bf 110,} 125304 (2013)}.




\bibitem{GN_staggered}
J. I. Cirac, P. Maraner, and J. K. Pachos, \href{https://journals.aps.org/prl/abstract/10.1103/PhysRevLett.105.190403}{Phys. Rev. Lett. {\bf 105,} 190403 (2010)}.

\bibitem{creutz_hubbard}
J. J\"unemann, A. Piga, S.-J. Ran, M. Lewenstein, M. Rizzi, and A. Bermudez, \href{https://journals.aps.org/prx/abstract/10.1103/PhysRevX.7.031057}{ Phys. Rev. X {\bf 7,} 031057 (2017)}.

\bibitem{wilson_kuno}
Y. Kuno, I. Ichinose, and Y. Takahashi, \href{https://arxiv.org/abs/1801.00439}{arXiv:1801.00439 (2018)}.


\bibitem{wilson_hauke}
T. V. Zache, F. Hebenstreit, F. Jendrzejewski, M. K. Oberthaler, J. Berges, and P. Hauke, \href{https://arxiv.org/abs/1802.06704}{arXiv:1802.06704 (2018)}.



\bibitem{Eguchi:1983gq}
  T.~Eguchi and R.~Nakayama, \href{https://www.sciencedirect.com/science/article/pii/0370269383900242}{  
  Phys.\ Lett.\  {\bf 126B} 89 (1983).} 

\bibitem{large_N_wilson_GN}
S. Aoki and K. Higashijima, \href{https://academic.oup.com/ptp/article/76/2/521/1848883/The-Recovery-of-the-Chiral-Symmetry-in-Lattice}{Prog. Theor. Phys. {\bf 76,} 521 (1986)}.


\bibitem{aoki_phases_finite_T}
S. Aoki, \href{https://journals.aps.org/prd/abstract/10.1103/PhysRevD.30.2653}{Phys. Rev. D {\bf 30,} 2653 (1984)}; T. Izubuchi, J. Noaki, and A. Ukawa, 	\href{https://journals.aps.org/prd/abstract/10.1103/PhysRevD.58.114507}{Phys. Rev. D {\bf 58,} 114507 (1998)}.

\bibitem{strong_coupling_wilson_2d_U(1)}
Y. Araki and T. Kimura, \href{https://journals.aps.org/prb/abstract/10.1103/PhysRevB.87.205440}{Phys. Rev. B {\bf 87,} 205440 (2013)}.

\bibitem{comment_grassmann}
We use the notation ${\Psi}^\star_k(\tau)$ for  the Grassmann variable associated to ${\Psi}^\dagger_k$, and refrain from   using $\overline{\Psi}_k(\tau)$ to avoid possible confusions when comparing the  functional approaches based on Hamiltonian and Euclidean lattice field theories.

\bibitem{special_functions}
{\it NIST Handbook of Mathematical Functions}, eds. F.W.J. Oliver, D. W. Lozier, 
R. F. Boisvert, and
Charles W. Clark, (Cambridge University Press, Cambridge, 2010).

\bibitem{detar_book}
T. DeGrand, and C. DeTar, {\it Lattice methods for quantum chromodynamics}, (World Scientific, New Jersey, 2006). 

\bibitem{aoki_phase_isotropic}
S. Aoki, A. Ukawa, and T. Umemura, \href{https://journals.aps.org/prl/abstract/10.1103/PhysRevLett.76.873}{Phys. Rev. Lett. {\bf 76,} 873 (1996)}.

\bibitem{Wolff:1985av}
  U.~Wolff, \href{https://www.sciencedirect.com/science/article/pii/0370269385906719}{
  Phys.\ Lett.\  {\bf 157B}  303 (1985).}
  
\bibitem{Hasenfratz:1983ba}
  P.~Hasenfratz and F.~Karsch, \href{https://www.sciencedirect.com/science/article/pii/037026938391290X}{
  Phys.\ Lett.\  {\bf 125B}  308 (1983)}.



  

  \bibitem{Hands:1992ck}
 I.~Barbour, S.~Hands, J.B.~Kogut, M.P.~Lombardo and S.~Morrison,\href{https://www.sciencedirect.com/science/article/pii/S0550321399004046}{Nucl.\ Phys.\ B {\bf 557} 327 (1999)}; S.~Hands, A.~Kocic and J.B.~Kogut, \href{https://www.sciencedirect.com/science/article/pii/0550321393904607/pdf?md5=884c2f4b3bc8f33c2ae46a75a56fc5c3&pid=1-s2.0-0550321393904607-main.pdf&_valck=1}{
  Nucl.\ Phys.\ B {\bf 390}  355 (1993).}



\bibitem{dmrg}
 S. R. White,  \href{https://journals.aps.org/prl/abstract/10.1103/PhysRevLett.69.2863}{Phys. Rev. Lett. {\bf 69,} 2863 (1992)}; {\it see }U. Schollw\"{o}ck, 
\href{https://journals.aps.org/rmp/abstract/10.1103/RevModPhys.77.259}{Rev. Mod. Phys. {\bf 77,} 259 (2005)}, {\it and references therein}.

\bibitem{thirring_model}
W. Thirring, \href{https://www.sciencedirect.com/science/article/pii/0003491658900150?via%3Dihub}{Ann. Phys. {\bf 3,} 91 (1958)}; S. Coleman, \href{https://journals.aps.org/prd/abstract/10.1103/PhysRevD.11.2088}{ Phys. Rev. D. {\bf 11,}  2088 (1975).}


\bibitem{kawa_smit}
N. Kawamoto, and J.Smit, \href{https://www.sciencedirect.com/science/article/pii/0550321381901966}{Nuc. Phys. B {\bf 192,}  100 (1981)}.

\bibitem{hubbard_model}
J. Hubbard, \href{http://rspa.royalsocietypublishing.org/content/276/1365/238}{Proc. R. Soc. London A {\bf 276,} 238 (1963)}.

\bibitem{facekas_book}
P. Fazekas, {\it Lecture Notes on Electron Correlation and Magnetism} (World Scientific Pub., London, 2003).


\bibitem{creutz_ladder}
M. Creutz, \href{https://journals.aps.org/prl/abstract/10.1103/PhysRevLett.83.2636}{Phys. Rev. Lett. {\bf 83,} 2636 (1999)}; 

\bibitem{creutz_ladder_refs}
A. Bermudez, D. Patane, L. Amico, and M. A. Martin-Delgado, \href{https://journals.aps.org/prl/abstract/10.1103/PhysRevLett.102.135702}{Phys. Rev. Lett. {\bf 102,} 135702 (2009)}; D. H\"{u}gel and B. Paredes, \href{https://journals.aps.org/pra/abstract/10.1103/PhysRevA.89.023619}{Phys. Rev. A {\bf 89,} 023619 (2014)}; N. Sun and L.-K. Lim, \href{https://journals.aps.org/prb/abstract/10.1103/PhysRevB.96.035139}{Phys. Rev. B {\bf 96,} 035139 (2017)}; M. Bischoff, J. J\"{u}nemann, M. Polini, and M. Rizzi, \href{https://journals.aps.org/prb/abstract/10.1103/PhysRevB.96.241112}{Phys. Rev. B {\bf 96,} 241112(R) (2017)}.





\bibitem{ent_spectrum}
H. Li and F. D. M. Haldane, \href{https://journals.aps.org/prl/abstract/10.1103/PhysRevLett.101.010504}{ Phys. Rev. Lett. {\bf 101,} 010504 (2015)}; F. Pollmann, A. M. Turner, E. Berg, and M. Oshikawa, \href{https://journals.aps.org/prb/abstract/10.1103/PhysRevB.81.064439}{Phys. Rev. B {\bf 81,} 064439 (2010)}.

\bibitem{ent_entropy}
G. Vidal, J. Latorre, E. Rico, and A. Kitaev, \href{https://journals.aps.org/prl/abstract/10.1103/PhysRevLett.90.227902}{Phys. Rev. Lett. {\bf 90,} 227902 (2003)}; P. Calabrese, and J. Cardy,  \href{http://iopscience.iop.org/article/10.1088/1742-5468/2004/06/P06002/meta}{J. Stat. Mech. P06002 (2004)}.

\bibitem{superexchange}
P.W. Anderson,  \href{https://journals.aps.org/pr/pdf/10.1103/PhysRev.115.2}{Phys. Rev. {\bf 115,} 2 (1959)}; A. H. MacDonald, S. M. Girvin, and D. Yoshioka, \href{https://journals.aps.org/prb/pdf/10.1103/PhysRevB.37.9753}{Phys. Rev. B {\bf 37,} 9753 (1988)}.

\bibitem{qim}
P. Pfeuty, \href{https://www.sciencedirect.com/science/article/pii/0003491670902708}{Ann. Phys. {\bf 57,} 79 (1970)}.

\bibitem{synthetic_dimension}
O. Boada, A. Celi, M. Lewenstein, and J. I. Latorre,  \href{https://journals.aps.org/prl/abstract/10.1103/PhysRevLett.108.133001}{Phys. Rev. Lett. {\bf 108,} 133001
(2012)}; A. Celi, P. Massignan, J. Ruseckas, N. Goldman, I.B. Spielman,
G. Juzeliunas, and M. Lewenstein, \href{https://journals.aps.org/prl/abstract/10.1103/PhysRevLett.112.043001}{Phys. Rev. Lett. {\bf 112,} 043001 (2014)}.

\bibitem{lattice_acceleration}
M. B. Dahan, E. Peik, J. Reichel, Y. Castin, and C. Salomon, Phys. Rev. Lett. {\bf 76,} 4508 (1996).

\bibitem{raman_assisted_scheme}
D. Jaksch and P. Zoller, \href{http://iopscience.iop.org/article/10.1088/1367-2630/5/1/356/meta}{New J. Phys. {\bf 5,} 56 (2003)}.

\bibitem{pat_gauge}
A. Bermudez, T. Schaetz, and D. Porras, \href{https://journals.aps.org/prl/abstract/10.1103/PhysRevLett.107.150501}{Phys. Rev. Lett. {\bf 107,} 150501 (2011)}; {\it ibid}. \href{http://stacks.iop.org/1367-2630/14/i=5/a=053049?key=crossref.d21f5876f060d0805821bef86dad9aa3}{New J. Phys. {\bf 14,} 053049 (2012)}.

\bibitem{pat_gauge_ol}
M. Aidelsburger, M. Atala, S. Nascimbene, S. Trotzky, Y.-A. Chen, and I. Bloch, \href{https://journals.aps.org/prl/abstract/10.1103/PhysRevLett.107.255301}{Phys. Rev. Lett. {\bf 107,} 255301 (2011)}; {\it ibid.} \href{https://link.springer.com/article/10.1007%2Fs00340-013-5418-1}{App. Phys.
B {\bf 113,} 1 (2013)}.

\bibitem{cdt}
D. H. Dunlap and V. M. Kenkre, \href{https://journals.aps.org/prb/abstract/10.1103/PhysRevB.34.3625}{Phys. Rev. B {\bf 34,} 3625 (1986)}; F. Grossman, T. Dittrich, P. Jung, and P. H\"{a}nggi, \href{https://journals.aps.org/prl/abstract/10.1103/PhysRevLett.67.516}{Phys. Rev. Lett. {\bf 67,}
516 (1991)}; M. Holthaus, \href{https://journals.aps.org/prl/abstract/10.1103/PhysRevLett.69.351}{Phys. Rev. Lett. {\bf 69,} 351 (1992)}.

\bibitem{ssh}
W. P. Su, J. R. Schrieffer, and A. J. Heeger, \href{https://journals.aps.org/prl/abstract/10.1103/PhysRevLett.42.1698}{Phys. Rev. Lett. {\bf 42,} 1698 (1979)}.

\bibitem{sp_creutz_atoms}
J. H. Kang, J. H. Han, and Y.-I. Shin, \href{https://arxiv.org/abs/1807.01444}{arXiv:1807.01444 (2018)}.

\bibitem{ghm_review}
{\it See} O. Dutta, M. Gajda, P. Hauke, M. Lewenstein, D.-S. L\"{u}hmann, B. A. Malomed, T. Sowi\'{n}ski, and J. Zakrzewski, \href{http://iopscience.iop.org/article/10.1088/0034-4885/78/6/066001/meta}{Rep. Prog. Phys. {\bf 78,} 066001 (2015)},  {\it and references
therein}.
  



\end{thebibliography}
\end{document}